%% file: MetastabilityOperational_arXiv2.tex
\documentclass[aps,pre,reprint,superscriptaddress,longbibliography,groupedaddress]{revtex4-1}

\usepackage{color}
\usepackage{amsmath}
\usepackage{bm}
\usepackage{amssymb}
\usepackage{wrapfig}
\usepackage{braket}
\usepackage{amsthm}
\usepackage{natbib}
\usepackage{dsfont}
\usepackage{graphicx}
\usepackage[colorlinks = true, citecolor = blue, linkcolor = blue, urlcolor=blue]{hyperref}

\definecolor{red}{rgb}{1.0, 0.0, 0.0}
\definecolor{green}{rgb}{0.0, 0.5, 0.0}
\definecolor{blue}{rgb}{0.0, 0.0, 0.5}

\DeclareFontFamily{OMX}{MnSymbolE}{}
\DeclareFontShape{OMX}{MnSymbolE}{m}{n}{
	<-6>  MnSymbolE5
	<6-7>  MnSymbolE6
	<7-8>  MnSymbolE7
	<8-9>  MnSymbolE8
	<9-10> MnSymbolE9
	<10-12> MnSymbolE10
	<12->   MnSymbolE12}{}
\DeclareSymbolFont{mnlargesymbols}{OMX}{MnSymbolE}{m}{n}
\SetSymbolFont{mnlargesymbols}{bold}{OMX}{MnSymbolE}{b}{n}
\DeclareMathDelimiter{\LL}{\mathopen}{mnlargesymbols}{'164}{mnlargesymbols}{'164}
\DeclareMathDelimiter{\rr}{\mathclose}{mnlargesymbols}{'171}{mnlargesymbols}{'171}

\usepackage{verbatim}
\usepackage[T1]{fontenc}
\usepackage[english]{babel}
\usepackage[utf8]{inputenc}

\newcommand{\Tr}{{\mathrm{Tr}}}

\renewcommand{\L}{\mathcal{L}}

\newcommand{\Pss}{\mathcal{P}_\text{\!ss}}
\newcommand{\rhoss}{\rho_\text{ss}}
\newcommand{\Oss}{O_\text{ss}}

\renewcommand{\P}{\mathcal{P}}

\renewcommand{\O}{\mathcal{O}}

\newcommand{\C}{\mathcal{C}}

\newcommand{\Cd}{\mathcal{C}_{\Delta}}

\newcommand{\I}{\mathcal{I}}

\newcommand{\T}{\mathcal{T}}

\makeatletter
\def\l@subsubsection#1#2{}
\makeatother
\makeatletter
\def\@bibdataout@aps{%
	\immediate\write\@bibdataout{%
		@CONTROL{%
			apsrev41Control%
			\longbibliography@sw{%
				,author="08",editor="1",pages="1",title="0",year="1"%
			}{%
				,author="08",editor="1",pages="1",title="",year="1"%
			}%
		}%
	}%
	\if@filesw \immediate \write \@auxout {\string \citation {apsrev41Control}}\fi 
}
\makeatother

\begin{document}
\title{
Operational approach to metastability
}

\author{Katarzyna Macieszczak}
\affiliation{TCM Group, Cavendish  Laboratory,  University  of  Cambridge,	J.  J.  Thomson  Ave.,  Cambridge  CB3  0HE,  United Kingdom}

\begin{abstract}
	In this work, we introduce an information-theoretic approach for considering changes in dynamics of finitely dimensional open quantum systems governed by master equations. This experimentally motivated approach arises from considering how the averages of system observables change with time and quantifies how non-stationary the system is during a given time regime. By drawing an analogy with the exponential decay, we are able to further investigate regimes when such changes are negligible according to the logarithmic scale of time, and thus the system is approximately stationary.  While this is always the case within the initial and final regimes of the dynamics, with the system respectively   approximated by its initial and asymptotic states, we show that a distinct regime of approximate stationarity may arise. In turn, we establish a quantitative description of the phenomenon of metastability in open quantum systems. The initial relaxation  occurring before the corresponding metastable regime and of the long-time dynamics taking place afterwards are also  characterised. Furthermore, we explain how metastability relates to the separation in the real part of the master equation spectrum and connect our approach to the spectral theory of metastability, 
	 clarifying when the latter follows.  	All of our general results directly translate to Markovian dynamics of classical stochastic systems.
\end{abstract}


\maketitle

\tableofcontents

\section{Introduction}

With continuing advances in the controllability of experimental
systems such as ultracold
atomic gases and Rydberg atoms as well as circuit quantum electrodynamics~\cite{Pritchard2010,Barreiro2011,Blatt2012,Britton2012,Dudin2012,Peyronel2012,Guenter2013,Schmidt2013}, it is now possible to observe a broad range of nonequilibrium
phenomena.
 In particular, open many-body quantum systems featuring both driving and dissipation give rise to stationary states no longer described by equilibrium distributions and phase diagrams featuring dissipative phase transitions~\cite{Tomadin2011,Diehl2008,Torre2013}. Furthermore, distinct timescales can arise in the relaxation towards stationary states, which are observed in experiments as plateaus in the dynamics of averages or time-correlations for system observables~\cite{Sciolla2015}. In equilibrium dynamics,  this phenomenon of metastability can be understood as a consequence of multiple local minima present in the system free energy function, which leads to the existence of multiple metastable phases different from stable phases. For non-equilibrium dynamics, however, such a general description linking dynamic and static properties is elusive. Nevertheless, the dynamics is governed by a master equation with stationary states corresponding to its zero eigenvalues, and if there exists a large enough separation in the real part of its spectrum, metastable states can be similarly associated with its small eigenvalues~\cite{Macieszczak2016a,Rose2016,Minganti2018,Macieszczak2020} (for classical stochastic systems, see Refs.~\cite{Gaveau1987,Gaveau1998,Bovier2002,Gaveau2006,Kurchan2016}).    
 
In this work, we propose  a general approach to metastability in finitely dimensional Markovian open quantum systems, which  does not assume a separation in the spectrum of the corresponding  master equation. Instead, motivated by experiments, we assume that changes in the system dynamics, quantified by changes in the averages of system observables, are negligible. This approach directly leads us to considering the distance in the space of states between system configurations at different times. Despite its abstractness, this perspective allows us to establish a simple but powerful analogy between open quantum dynamics and single-mode dynamics. 

First, we argue how considering changes with respect to the logarithmic rather than linear scale of time, as already typically done, is in fact necessary due to Markovianity of the dynamics. Second, we show that an open quantum system is  approximately stationary in the initial and final regimes of its dynamics, but there may exist a distinct time regime when the system changes negligibly and cannot be approximated by its initial or asymptotic states. This is achieved by obtaining bounds on the distance to initial and final regimes, which  do not rely on perturbative arguments nor depend on the dimension of the system space. Therefore, we arrive at a general quantitative description of the phenomenon of metastability.

It is important to note that, albeit indirectly, the spectrum of the master equation can be accessed experimentally by measuring observable averages. Thus,  our approach is also connected to the existence of a separation in the real part of the spectrum. In fact, we are able to derive general conditions on when the spectral theory as introduced in Refs.~\cite{Macieszczak2016a,Rose2016,Minganti2018,Macieszczak2020} is valid.

This paper is organized as follows. In Sec.~\ref{sec:master}, we review Markovian dynamics of open quantum systems. In Sec.~\ref{sec:single}, we consider single-mode dynamics and discuss regimes of its approximate stationary. 
In Sec.~\ref{sec:meta}, we introduce the approach for quantifying changes in the dynamics of an open quantum system. We then define approximate stationarity and show how it may also correspond to metastability, in which case we characterise resulting metastable states, as well as, discuss the initial relaxation and the long-time dynamics.
In Sec.~\ref{sec:spectral}, we further connect metastability to the existence of a separation in the spectral decomposition of the system evolution. Finally, in Sec.~\ref{sec:spectral_theory}, we clarify the relation between the operational approach and the spectral theory of metastability.

\section{Dynamics of open quantum systems}\label{sec:master}

In this work, we consider time-homogeneous Markovian dynamics~\cite{Lindblad1976,Gorini1976} of finitely dimensional open quantum systems. Here, we briefly review a general structure of such dynamics and discuss the corresponding spectral decomposition.  

\subsection{Open quantum system}

\subsubsection{Master operator}
An average state of the system at time $t$, described by a density matrix $\rho_t$, evolves according to a master equation 
\begin{equation}\label{eq:master}
	\frac{d}{dt}\rho_t=\L(\rho_t),
\end{equation}
 where the \emph{master operator}~\cite{Lindblad1976,Gorini1976}
\begin{equation}\label{eq:L}
\L(\rho) \equiv -i[H,\rho]+\sum_{j}\left[{J}_{j}\rho{J}_{j}^{\dagger}-\frac{1}{2}\left\{{J}_{j}^{\dagger}{J}_{j},\rho\right\}\right].
\end{equation}
Equation~\eqref{eq:L}  arises for systems interacting weakly with an effectively memoryless environment (see, e.g., Ref.~\cite{Gardiner2004}),  with $H$ denoting the system Hamiltonian and the jump operators ${J}_{j}$ providing coupling to the surrounding environment. 

\subsubsection{Dynamics}
Since the master operator $\L$ acts linearly on $\rho$, the system state at time $t$ in Eq.~\eqref{eq:master} is given by 
\begin{equation}\label{eq:rho_t}
\rho_t = e^{t\L}(\rho_{0}),
\end{equation}
where $\rho_0$ denotes an initial state and we refer to  $e^{t\L}$ an evolution operator.
As at any time $t$, $\rho_t$ is a density matrix, this requires the dynamics to preserve the Hermiticity  [$[\L(\rho)]^\dagger=\L(\rho^\dagger)$], the trace 
[$\L^\dagger(\mathds{1})=0$] and the positivity of operators, which (together with complete positivity; cf., e.g., Ref.~\cite{Wolf2012}) determines the structure of the master operator as that in Eq.~\eqref{eq:L}.

In this work we \emph{assume} that the system asymptotically reaches a time-independent state,
\begin{equation}\label{eq:rho_ss}
		\rhoss=\lim_{t\rightarrow\infty}\rho_t\equiv\Pss (\rho_{0}),
\end{equation}
which we refer to as a stationary state. As a stationary state depends linearly on the initial conditions, we introduce the projection $\Pss$ on the set of stationary states. When $\rhoss$ is unique, a generic situation in finite open many-body systems~\cite{Spohn1977,Evans1977,Schirmer2010,Nigro2019}, we simply have $\Pss(\rho)=\rhoss\Tr(\rho)$. In general, however, stationary states correspond to decoherence free subspaces~\cite{Zanardi1997,Zanardi1997a,Lidar1998}, noiseless subsystems~\cite{Knill2000,Zanardi2000} and disjoint stationary states (cf.~Ref.~\cite{Baumgartner2008}).

If asymptotic states are time dependent, the general structure of their manifold remains the same, while their evolution is that of unitary dynamics within decoherence free subspaces and noiseless subsystems. This case 
will be discussed elsewhere.

\section{Single-mode dynamics}\label{sec:single}

Before introducing the description of metastability in open quantum systems, 
we consider the dynamics of a single mode and discuss regimes of its approximate stationarity. 
Despite the simplicity of the considered dynamics, it is closely connected to the general evolution of an open quantum system as it shares its two crucial properties: the contractivity and its exponential form. Thus, the methods introduced here are directly applicable to the discussion of metastability that follows in Sec.~\ref{sec:meta}. 
Furthermore, such a single-mode dynamics can be observed in an open quantum system by measuring system observables corresponding to eigenmodes of the master operator, which links the phenomenon of metastability to the spectral properties of the evolution, as we demonstrate in  Sec.~\ref{sec:spectral}.

While the regimes of approximate stationarity of a single mode correspond to times 'before' and 'after' the dynamics and thus  can be considered trivial,  they highlight a key aspect of the characterising the system evolution with respect to the logarithmic rather than linear scale of time, which directly translates to the formal definition of metastability in Sec.~\ref{sec:meta}. 
Actually, as we show  in Sec.~\ref{sec:spectral},  metastability in an open quantum system  may only occur at the overlap of either of these regimes for all of eigenmodes present in the dynamics.

\begin{figure}[t!]
	\begin{center}
		\includegraphics[width=\columnwidth]{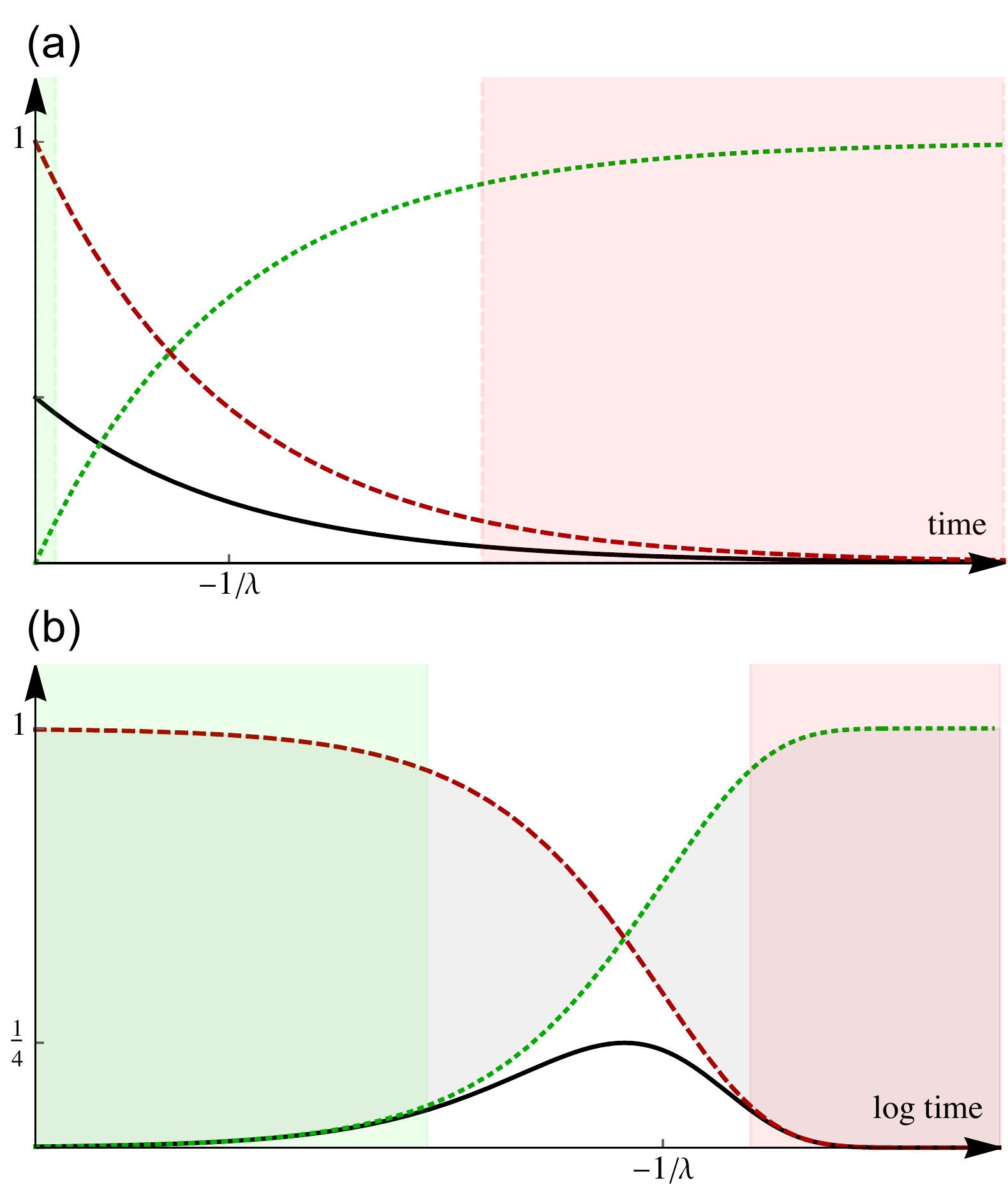} 
		\caption{
			\textbf{Single mode and its approximate stationarity}. A real single-mode dynamics is described by $e^{t\lambda}$ 	(red dashed) which decays with time $t$ for $\lambda<0$, so that the value of the function itself equals the distance to its asymptotic limit. As a consequence, the distance to its initial value of $1$ (green dotted) increases with time. In the initial and final regimes  $e^{t\lambda}$ is approximated by $1$ and $0$ respectively [areas shaded green and red, respectively, for $c=0.1$ in Eq.~\eqref{eq:t_0_ss}]. \textbf{(a)} The changes in the function (black solid) with respect to the linear scale of time, i.e., $e^{t\lambda}-e^{(t+\delta t)\lambda}$ with a fixed $\delta t\geq 0$, decay exponentially at the rate $\lambda$ from the initial value $1-e^{\delta t\lambda}$, so that  they are negligible only in the final regime. \textbf{(b)} In contrast, the changes with respect to the logarithmic scale of time, e.g., $e^{t\lambda}-e^{2t\lambda}$ (black solid), are negligible both in the initial and final regimes. Furthermore, they equal the product of the distances to the initial and asymptotic values (green dotted and red dashed) and are bounded by $1/4$. It follows that for $c_\Delta=e^{t\lambda}-e^{2t\lambda}$ we have $E_+(c_\Delta)=\max(e^{t\lambda},1-e^{t\lambda})$ and $E_-(c_\Delta)=\min(e^{t\lambda},1-e^{t\lambda})$, and thus the bounds in Eq.~\eqref{eq:change2_+-} saturate [excluded values shaded grey; cf.~Fig.~\ref{fig:example}(a)]. 
		}\vspace*{-7mm}
		\label{fig:exp}
	\end{center}
\end{figure}

\subsection{Dynamics} 

We now introduce the dynamics of a real single mode and identify initial and final regimes of its approximate stationarity. The dynamics is one dimensional and describes an exponential decay (see Figs.~\ref{fig:exp}). We discuss the case of a complex mode, which corresponds to a spiral motion in a plane, in Appendix~\ref{app:single}.

We consider the dynamics given by $e^{t\lambda}$, where we assume a negative rate  $\lambda<0$. The exponential behaviour is analogous to the open quantum system dynamics in Eq.~\eqref{eq:rho_t}, while the rate condition 
ensures a constant asymptotic limit as in Eq.~\eqref{eq:rho_ss}; cf.~Fig.~\ref{fig:exp}(a).

As the only timescale  is given by $-1/\lambda$, there are two regimes of approximate stationarity in the dynamics of a single mode (see Fig.~\ref{fig:exp}).
The \emph{initial regime} takes place when the dynamics can be neglected and the function is approximated by its initial value $1$. The \emph{final regime} occurs for large enough times, so that the function has decayed close to its asymptotic limit of $0$. 
We formally define the regimes by requiring that the difference of  function values to its initial and asymptotic limit is not larger than 
\begin{equation}\label{eq:c}
c\ll1,
\end{equation}
that is,
\begin{equation}\label{eq:t_0_ss}
	e^{t\lambda}\geq 1-c\quad\text{or}\quad 	e^{t\lambda}\leq c.
\end{equation}
Therefore, the approximate stationarity occurs for
\begin{equation}\label{eq:t_0_ss2}
 t(-\lambda)\leq-\ln(1-c)=c+...\quad\text{or}\quad  t(-\lambda)\geq -\ln(c).
\end{equation}
Here, the inequalities are valid for any $0<c<1$ and correspond to distinct regimes for $c<1/2$; we expanded up to linear in $c$ using Eq.~\eqref{eq:c}.

\subsection{Changes in time}\label{sec:single_AS}

The initial regime is finite, while the final regime is unbounded; cf.~Fig.~\ref{fig:exp}.
Furthermore,  in the initial regime corrections to their initial value grow linearly with time and proportionally to $\lambda$, while in the final regime the mode decays to its limit exponentially with time at this rate.
Nevertheless, we now introduce a way to quantify how non-stationary is the dynamics, which allows to identify the initial and final regimes 
despite their differences. 

To this aim we consider how the function changes with time, that is, $|e^{t_1\lambda}-e^{t_2\lambda}|$ for $t_1\neq t_2$, and view it as \emph{approximately stationary when changes} during a given time regime are \emph{negligible}. In particular, we show that negligible changes with respect to the \emph{logarithmic scale of time} for a real mode occur only in the initial and the final regimes of its dynamics [cf.~Fig.~\ref{fig:exp}(b)].

\subsubsection{Linear scale}\label{sec:single_AS_lin}
 A crucial property of the dynamics of a single mode is its \emph{contractivity}, that is, the decay of the distance between function values at different times when shifted forward in time,
$|e^{ (t_1+\delta t)\lambda}-e^{(t_2+\delta t)\lambda}|\leq |e^{t_1 \lambda }-e^{t_2\lambda}|$
for $\delta t\geq 0$; cf.~Fig.~\ref{fig:exp}(a). Thus, requiring that the function changes no more than a given distance, does not
uniquely determine a time regime. In fact, a negligibly small change can be achieved after an arbitrary time, simply by considering time difference  within the initial regime as $| e^{t_1\lambda }-e^{t_2\lambda}|\leq |e^{|t_1-t_2|\lambda}-1|$, and such approximate stationarity can be seen as inherited from the initial regime.
This issue remains when normalising $| e^{t_1\lambda }-e^{t_2\lambda}|$ by the bigger (or the smaller) of the function absolute values, as this yields $|e^{|t_1-t_2|\lambda}-1|$ (or $|e^{-|t_1-t_2|\lambda}-1|$) dependent solely on the time difference.

\subsubsection{Logarithmic scale}
We now show that when the mode changes negligibly during a time interval at least of the order of the earlier time, it belongs either to the initial regime or the final  regime; cf.~Fig.~\ref{fig:exp}(b). 
This is possible, as we are able to consider how the dynamics changes with respect to the logarithmic rather than the linear scale.

Indeed, for a real mode and
\begin{equation}\label{eq:c_delta}
	c_{\Delta}\ll 1,
\end{equation}
we consider time $t$  such that
\begin{equation}\label{eq:change2}
	e^{t\lambda}-e^{2t\lambda}\leq   c_{\Delta}.
\end{equation}
With  $E_\pm(c_{\Delta})$ denoting two solutions  for the quadratic equation  $e^{2t\lambda}-e^{t\lambda}+c_{\Delta}=0$ with respect to $e^{t\lambda}$, Eq.~\eqref{eq:change2} then holds if and only if [see~Fig.~\ref{fig:exp}(b)]
\begin{subequations}\label{eq:change2_+-}
	\begin{align}
		\label{eq:change2_+}
		&e^{t \lambda}\geq E_+(c_{\Delta})\equiv\frac{1+\sqrt{1-4c_{\Delta}}}{2}=1-c_{\Delta}+...\quad \text{or}\quad	\\	\label{eq:change2_-}
		&e^{t \lambda}   \leq E_-(c_{\Delta})\equiv\frac{1-\sqrt{1-4c_{\Delta}}}{2}=c_{\Delta}+...,
	\end{align}
\end{subequations}
where in the inequalities we assumed $c_{\Delta}\leq1/4$; we expanded up to linear order in $c_{\Delta}$  using Eq.~\eqref{eq:c_delta}.

Note that $1-E_+(c_{\Delta})=E_-(c_{\Delta})$, so that Eq.~\eqref{eq:change2} holds if and only if $t$ belongs to the initial and final regimes in Eq.~\eqref{eq:t_0_ss} with $c=E_-(c_{\Delta})$. 
Furthermore, as $e^{t\lambda}-e^{2t\lambda}$ is the maximal change for times between $t$ and $2t$, we have that all times between $t$ and $2t$ belong either to the initial regime with $c=1-[E_+(c_\Delta)-c_\Delta]=E_-(c_{\Delta})+c_\Delta$ or to the final regime with $c=E_-(c_{\Delta})$. These regimes are distinct for $c_\Delta<-2+\sqrt{5}=0.236...$.

\section{Operational approach to metastability}\label{sec:meta}

Metastability is typically considered to occur when there exists a \emph{pronounced time regime} during which system states \emph{appear stationary}. In this work, we introduce an operational approach to  metastability in open quantum systems, which among others, allows us to formalize this definition.

We consider accessing a state of an open quantum system by measuring observables averages. This leads to a quantitative description of changes in the system dynamics  in terms of the trace  norm. In turn, this allows us to introduce initial and final regimes of the dynamics in analogy to the dynamics of a single mode discussed earlier. In contrast to that case, however, we find that a distinct regime of approximate stationarity may arise in the dynamics of an open quantum system, which we recognise as the phenomenon of metastability. 

Using the operational approach, we are able to identify the set of metastable states and characterise both the timescales and the structure of dynamics before and after the metastable regime.

\subsection{Changes in time}\label{sec:meta_changes}
We begin by discussing how an  open quantum system changes during its evolution. We consider accessing a state of the system by measuring observable averages, since a state of a quantum system of with the space dimension $D$ can be reconstructed by measuring $D^2$ linearly independent observables.  \\

For an observable $O$, the  maximal change in the average of an observable $O$ within a given time regime $t'\leq t \leq t''$ is given by
\begin{eqnarray}\label{eq:meta_state_obs_Delta}
	\Cd(\rho_0;O;t'',t')&\equiv&\!  \sup_{t''\leq t_1,t_2\leq t' }\!\! \frac{|\Tr(O\rho_{t_1})-\Tr(O\rho_{t_2})|}{\lVert O\rVert_{\max}},\qquad
\end{eqnarray}
where $\rho_0$ is initial system state. Here, we normalise by  the max norm of the observable, $\lVert O\rVert_{\max}\equiv \sup_{\rho_0}\Tr(O\rho_0)$  (see Appendix~\ref{app:norm_def} and~\footnote{Normalising changes of the observable average by $\sup_{t\geq 0}|\Tr(O\rho_t)|$ leads to larger values than in Eq.~\eqref{eq:meta_state_obs_Delta} and thus more restrictive definitions of the initial and final regimes as well as that of the metastability.}). 

Assuming that measuring of all observables is possible (or, equivalently, $D^2$ linearly independent observables can be measured),
the maximal change corresponds to the maximal distance between system states measured in the trace norm 
 \begin{eqnarray}\label{eq:meta_state_Delta}
 	\Cd(\rho_0;t'',t')&\equiv& \,\,\sup_O \,\,\Cd(\rho_0;O;t'',t')\\\nonumber
 	&=& \sup_{t''\leq t_1,t_2\leq t' } \lVert\rho_{t_1}-\rho_{t_2}\rVert.
 \end{eqnarray}
Indeed, $ |\Tr(O\rho_{t_1})-\Tr(O\rho_{t_2})|\leq \lVert O\rVert_{\max}\lVert\rho_{t_1}-\rho_{t_2}\rVert$, with the inequality saturated for $O=\Pi_+-\Pi_-$, where $\Pi_+$ and $\Pi_-$ are the projections on the direct sum of eigenspaces of $\rho_{t_1}-\rho_{t_2}$ corresponding to positive and negative eigenvalues, respectively (see Appendix~\ref{app:norm_def}). 
We note that there is no additional normalisation needed in Eq.~\eqref{eq:meta_state_Delta}, as for the system states $\lVert\rho_t\rVert=1$.

Assuming further the system can be initialised in any state (or, equivalently, $D^2$ linearly independent initial states can be prepared), the maximal change
in the dynamics	$\Cd(t'',t')\equiv \sup_{\rho_0} \sup_{O}	\Cd(\rho_0;O;t'',t')$ captures by the distance between  the corresponding evolution operators in the norm induced by the trace norm (see Appendix~\ref{app:norm_ind}),
	\begin{eqnarray}\label{eq:meta_Delta}
	\Cd(t'',t')
	&=&\,\, \sup_{\rho_0} \,\,\sup_{O}\,\,	\Cd(\rho_0;O;t'',t')\\\nonumber
	&=&\sup_{t''\leq t_1,t_2\leq t' } \lVert e^{t_1\L}-e^{t_2\L}\rVert.
\end{eqnarray}
Here, the supremum is achieved for a pure initial state, $\rho_0=\rho_0^2$.\\

We have that [cf.~Eq.~\eqref{eq:meta_state_obs_Delta}]
\begin{eqnarray}\label{eq:meta_state_obs_Delta1}
	\Cd(\rho_0;O;t'',t')
	&=&\sup_{t''\leq t\leq t' }\frac{ \Tr(O\rho_{t})}{\lVert O\rVert_{\max}} \!-\!\!\!\inf_{t''\leq t\leq t' }\frac{ \Tr(O\rho_{t})}{\lVert O\rVert_{\max}} .\qquad
\end{eqnarray}
Furthermore, as $\lVert e^{t\L}\rVert=1$, the dynamics is \emph{contractive}  with respect to the trace norm, $\lVert\rho_{t_1+t}-\rho_{t_2+t}\rVert\leq \lVert e^{t\L}\rVert \lVert\rho_{t_1}-\rho_{t_2}\rVert=\lVert\rho_{t_1}-\rho_{t_2}\rVert$ (see, e.g., Refs.~\cite{Watrous2005,Wolf2012} or Appendix~\ref{app:norm_master}). It then follows that [cf.~Eqs.~\eqref{eq:meta_state_Delta} and~\eqref{eq:meta_Delta}]
	\begin{eqnarray}
		  \label{eq:meta_state_Delta1}
	 	\Cd(\rho_0;t'',t')	&=& \sup_{t''\leq t\leq t' } \lVert\rho_{t''}-\rho_{t}\rVert,\\
	\label{eq:meta_Delta1}
	\Cd(t'',t')	&=&\sup_{t''\leq t\leq t' } \lVert e^{t''\L}-e^{t\L}\rVert.
\end{eqnarray}

The quantities in Eqs.~\eqref{eq:meta_state_obs_Delta1},~\eqref{eq:meta_state_Delta1}, and~\eqref{eq:meta_Delta1} grow when the time regime is enlarged, that is, when $t'$ increases or $t''$ decreases (see Fig.~\ref{fig:example}). In particular, for Eqs.~\eqref{eq:meta_state_Delta1} and~\eqref{eq:meta_Delta1}, extending the length of the considered regime  leads to at most a linear increase,
\begin{eqnarray}	\label{eq:meta_state_lin}
		\Cd[\rho_0;t'',t''+n(t'-t'')]&\leq& n \,\Cd(\rho_0;t'',t'),
	\\\label{eq:meta_lin}
\Cd[t'',t''+n(t'-t'')]&\leq& n \,\Cd(t'',t').
\end{eqnarray}
This follows from the contractivity of the dynamics; for $t_1,t_2\geq t''$ such that $(n\!-\!1)(t''-t')\leq t_1-t_2\leq n (t''-t') $ we have $\lVert \rho(t_1)-\rho(t_2)\rVert\leq\sum_{l=1}^{n-1}\lVert \rho[t_2+ l(t'-t'')]-\rho[t_2+ (l-1) (t'-t'')]\rVert +\lVert \rho(t_1)-\rho[t_2+ (n-1)(t'-t'')]\rVert\leq n  \Cd(\rho_0;t'',t')$.

Finally, we note that for any time regime changes in the system state are bounded by
\begin{eqnarray}\label{eq:meta_state_Delta3}
	\Cd(\rho_0;t'',t')
	&\leq&2,
\end{eqnarray}
where the bound corresponds to the distance between any two orthogonal states. 
From Eq.~\eqref{eq:meta_state_Delta3} we obtain (cf.~Fig.~\ref{fig:example})
\begin{eqnarray}\label{eq:meta_Delta3}
	\Cd(t'',t')
	&\leq&2.
\end{eqnarray}

	\subsection{Initial and final regimes}\label{sec:initial_final}
	
		For any open quantum system, there always exist two time regimes where changes can be neglected. These are initial and final regimes, which can be defined in a direct analogy to the dynamics of a single mode.

	\subsubsection{Definition}

	For a chosen observable $O$ and an initial state $\rho_0$, we consider the initial and final regimes by requiring that the normalised difference of the observable average $\Tr(O\rho_t)$ to its initial value $\Tr(O\rho_0)$ or its asymptotic limit $\Tr(O\rhoss)$, respectively, is not larger than 
	[cf.~Eq.~\eqref{eq:c}]
	\begin{equation}
		\label{eq:C}
		\C\ll 1.
	\end{equation}
	The overlap of  initial or final regimes for all observables and initial system states [cf.~Eqs.~\eqref{eq:meta_state_obs_Delta},~\eqref{eq:meta_state_Delta}, and~\eqref{eq:meta_Delta}] leads to the following.

	The  \emph{initial regime} of the dynamics holds for $t\leq t_0(\C)$, where $t_0(\C)$ is the shortest time such that $\lVert e^{t_0(\C)\L}-\I\rVert =\C$ with $\I(\rho)\equiv\rho$, and thus 
	\begin{equation}\label{eq:initial}
	\lVert e^{t\L}-\I\rVert \leq \C.	
	\end{equation}
	The \emph{final regime} of the dynamics holds for times
	\begin{equation}\label{eq:final}
		\lVert e^{t\L}-\Pss\rVert \leq \C	
	\end{equation}
	[due to the contractivity of the dynamics $\lVert e^{t\L}-\Pss\rVert$ decays with time $t$, so that Eq.~\eqref{eq:final} holds for all $t\geq t_\text{ss}(\C)$ where $t_\text{ss}(\C)$ is the shortest time such that $\lVert e^{t_\text{ss}(\C)\L}-\Pss\rVert =\C$]. 
	Note that the initial and final regimes are distinct for $\C<1/2$ [note that $\lVert \I-\Pss\rVert\leq \lVert e^{t\L}-\I\rVert+\lVert e^{t\L}-\Pss\rVert$ and cf.~Eq.~\eqref{eq:IPss} below; see Fig.~\ref{fig:example}(a)].

		\begin{figure}[t!]
		\begin{center}
			\includegraphics[width=\columnwidth]{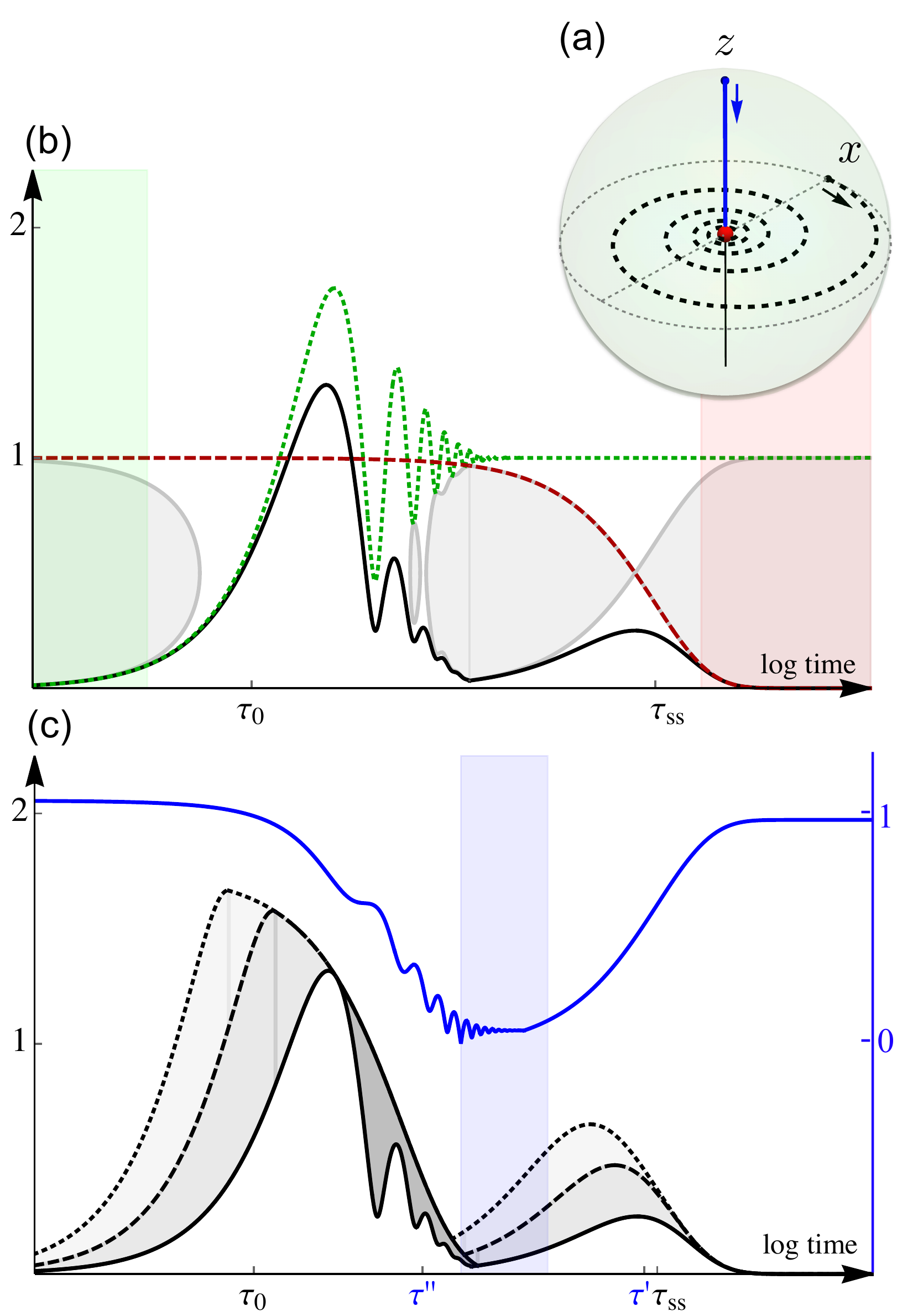} 
			\vspace*{-5mm}
			\caption{
				\textbf{Minimal open quantum system with metastability}. We consider a spin-$1/2$ in the presence of magnetic field, $H\!=\!-\omega S_z$, whose uncorrelated-in-time fluctuations amount to dephasing, $J_z\!=\!\sqrt{\gamma}S_z$, and in contact with an infinite-temperature bath, $J_+\!=\!\sqrt{\kappa/2}\,S_{+}$ and $J_-\!=\!\sqrt{\kappa/2}\,S_{-}$, where $S_{\pm}\!=\!S_{x}\pm S_{y}$ and $S_x$, $S_y$, $S_z$ are the spin operators along $x$, $y$, $z$ axes, respectively. \textbf{(a)} The magnetisation along $z$ axis (blue solid) decays exponentially to $0$ in correspondence with a real mode ($\lambda\!=\!\kappa$; cf.~Fig.~\ref{fig:exp}), while along $x$ and $y$ axes (black dashed) its dynamics features additional oscillations in analogy to a complex mode ($\lambda=(\kappa\!+\!\gamma)/2\!-\!i\omega$; cf.~Appendix~\ref{app:single}).
				\textbf{(b)} The distances of the evolution operator to its initial limit, $\lVert e^{t\L}\!-\!\I\rVert$ (green dotted),  and its asymptotic limit, $\lVert e^{t\L}\!-\!\Pss\rVert$ (red dashed) are restricted by $E_\pm(\lVert e^{t\L}\!-e^{2t\L}\rVert)$ [grey solid lines, excluded area shaded grey, cf.~Eqs.~\eqref{eq:all} and~\eqref{eq:ss_IP}] and are negligible in the initial and final regimes, respectively [shaded green and red for $\C=0.1$ in Eqs.~\eqref{eq:initial} and~\eqref{eq:final}]. \textbf{(c)} In general, $\Cd(t,2t)\geq \lVert e^{t\L}\!-\!e^{2t\L}\rVert$ [black solid lines with the dark grey shaded area in between; cf.~panel (b)], and the changes increase with the length of considered regime [$\Cd(t,4t)$ and $\Cd(t,8t)$ shown as black dashed and black dotted lines]. Nevertheless, in the weak coupling limit, $\kappa/\gamma\ll 1$, a distinct time regime arises when changes are negligible [area shaded blue for $\Cd(t'',t')=0.1$], while $\lVert e^{t\L}\!-\!\I\rVert$ and $\lVert e^{t\L}\!-\!\Pss\rVert$ are close to $1$ -- as (at least) required by the bounds in Eqs.~\eqref{eq:all} and~\eqref{eq:ss_IP} [cf.~panel (a)]. $\lVert e^{t\L}\!-\!e^{t''\L}\rVert$ is also shown (blue solid; note the shifted vertical axis).
				The parameters are $\kappa/\gamma=0.005$ and $\omega/(\gamma+\kappa)=5$.
			}\vspace*{-10mm}
			\label{fig:example}
		\end{center}
	\end{figure}

	\subsubsection{Relation to dynamics timescales}
	We now discuss the relation of the initial and final regimes to the shortest and longest timescales in the dynamics. The relation to the spectrum of the master operator is discussed in Sec.~\ref{sec:spectral}. \\

	We begin by noting that 
	\begin{equation}\label{eq:IPss}
		1\leq 	\lVert \I-\Pss \rVert\leq 2
	\end{equation}
	for non-trivial dynamics, $\L\neq 0$. The upper bound is saturated when there exists a decay subspace~\cite{Albert2016} in the dynamics [for the saturated lower bound, see, e.g.,  Fig.~\ref{fig:example}(a)]. 
	For derivation of Eq.~\eqref{eq:IPss} and a lower bound dependent the rank of stationary states, see Appendix~\ref{app:dynamics0}.\\
	
	\emph{Shortest timescale}. In analogy to the dynamics of a single mode, we define the shortest timescale $\tau_0$ of the dynamics as the shortest time such that~\footnote{Here, $1/e$ could be replaced by any positive number less than $1$; cf.~Eq.~\eqref{eq:IPss}.} 
	\begin{equation}\label{eq:tau_0}
	\lVert e^{ \tau_0\L}-\I \rVert=1-\frac{1}{e}.
	\end{equation}
    Equation~\eqref{eq:tau_0} is well defined as $\lVert e^{ t\L}-\I \rVert$ is a continuous function of time (see Appendix~\ref{app:norm_master}) initially equal zero and with the asymptotic limit bounded as in Eq.~\eqref{eq:IPss}.
    
	By the triangle inequality and the contractivity of the dynamics, we have that $\lVert e^{ (t_1+t_2)\L}-\I \rVert\leq 	\lVert e^{(t_1+t_2)\L}-e^{ t_2\L} \rVert+	\lVert e^{ t_2\L}-\I \rVert \leq	\lVert e^{ t_1\L}-\I \rVert+	\lVert e^{ t_2\L}-\I \rVert$,
	so that the distance to its initial state \emph{increases at most linearly} with time,
	\begin{equation}\label{eq:0_lin}
		\lVert e^{ n t\L}-\I \rVert\leq n	\lVert e^{ t\L}-\I \rVert.
	\end{equation}
	Therefore, time required to achieve the distance from an initial state larger than $1-1/e$ is bounded from below
	\begin{eqnarray}
		\frac{t}{\tau_0}&\geq& \left\lfloor\frac{	\lVert e^{t\L}-\I \rVert}{\lVert e^{ \tau_0\L}-\I \rVert}\right\rfloor
= \left\lfloor\frac{	\lVert e^{t\L}-\I \rVert}{1-\frac{1}{e}}\right\rfloor,
	\end{eqnarray} 
 while for times $t\leq t_0$, the bound holds for the inverted ratio. In particular, times within the initial regime in Eq.~\eqref{eq:initial} are bounded as $t\leq t_0(\C)\leq \tau_0/\lfloor (1-1/e)/C\rfloor$ [cf.~Fig.~\ref{fig:example}(b)].
 
 	Alternatively, we can consider $1/\lVert \L\rVert$ as the unit of time. Since 
 	\begin{equation}\label{eq:0_exp}
 		\lVert e^{t\L}-\I\rVert\leq e^{t\lVert\L\rVert}-1,
 	\end{equation}  
 	 we obtain $\tau_0\lVert \L\rVert\geq \ln(2-1/e)$.  It also follows that $t_0(\C)\lVert \L\rVert\geq \ln(1+\C)$, so that times $t\leq \ln(1+\C)/\lVert \L\rVert=(\C+...)/\lVert \L\rVert$ belong to the initial regime in Eq.~\eqref{eq:initial} [cf.~Eq.~\eqref{eq:t_0_ss2}]. 	
 	Furthermore, 
 	as
 	 \begin{equation}\label{eq:0_exp2}
 		\lVert e^{t\L}-\I\rVert\geq 	2 t\lVert\L\rVert	-e^{t\lVert\L\rVert}+1,
 	\end{equation}   
     we also have that $t_0(\C)\lVert \L\rVert\leq E_1(\C)=\C+...$, where $E_1(C)$ is the inverse to the function $2 t\lVert\L\rVert	-e^{t\lVert\L\rVert}+1$ for $t\lVert\L\rVert\leq\ln (2)$, so that $E_1(C)$ is well defined for $\C\leq 2\ln(2)-1=0.386...$, in which case $1\leq E_1(\C)/\C\leq \ln(2)/[2\ln(2)-1]=1.794...$; in its expansion around $0$ we assumed that $\C$ is within the non-zero convergence radius (cf.~the inverse Lagrange theorem). Therefore, the distance to initial state \emph{increases approximately linearly} within the initial regime.
 	 \\

	\emph{Longest timescale}. The longest timescale in the dynamics is given by the timescale $\tau_\text{ss}$ of the system relaxation towards its stationary states, which is usually defined as the shortest time such that~\cite{Note2}
	\begin{equation}\label{eq:tau_ss}
		\lVert e^{\tau_\text{ss}\L}-\Pss \rVert=\frac{1}{e}
	\end{equation}
    and referred to as the \emph{(final) relaxation time}. 
	Equation~\eqref{eq:tau_ss} is well defined as 	$\lVert e^{ t\L}-\Pss\rVert$ is a continuous function of time (see Appendix~\ref{app:norm_master}) with the initial value bounded as in Eq.~\eqref{eq:IPss} and zero asymptotic limit. Moreover, from Eq.~\eqref{eq:IPss} it follows that 	$\lVert e^{\tau_\text{ss}\L}-\I \rVert\geq 1-1/e$, so that $\tau_0\leq \tau_\text{ss}$ [cf.~Fig.~\ref{fig:example}(b)].

	The process of (final) relaxation can be understood by the distance of a system state to the stationary state, which  for times $t\geq \tau$ \emph{decays at least exponentially}, as $\lVert e^{(t_1+t_2)\L}-\Pss\rVert \leq \lVert e^{ t_1\L}- \Pss\rVert \lVert e^{ t_2\L}- \Pss\rVert$ so that 
	\begin{equation}\label{eq:ss_exp}
    \lVert e^{n t\L}-\Pss\rVert \leq \lVert e^{ t\L}- \Pss\rVert^n .
    \end{equation}
     Therefore, time required to achieve the distance from a stationary state smaller than $1/e$ can be bounded from above as
	\begin{eqnarray}
		\frac{t}{\tau_\text{ss}}&\leq& \lceil\log_{\lVert e^{\tau_\text{ss}\L}-\Pss\rVert } (\lVert e^{t\L}-\Pss\rVert)\rceil
		\\\nonumber&=& -\lceil \ln (\lVert e^{t\L}-\Pss\rVert)\rceil,
	\end{eqnarray} 
	where $t\geq \tau$. In particular, we obtain $t_\text{ss}(\C)/\tau_\text{ss}
	\leq- \lceil \ln (\C)\rceil $ for $\C<1/e$, so that times $t\geq -\lceil \ln (\C)\rceil \tau_\text{ss}$ belong to the final regime in Eq.~\eqref{eq:final} [cf.~Fig.~\ref{fig:example}(b)].

	\subsection{Metastability}\label{sec:meta_def}
	
	In the initial and final regimes in Eqs.~\eqref{eq:initial} and~\eqref{eq:final}, the system changes are bounded by $2\C$, 
	\begin{eqnarray}\label{eq:change_all0}
		\lVert e^{t_1\L}-e^{t_2\L}\rVert	&\leq& 	\lVert e^{t_1\L}-\I\rVert+ 	\lVert e^{t_2\L}-\I\rVert,\\
		\label{eq:change_ss0}
		\lVert e^{t_1\L}-e^{t_2\L}\rVert	&\leq& 	\lVert e^{t_1\L}-\Pss\rVert+ 	\lVert e^{t_2\L}-\Pss\rVert,
	\end{eqnarray}
	and thus negligible (in fact, the changes in the initial regime can be bounded by $\C$ due to the contractivity of the dynamics).
	We now argue that an open quantum system may change negligibly also beyond the initial and final regimes of its dynamics, which corresponds to the phenomenon of metastability.

\begin{figure*}[t!]
	\begin{center}
		\includegraphics[width=\textwidth]{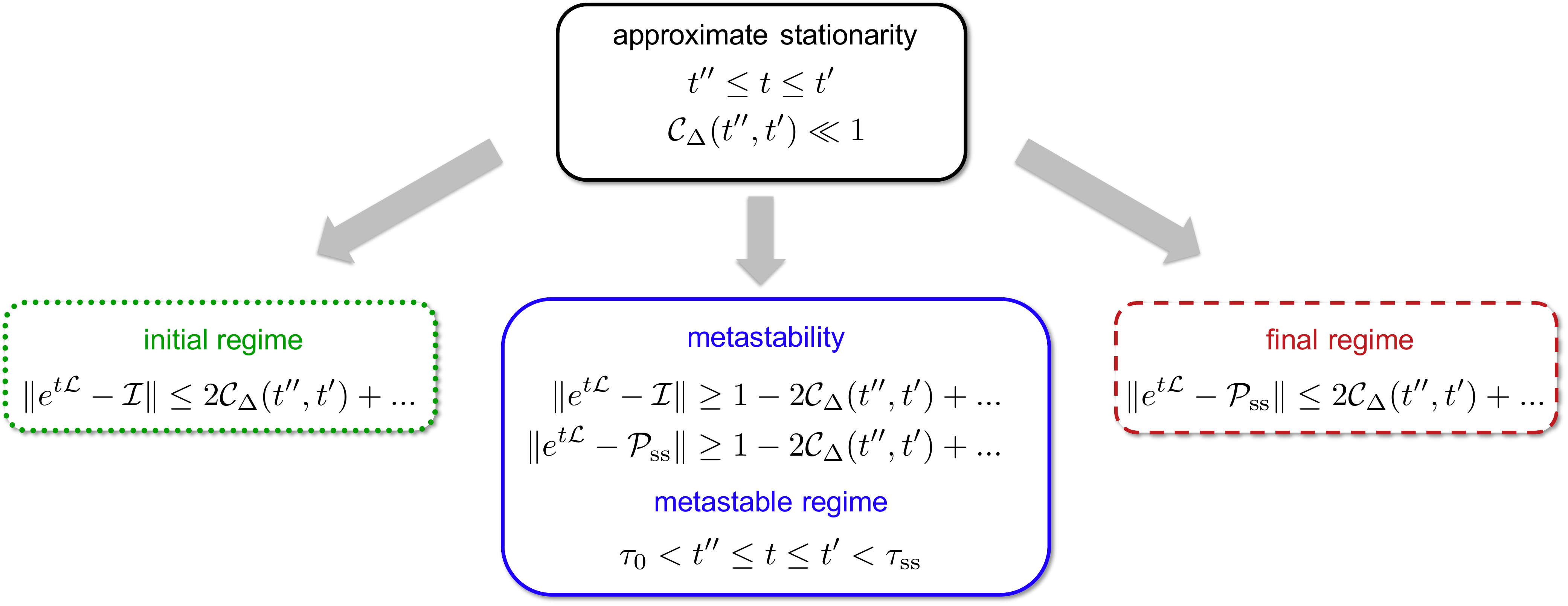} 
		\caption{
			\textbf{Regimes of approximate stationarity in open quantum dynamics}. 	In this work, we show that approximate stationarity in Markovian open quantum dynamics, which is defined in Eqs.~\eqref{eq:meta_cond} and~\eqref{eq:meta_Delta2}, can only occur in the initial regime [Eq.~\eqref{eq:initial}], the final regime [Eq.~\eqref{eq:initial}], or in a distinct time regime, which we refer to as the metastable regime, occurring after the shortest timescale of the dynamics [Eq.~\eqref{eq:tau_0}] and before the final relaxation [Eq.~\eqref{eq:tau_ss}].
			In this last case, the dynamics non-negligibly differs both from its initial and asymptotic limits, which we recognise as the phenomenon of metastability. This general result follows from the analogy of open quantum dynamics to single-mode dynamics [cf.~Eqs.~\eqref{eq:all} and~\eqref{eq:ss_IP}]. 
		}\vspace*{-7mm}
		\label{fig:meta}
	\end{center}
\end{figure*}

 	\subsubsection{Approximate stationarity}

 	 We define the system to be \emph{approximately stationary}  when its changes quantified by Eq.~\eqref{eq:meta_Delta} are negligible,
 	\begin{equation}\label{eq:meta_Delta2}
 		\Cd(t'',t')\ll 1,
 	\end{equation}
   during a  \emph{pronounced time regime} $t'\leq t \leq t''$, where
 	\begin{equation}\label{eq:meta_cond}
 	2t''\leq t'.
 	\end{equation}

    For any quantum system its changes are bounded from above by a constant independent from its dimension; see
    Eq.~\eqref{eq:meta_state_Delta3}. Moreover,  they are non-zero whenever $t'>t''$, $\Cd(t'',t')>0$ unless $\L=0$ or $(\I-\Pss)\L=-\infty$ (cf.~Sec.~\ref{sec:spectral_mode}). This motivates Eq.~\eqref{eq:meta_Delta2}.  We would like to emphasise, however, that methods introduced in this work are not based on any perturbative approach and Eq.~\eqref{eq:meta_Delta2} is simply used to consider a limit of our general results.

    Due to the contractivity of the dynamics, changes in the system only decay when shifted forward in time, and thus considering changes within the regime with respect to the \emph{logarithmic scale of time} rather than the linear scale of time is necessary. In particular, Eq.~\eqref{eq:meta_cond} allows for considering changes with the time ratio equal $2$. While changes with respect to the logarithmic scale of time correspond more generally to any constant ratio of times, a time regime can be extended to fulfil Eq.~\eqref{eq:meta_cond}
	provided that $C_\Delta(t'',t') \lceil t''/(t'-t'')\rceil  \ll 1$ [cf.~Eq.~\eqref{eq:meta_lin}]. In Appendix~\ref{app:dynamics1}, we show that this condition in fact excludes approximate stationarity inherited from the initial regime. \\

In order to characterise the origin of the approximate stationarity, we exploit the following results on the changes in the dynamics  and its distance to the initial and asymptotic limits:
\begin{equation}
	\label{eq:change2_all}
	\lVert e^{t\L}-\I\rVert  \left(1-	\lVert e^{t\L}-\I\rVert   \right)\leq 	\lVert e^{t\L}-e^{2t\L}\rVert
\end{equation}
and
\begin{equation}
	\label{eq:change2_ss}
	\lVert e^{t\L}-\Pss\rVert \left(1-\lVert e^{t\L}-\Pss\rVert\right)\leq	\lVert e^{t\L}-e^{2t\L}\rVert.
\end{equation}
For derivation, see Appendix~\ref{app:dynamics2}.

From Eq.~\eqref{eq:change2_all},   it follows that [cf.~Eqs.~\eqref{eq:change2} and~\eqref{eq:change2_+-}; we assume Eq.~\eqref{eq:meta_cond} and $C_\Delta(t'',t')\leq 1/4$]
\begin{subequations}\label{eq:all} 
	\begin{align}
	\label{eq:all+}	
	&\lVert e^{t\L}-\I\rVert \geq	E_+[\Cd(t'',t')]
		\quad\text{or}
		\\
		\label{eq:all-}
		&\lVert  e^{t\L}-\I\rVert\leq  E_-[\Cd(t'',t')]
	\end{align}
\end{subequations}
for $t''\leq t\leq t'/2$~\footnote{Here, $E_\pm[\Cd(t'',t')]$ could be further replaced by $E_\pm[\C^{(2)}_\Delta\!(t'',t')]$ with $\C^{(2)}_\Delta\!(t'',t')\equiv\sup_{t''\leq t\leq t'/2}\lVert e^{t\L}-e^{2t\L}\rVert$.}. Since $\lVert  e^{t\L}-\I\rVert$ is a continuous function of time (cf.~Appendix~\ref{app:norm_master}), the same bound, either Eq.~\eqref{eq:all+} or Eq.~\eqref{eq:all-}, holds for all such $t$ when  $E_+[C_\Delta(t'',t')]>E_-[C_\Delta(t'',t')]$, i.e.. for $C_\Delta(t'',t')< 1/4$.
For $ t'/2< t\leq t'$ , the bounds $ E_\pm[C_\Delta(t'',t')]$ are further replaced by $ E_\pm[C_\Delta(t'',t')]\mp\Cd(t'',t')$, and we have $ E_+[C_\Delta(t'',t')]-\Cd(t'',t')> E_-[C_\Delta(t'',t')]+\Cd(t'',t')$ for $C_\Delta(t'',t')< (-1+\sqrt{2})/2=0.207...$.

Similarly, from Eq.~\eqref{eq:change2_ss} we have
\begin{subequations}\label{eq:ss_IP}
	\begin{align}
		\label{eq:ss_IP+}
		&\lVert e^{t\L}-\Pss\rVert \geq	E_+[C_\Delta(t'',t')]
		\quad\text{or}
		\\	\label{eq:ss_IP-}
		&\lVert  e^{t\L}\!-\Pss\rVert\leq  E_-[C_\Delta(t'',t')]
	\end{align}
\end{subequations}
for $t''\leq t\leq t'/2$~\cite{Note3}. As  $\lVert  e^{t\L}-\Pss\rVert$ is a continuous function of time (cf.~Appendix~\ref{app:norm_master}), for $C_\Delta(t'',t')<1/4$, the same bound, either Eq.~\eqref{eq:ss_IP+} or Eq.~\eqref{eq:ss_IP-},  holds for all such $t$. For $ t'/2< t\leq t'$, $E_+[C_\Delta(t'',t')]$ is replaced by $E_+[\Cd(t'',t')]-\Cd(t'',t')$, and we have $ E_+[C_\Delta(t'',t')]-\Cd(t'',t')> E_-[C_\Delta(t'',t')]$ for $C_\Delta(t'',t')< -2+\sqrt{5}=0.236...$.

	\subsubsection{Initial and final approximate stationarity}
	
	Using Eqs.~\eqref{eq:all} and~\eqref{eq:ss_IP}, we are now ready to identify distinct regimes of approximate stationarity in an open quantum system (see Fig.~\ref{fig:meta}). We first consider Eqs.~\eqref{eq:all-}  and~\eqref{eq:ss_IP-} when the system changes negligibly, that is, Eq.~\eqref{eq:meta_Delta2} holds. We show that in this case, they respectively correspond to the initial and final regimes of the dynamics (cf.~Fig.~\ref{fig:example}).\\

\emph{Initial regime}.  Eq.~\eqref{eq:all-} describes the approximate stationarity with the system approximated by its initial state during the considered time regime. In fact, it can be shown that Eq.~\eqref{eq:all-} corresponds to the initial regime in Eq.~\eqref{eq:initial} with $\C= 2\Cd(t'',t') +...$.

We have (see Appendix~\ref{app:dynamics2})
\begin{equation}
	\label{eq:all_lin}
	\lVert e^{t\L}\!-\! e^{(t+\delta t)\L}\rVert\geq (2\!-\!\lVert e^{t\L}\!-\!\I\rVert)\,\delta t \lVert \L \rVert - e^{\delta t\lVert \L\rVert }+1.
\end{equation}
Therefore, from Eq.~\eqref{eq:all-} it follows  for $t''\leq t\leq t'/2$  that [cf.~Eq.~\eqref{eq:0_exp2}]
\begin{equation}
	\label{eq:all_lin2}
	\lVert e^{t\L}\!-\! e^{(t+\delta t)\L}\rVert\geq \frac{3}{2}\,\delta t \lVert \L \rVert - e^{\delta t\lVert \L\rVert }+1.
\end{equation}
Furthermore, for $t=t''$ and $\delta t\leq t'-t''$, we obtain that the left-hand side is bounded from above by $ \Cd(t'',t')$, so that  
\begin{eqnarray}\label{eq:all_lin3}
	&&(t'-t'')\lVert  \L \rVert\leq E_2[\Cd(t'',t')]=2\Cd(t'',t')+..., \qquad
\end{eqnarray}
where $E_{2}[\Cd(t'',t')]$ denotes the inverse of the function  $	3\delta t\lVert  \L \rVert/2-e^{\delta t\lVert  \L \rVert}+1$ for $ \delta t\lVert  \L \rVert\leq\ln(3/2)$; 
here, we assume $\Cd(t'',t')\leq[3\ln(3/2)-1]/2=0.108...$, in which case $1\leq E_2(\C)/\C\leq 2\ln(3/2)/[3\ln(3/2)-1]=3.747...$, while in the expansion around $0$ we assume that $\Cd(t'',t')$ is within the non-zero convergence radius of the inverse (cf.~the inverse Lagrange theorem). 
From Eq.~\eqref{eq:0_exp}, for $0\leq t\leq t'-t''$ we  have $\lVert e^{t\L}-\I\rVert\leq e^{E_2[\Cd(t'',t')]}-1=2\Cd(t'',t') +...$, so that together with  Eq.~\eqref{eq:all-}, we obtain that  $t''\leq t\leq t'$ belongs to the initial regime
$\C=\max\{e^{E_2[\Cd(t'',t')]}\!-\!1, E_-[\Cd(t'',t')]+\Cd(t'',t')\}=2\Cd(t'',t') +...\leq 1/2$.\\

\emph{Final regime}.  Eq.~\eqref{eq:ss_IP-} describes the approximate stationarity with system states approximated by the corresponding  stationary state.  In fact, by definition, the considered regime belongs to the {final regime in Eq.~\eqref{eq:final} with $\C=E_-[\Cd(t'',t')]$.\\

It is important to note that because of Eq.~\eqref{eq:IPss}, Eqs.~\eqref{eq:all-} and~\eqref{eq:ss_IP-} cannot hold simultaneously when $C_\Delta(t'',t')< 1/4$.
Therefore, we have that Eq.~\eqref{eq:all-} implies Eq.~\eqref{eq:ss_IP+}, 
while Eq.~\eqref{eq:ss_IP-} implies Eq.~\eqref{eq:all+}. 
It may happen, however,  that Eqs.~\eqref{eq:all+} and~\eqref{eq:ss_IP+} hold together. This last case is in fact the main focus of this work.

	\subsubsection{Metastability}

	We define \emph{metastability} to take place when  the approximate stationarity defined in Eqs.~\eqref{eq:meta_cond} and~\eqref{eq:meta_Delta2} corresponds to  (see Fig.~\ref{fig:meta})
	\begin{eqnarray}\label{eq:dist_0}
		\lVert e^{t\L} -\I \rVert&\geq& E_+[\Cd(t'',t')],
		\\\label{eq:dist_ss}
		\lVert e^{t\L} -\Pss \rVert&\geq&  E_+[\Cd(t'',t')],
	\end{eqnarray}
	where $t''\leq t\leq t'/2$~\cite{Note3}, while for $t'/2< t\leq t'$ both bounds are replaced by $E_+[\Cd(t'',t')]-\Cd(t'',t')$ [this requires $\Cd<(-1+\sqrt{2})/2$ and $\Cd<-2+\sqrt{5}$, respectively].\\
	
	We refer to a time regime when metastability occurs as a \emph{metastable regime}. It is important to note that it is distinct both from the initial regime in Eq.~\eqref{eq:initial} and the final regime in Eq.~\eqref{eq:final}. Indeed, from Eq.~\eqref{eq:dist_0} it follows that $t''/t_0(\C)\geq \lfloor  E_+[\Cd(t'',t')]/\C \rfloor$  [we also have $t''\lVert \L\rVert\geq \ln \{1+ E_+[\Cd(t'',t')]\}$]. Similarly, from Eq.~\eqref{eq:dist_ss} we have $t'/t_\text{ss}(\C)\leq 1/\lceil\ln(\C) \rceil$ with respect to the final regime in Eq.~\eqref{eq:final}. Furthermore, we also have 
	 	\begin{equation}\label{eq:meta_cond2}
		\tau_{0} < t'' \leq t\leq 	t'< \tau_\text{ss},
	\end{equation} 
	where the first inequality follows from Eq.~\eqref{eq:dist_0} for $\Cd(t'',t')<(1-1/e)/e=0.233...$ and the last inequality from Eq.~\eqref{eq:dist_ss} for $\Cd(t'',t')<-2+\sqrt{5}$; see Appendix~\ref{app:def} and cf.~Fig.~\ref{fig:example}. \\

	We note that Eq.~\eqref{eq:meta_cond2} together with Eqs.~\eqref{eq:meta_cond} and~\eqref{eq:meta_Delta2} can be considered as \emph{equivalent definition} of the phenomenon of metastability. That is, the metastability can be identified as the approximate stationarity which takes place after the shortest timescale $\tau_0$ and before the final relaxation time $\tau_\text{ss}$. 
	
	Indeed,    the condition in Eq.~\eqref{eq:meta_cond2} guarantees that if such a  regime exists, the approximate stationarity does not correspond to the trivial approximate stationarity of the initial and final regimes.  We have that Eq.~\eqref{eq:dist_0} follows from $t''> \tau_0$  for $\Cd(t'',t')\leq[3\ln(3/2)-1]/2$ and Eq.~\eqref{eq:dist_ss} from $t'< \tau_\text{ss}$ for $\Cd(t'',t')\leq(1-1/e)/e$ (see Appendix~\ref{app:def}). The condition in Eq.~\eqref{eq:meta_cond} further excludes approximate stationarity inherited from a metastable regime (see Appendix~\ref{app:def}).

It is important to stress that there exists no non-trivial regime of approximate stationarity beyond Eq.~\eqref{eq:meta_cond2}. Indeed, the regime of approximate stationarity with $ t''\leq \tau_0$ necessarily belongs to the initial regime with $\C=2\Cd(t'',t')+...\leq 1/2$ for $\Cd(t'',t')\leq[3\ln(3/2)-1]/2$; see Appendix~\ref{app:def}. Similarly,   the regime of approximate stationarity with  $t'\geq\tau_\text{ss}$ belongs to the final regime with $\C=E_-[\Cd(t'',t')]$ for $\Cd(t'',t')< -2+\sqrt{5}=0.236...$; see Appendix~\ref{app:def}.  Therefore, the condition in Eq.~\eqref{eq:meta_cond2} does not restrict the generality of metastability phenomenon.

\subsubsection{Metastable states}

Having defined the presence of metastability in the system dynamics, we now characterise  system states during the metastable regime. In Appendix~\ref{app:Heisenberg}, by considering the dynamics in the Heisenberg picture, we show that metastability implies the existence of quasi-conserved observables.

 During the metastable regime, the system changes negligibly for any initial state and thus we refer to its state as a \emph{metastable state}. In fact, it follows that the system initialised in such a state, $\rho_0=\rho_{t_0}'$ where  $\rho_0'$ is an initial state and $t''\leq t_0\leq t'$, changes negligibly for all times up to the end of metastable regime, $\lVert \rho_t-\rho_0 \rVert\leq 3\Cd(\rho_0';t'',t')\leq 3\Cd(t'',t')$ for $ t\leq t'$ [cf.~Eq.~\eqref{eq:meta_state_lin}]. While this is true at all times for initial states negligibly close to the set of stationary states, from Eqs.~\eqref{eq:dist_0} and~\eqref{eq:dist_ss}, there exist metastable states which are non-negligibly different both from the corresponding initial and stationary states. 

For example, for time $t''\leq t_0\leq t'/2$, consider  initial states $\rho_0'$ and $\rho_0''$ such that 	$\lVert \rho_{t_0}'-\rho_0'\rVert=\lVert e^{t_0\L} -\I \rVert$ and $\lVert \rho_{t_0}''-\rhoss''\rVert=\lVert e^{t_0\L} -\Pss \rVert$. For their mixture $\rho_0=(\rho_0'+\rho_0'')/2$, from Eqs.~\eqref{eq:dist_0} and~\eqref{eq:dist_ss} we obtain $\lVert \rho_{t}-\rho_0\rVert,\lVert \rho_{t}-\rhoss\rVert\geq E_+[\Cd(t'',t')]/2-\Cd(t'',t')/2$ for $t''\leq t\leq t'$. In the model in  Fig.~\ref{fig:example}, the states of spin up or down along $z$-axis are metastable.

\subsection{Initial relaxation and long-time dynamics}

Having discussed the system dynamics during the metastable regime,  we now characterise the  initial relaxation taking place before  and the long-time dynamics afterwards. 

\subsubsection{Initial relaxation}

In  Eq.~\eqref{eq:tau_0}, we introduced the shortest timescale $\tau_0$ of the system dynamics, which, by definition, is shorter than the metastable regime [cf.~Eq.~\eqref{eq:meta_cond2}]. We now discuss the longest timescale and the structure of the dynamics taking place before the metastable regime. 
\\  

\emph{Longest timescale in short-time dynamics}. 
	From Eqs.~\eqref{eq:dist_0}, we can analogously to Eq.~\eqref{eq:tau_ss} define the \emph{initial relaxation time} $\tau''$, i.e., timescale of relaxation of the system towards its metastable states, as the shortest time such that~\footnote{Here, time $t''$ could be replaced by any time within the metastable regime, but for times longer than  $t'/2$, $E_-[\Cd(t'',t')]$ should be replaced by $E_-[\Cd(t'',t')]+\Cd(t'',t')$ in order to keep Eq.~\eqref{eq:spectral_tau2} unchanged.}
   \begin{equation}\label{eq:tau''}
   \lVert e^{\tau''\L}-e^{t''\L}\rVert =\frac{1}{e}-E_-[\Cd(t'',t')],
   \end{equation}
where we consider $\Cd(t'',t')\leq (1\!-\!1/e)/e=0.233...$.
It follows that $\tau_0\leq \tau''<t ''$ [cf.~Eq.~\eqref{eq:tau_0}  and see Fig.~\ref{fig:example}(c)]. Indeed, $\lVert e^{\tau''\L}-\I \rVert\geq \lVert e^{t''\L} -\I \rVert-\lVert e^{\tau''\L}-e^{t''\L} \rVert\geq 1- 1/e$; cf.~Eq.~\eqref{eq:dist_0} and recall that $E_+[\Cd(t'',t')]=1-E_-[\Cd(t'',t')]$.\\

\emph{Initial relaxation}. More generally, $\lVert e^{t\L}-e^{t''\L}\rVert$  describes the relaxation of the system towards the metastable regime.
While, in contrast to $\lVert e^{t\L}-\Pss \rVert$,  it is in general  not a decaying function of  time  $t\leq t''$ [cf.~Fig.~\ref{fig:example}(c)], it cannot increase more than $\Cd(t'',t')$  as
\begin{equation}
\lVert e^{(t+\delta t)\L}-e^{t''\L} \rVert \leq \lVert e^{t\L}-e^{t''\L} \rVert+\Cd(t'',t')
\end{equation}
 for $\delta t\leq t''-t'$. Furthermore, it \emph{decays at least exponentially} up to a constant correction proportional to $\Cd(t'',t')$, 
	\begin{equation}\label{eq:''_exp}
	\lVert e^{n t\L}-e^{t''\L}\rVert  \leq \lVert e^{ t\L}-e^{t''\L}\rVert^n +\frac{2\Cd(t'',t')}{1-\lVert e^{ t\L}-e^{t''\L}\rVert}
\end{equation}
 for $(n-1)t\leq t'$ and $\lVert e^{ t\L}-e^{t''\L}\rVert<1$ [cf.~Eq.~\eqref{eq:ss_exp} and see Appendix~\ref{app:time''}].

\subsubsection{Long-time dynamics}

In Eq.~\eqref{eq:tau_ss}, we introduced the longest timescale in the system dynamics given by the final relaxation time $\tau_\text{ss}$, which, by definition, occurs after the metastable regime.  We now discuss the shortest timescale and the structure of the long-time dynamics taking place after the metastable regime.
\\

\emph{Shortest timescale in long-time dynamics}. From Eqs.~\eqref{eq:dist_ss}, we can analogously to Eq.~\eqref{eq:tau_0}  define the timescale  as the shortest time  $\tau'\geq t''$ that fulfils~\cite{Note4}
\begin{equation}\label{eq:tau'}
 \lVert e^{\tau'\L}-e^{t''\L}\rVert =1-\frac{1}{e}-E_-[\Cd(t'',t')].
\end{equation}
It follows that $t'< \tau'\leq \tau_\text{ss}$ [cf.~Eq.~\eqref{eq:tau_ss} and see Fig.~\ref{fig:example}(c)]. Indeed, $\lVert e^{\tau'\L}-\Pss \rVert\geq \lVert e^{t''\L} -\Pss \rVert-\lVert e^{\tau'\L}-e^{t''\L} \rVert\geq 1/e$ [cf.~Eq.~\eqref{eq:dist_ss}].\\

\emph{Long-time dynamics}. The distance from metastable states \emph{increases at most linearly} with time up to a constant correction $\Cd(t'',t')$ as
\begin{equation}\label{eq:'_lin}
	\lVert e^{n t\L}-e^{t''\L}\rVert +\Cd(t'',t')  \leq n[\lVert e^{ t\L}-e^{t''\L}\rVert +\Cd(t'',t')]
\end{equation}
 [cf.~Eq.~\eqref{eq:0_lin} and Appendix~\ref{app:time'}].
Therefore, time required to achieve the distance larger than in Eq.~\eqref{eq:tau'} is bounded from below as
\begin{eqnarray}
	\frac{t}{\tau'}&\geq& 
	 \left\lfloor\frac{	\lVert e^{t\L}-e^{t''\L}\rVert+\Cd(t'',t')}{1-\frac{1}{e}-E_-[\Cd(t'',t')]+\Cd(t'',t')}\right\rfloor,
\end{eqnarray} 
while for times $t\leq \tau'$, the bound holds for the inverted ratio of times, and with $\lVert e^{t\L}-e^{t''\L}\rVert$ and $1-1/e-E_-[\Cd(t'',t')]$ exchanged.
In particular, we obtain 
\begin{equation}\label{eq:tau'2}
	\frac{\tau'}{t''}\geq \left\lfloor \frac{1-1/e-E_-[\Cd(t'',t')]}{C_\Delta(t'',t')}\right\rfloor+1.
\end{equation}
(Appendix~\ref{app:time'}). \\

\emph{Changes in long-time dynamics}.
	 From the contractivity of the dynamics, we have that system states for times separated by less than the length of metastable regime do not differ by more than $\Cd(t'',t')$ also after the metastable regime. In fact,
	\begin{equation}\label{eq:meta_length_state}
	\lVert \rho_{t_1}-\rho_{t_2}\rVert\leq n\,\Cd(\rho_0;t'',t')
	\end{equation}
	for $t_1,t_2\geq t''$ such that $|t_1-t_2|\leq n(t'-t'')$ [cf.~Eq.~\eqref{eq:meta_state_lin}].
	 It also follows that for times during and after the metastable regime, $t\geq t''$, the dynamics is \emph{effectively restricted to the set of metastable states}. In fact, the dynamics of a system state is simply approximated by the dynamics of the corresponding metastable state as from Eq.~\eqref{eq:meta_length_state} we have $\lVert \rho_t-\rho_{t+t''}\rVert\leq \C(\rho_0;t'',t')$ [here, $t''$ can be replaced by any time $\leq t'-t''$, or by any time $\leq t'$ provided that $\C(\rho_0;t'',t')$ is replaced by  $2\C(\rho_0;t'',t')$].

Similarly,  the observable \emph{averages}  and \emph{higher order correlations} change negligibly when measured at times separated by intervals comparable to the length of the metastable regime,
	\begin{eqnarray}\label{eq:meta_corr}
		&&\frac{\big\lVert \O^{(n)}\!e^{t_1^{(n)}\!\L}\!\cdot\!...\!\cdot\! \O^{(1)}\!e^{t_1^{(1)}\!\L}- \O^{(n)}\!e^{t_2^{(n)}\!\L}\!\cdot\!...\!\cdot\! \O^{(1)}\!e^{t_2^{(1)}\!\L}\big\rVert}{\lVert O^{(n)}\!\rVert_{\max}\!\cdot\!...\!\cdot\! \lVert O^{(1)}\!\rVert_{\max}}\qquad
		\\\nonumber
		&&\leq \sum_{l=1}^{n} n_l\, \mathcal{C}_\Delta(t'',t') ,
	\end{eqnarray}
	where  we consider times $t_1^{(l)},t_2^{(l)}\geq t''$ and $|t_1^{(l)}-t_2^{(l)}|\leq {n_l} (t'-t'')$ for $l=1,...,n$ and $\mathcal{O}$ denotes either a superoperator encoding outcomes of measuring $O$ in the corresponding conditional system states, or $\mathcal{O}(\rho)\equiv(O\rho+\rho O)/2$;  in both cases $\lVert \O\rVert=\lVert O\rVert_{\max}$  (see Appendix~\ref{app:norm_obs}), so that $\lVert\O e^{t_1\L}-\O e^{t_2\L}\rVert\leq \lVert O\rVert_{\max} \lVert e^{t_1\L}-e^{t_2\L}\rVert$.	Whenever $\sum_{l=1}^{n} n_l  \mathcal{C}_\Delta(t'',t')\ll 1$,  considered times correspond to a plateau (cf.~Appendix~\ref{app:Heisenberg_meta}).

    \section{Spectral decomposition of metastability}\label{sec:spectral}

We now discuss spectral decomposition of the dynamics  and how changes of individual eigenvalues can be observed in experiments. This provides a direct connection to the single-mode dynamics, so  that approximate stationarity can only occur within the overlap of either initial or final regimes for individual eigenvalues.  Furthermore, whenever this overlap is non-trivial, there exists a separation in the real part of the master operator spectrum and approximate stationarity necessarily corresponds to metastability.  We also show how timescales of the initial and long-time dynamics can be bounded in terms of the master operator eigenvalues.

\subsection{Spectral decomposition of dynamics}\label{sec:spectral_mode}

\subsubsection{Spectral decomposition of master and evolution operators}

Although $\L$ is Hermiticity preserving it is in general not Hermitian ($\L\neq\L^\dagger$) or normal ($\L\L^\dagger=\L^\dagger\L$ requires the existence of the uniform stationary state $\rhoss=\mathds{1}/D$, where $D$ is the system Hilbert space dimension)~\footnote{Indeed, for a density matrix $\rho$ expressed as a  vector in an orthogonal basis of $D^2$ Hermitian operators acting on the system space, $\L$ in Eq.~\eqref{eq:L} corresponds to a $D^2\times D^2$ matrix, which is real, but not necessarily symmetric or normal.}. Nevertheless, we can consider its Jordan normal form, leading to the spectral decomposition
\begin{equation}\label{eq:L_spectral}
	\L(\rho)= \sum_{k=1}^{D^2}  \lambda_k  \bigg[ \Tr(L_k \rho)+(1-\delta_{d_k,D_k}) \Tr(L_{k+1} \rho)\bigg] R_{k}.
\end{equation}
Here, $k$ indexes the eigenvalues ${\lambda}_{k}={\lambda}_{k}^{R}+i\,{\lambda}_{k}^{I}$ of $\L$ together with their algebraic multiplicities,  while the corresponding (generalised) left and right eigenmatrices $L_k$ and $R_k$, which we will refer to as left and right (generalised) eigenmodes, respectively,  
are normalised so that $\Tr(L_k R_l)=\delta_{kl}$. For $L_k$ and $R_k$ corresponding to a Jordan block of size $D_k$, we have introduced \emph{additional rescaling} by  $1/\lambda_k^{d_k-1}$ and $\lambda_k^{d_k-1}$, respectively, where $d_k$ is their position within the block, in order to ensure that the eigenmodes can be chosen unitless. 

Using Hermiticity-preservation of $\L$,  for a real $\lambda_k$, we choose $L_k$ and $R_k$ as Hermitian operators, while for a complex $\lambda_k$, $L_k^\dagger$ and $R_k^\dagger$ are chosen as (generalised) eigenmodes for $\lambda_k^*$. Furthermore, from the positivity and trace preservation of the dynamics generated by $\L$, we have ${\lambda}_{k}^{R}\leq 0$ and ${\lambda}_{k}^{R}= 0$ necessarily corresponds to $D_k=1$ (cf.~Ref.~\cite{Wolf2012}).  We order eigenvalues to decrease in their real part and define $m_\text{ss}$ such that ${\lambda}_{k}^{R}= 0$ for $k\leq m_\text{ss} $ and ${\lambda}_{k}^{R}< 0$ for $k>m_\text{ss} $  ($m_\text{ss}\geq 1$ follows from the trace preservation). 

The system evolution can now be understood in terms of the generalised spectrum of the master operator as [cf.~Eq.~\eqref{eq:rho_t}]
\begin{eqnarray}\label{eq:rho_t_spectral}
	\rho_t&=&  \sum_{k=1}^{m_\text{ss}}  \Tr(L_k \rho_0)  R_{k} \\\nonumber&&
	+\!\!\!\sum_{k> m_\text{ss}}^{D^2}\!\! e^{t\lambda_k}  \bigg[ \Tr(L_k \rho_0) +\!\!\!\sum_{l=1}^{D_k-d_k} \frac{( t\lambda_k)^l}{l!}\Tr(L_{k+l} \rho_0) \bigg]R_{k}.
\end{eqnarray}
From Eq.~\eqref{eq:rho_ss} we have that ${\lambda}_{k}^{I}= 0$ for $k\leq m_\text{ss}$ and the zero right eigenmatrices directly correspond to stationary states~\cite{Baumgartner2008,Albert2016}, while zero left eigenmatrices to conserved observables~\cite{Albert2014,Gough2015} (see also Appendix~\ref{app:Heisenberg} and,  for the relation to dynamical symmetries, Ref.~\cite{Buvca2012}), with the projection on the set of stationary states given by
\begin{eqnarray}\label{eq:rho_ss_spectral}
	\Pss(\rho_0)&=&  \sum_{k=1}^{m_\text{ss}}  \Tr(L_k \rho_0)  R_{k}.
\end{eqnarray}
Thus, the dimension of the set of stationary states is $m_\text{ss}-1$ [note the constraint $\Tr(\rhoss)=1$]. For a generic case of a unique stationary state~\cite{Spohn1977,Evans1977,Schirmer2010,Nigro2019}, that is $m_\text{ss}=1$, we have $R_1={\rho}_{\rm ss}$ with the corresponding conserved observable $L_1=\mathds{1}$ (here, we use the remaining freedom to rescale all right and all left generalized eigenmodes within a Jordan block by a constant and its inverse, respectively).

\begin{figure*}[t!]
	\begin{center}
		\includegraphics[width=\textwidth]{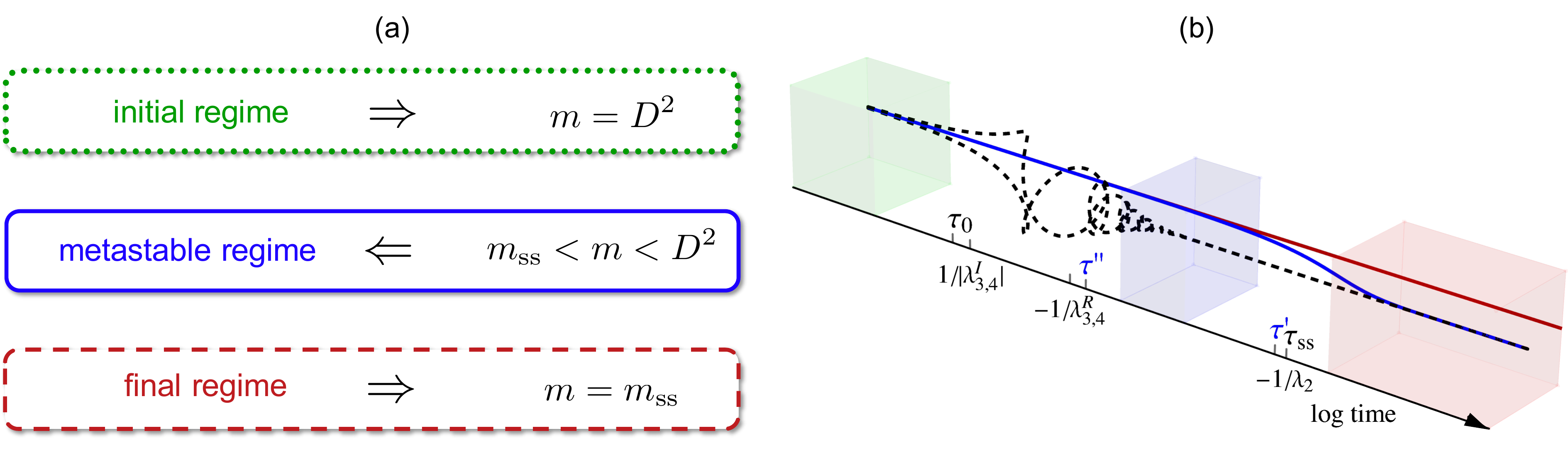} 
		\vspace*{-7mm}
		\caption{
			\textbf{Separation in evolution operator spectrum from approximate stationarity}. 	\textbf{(a)} Approximate stationarity requires a separation in  the spectrum of the evolution operator between $k\leq m$ and $k>m$ [cf.~Eqs.~\eqref{eq:meta_lambda} and~\eqref{eq:meta_lambda'''}].
			When the approximate stationarity corresponds to the initial regime, we have $m=D^2$. Similarly, when the approximate stationarity corresponds to the final regime, we have $m=m_\text{ss}$. Therefore,  the case $m_\text{ss}<m<D^2$, which implies also the separation in the master operator  spectrum [cf.~Eqs.~\eqref{eq:separation}], requires that the approximate stationarity corresponds to metastability.
			\textbf{(b)} The master operator spectrum for the model in Fig.~\ref{fig:example} is: $\lambda_1=0$, $\lambda_2=\kappa$, $\lambda_{3,4}=(\gamma+\kappa)/2\mp i\omega$, with $\rhoss=\mathds{1}/2$, $R_2=S_z$, $R_{3,4}=S_\pm$, and $L_k=2R_k^\dagger$ for $k=1,2,3,4$ [cf. Eq.~\eqref{eq:L_spectral} and note that $\L^\dagger\L=\L\L^\dagger$; the order of eigenvalues is given for $\kappa/\gamma\leq 1$]. In the limit $\kappa/\gamma\ll 1$, there exists a separation in the spectrum with $m=2$ (cf.~the time axis). 
			The depicted spectrum of the evolution operator, $e^{t\lambda_k}$, within the metastable regime (shaded blue)  is approximated by $1$ for $k\leq m$ ($k=1$ red solid, $k=2$ blue solid), while for $k<m$ by $0$ ($k=3,4$ black dashed). In contrast,  in the initial regime (shaded green), all eigenvalues are approximated by $1$, while 
			in the final regime (shaded red), all non-stationary eigenvalues ($k>1$) are approximated by $0$.
		}\vspace*{-7mm}
		\label{fig:separation}
	\end{center}
\end{figure*}

\subsubsection{Eigenmodes as observables} 
We now discuss how the spectrum of evolution operator can be accessed by considering  averages of observables related to left (generalised) eigenmodes. It then follows that its changes are bounded from above by the changes in the dynamics. \\

\emph{Hermitian eigenmodes}.
For a real eigenvalue $\lambda_k=\lambda_k^R$ of the master operator $\L$ in Eq.~\eqref{eq:L_spectral}, the corresponding Hermitian left eigenmode $L_k$ can be chosen as an observable, in which case the average decays exponentially,
\begin{equation}\label{eq:mode_real}
\Tr(L_k\rho_t)=\Tr(L_k\rho_0)\, e^{t\lambda_k}
\end{equation}
 [cf.~Eqs.~\eqref{eq:L_spectral} and~\eqref{eq:rho_t_spectral}].
In particular, for $\rho_0$ supported in the eigenspace of $L_k$ with the absolute value of the corresponding  eigenvalue equal $\lVert L_k \rVert_{\max}$, the normalised average corresponds to the dynamics of a single real mode, $|\Tr(L_k\rho_t)|/\lVert L_k \rVert_{\max}=e^{t\lambda_k}$.
\\

\emph{Non-Hermitian eigenmodes}. For a complex eigenvalue $\lambda_k$ and the corresponding  non-Hermitian left eigenmode $L_k$,  we consider observables
\begin{equation}\label{eq:L_k_phi}
	L_k(\phi)\equiv\frac{e^{i\phi}L_k+e^{-i\phi}L_k^\dagger}{2},
\end{equation}
so that
\begin{equation}\label{eq:mode_complex}
\Tr[L_k(\phi)\rho_t]=
\Tr(L_k\rho_0)\,e^{t\lambda_k^R}\cos[\lambda_I t+\phi+\phi_{k}(\rho_0)],
\end{equation}
where $\Tr(L_k\rho_0)=e^{i\phi_{k}(\rho_0)}|\Tr(L_k\rho_0)|$. In particular, for $\rho_0$ corresponding to the maximum $|\Tr(L_k\rho_0)|$ (which can be chosen pure; cf.~Appendix~\ref{app:norm_def}), we have $|\Tr(L_k\rho_0)|=\lVert L_k[-\phi_{k}(\rho_0)] \rVert_{\max}\geq \lVert L_k(\phi) \rVert_{\max}$, so that $|\Tr[L_k(\phi)\rho_t]|/\lVert L_k(\phi) \rVert_{\max}\geq e^{t\lambda_k^R}|\cos[\lambda_I t+\phi+\phi_{k}(\rho_0)]|$, with the equality for $\phi=-\phi_{k}(\rho_0)$. 
\\

\emph{Generalised eigenmodes}. When an eigenvalue corresponds to a Jordan block, let $L_k$ be the last from the corresponding Hermitian left generalised eigenmodes, that is, $k$ such that $d_k=D_k$ [cf.~Eqs.~\eqref{eq:L_spectral} and~\eqref{eq:rho_t_spectral}]. Then, when  $\lambda_k$ is real, Eq.~\eqref{eq:mode_real} holds,  or, when it is complex, we have Eq.~\eqref{eq:mode_complex}. \\

\emph{Changes in time}.
 As changes in observable averages are bounded by the changes in the dynamics [cf.~Eqs.~\eqref{eq:meta_state_obs_Delta}-\eqref{eq:meta_Delta}], by considering observables related to left eigenmodes of the master operator [cf.~Eqs.~\eqref{eq:mode_real}-\eqref{eq:mode_complex}], it can be shown that 
\begin{equation}\label{eq:change_spectral}
	|e^{t_1\lambda_k}-e^{t_2\lambda_k}|\leq   \lVert e^{t_1\L}-e^{t_2\L}\rVert
\end{equation}
 (see Appendix~\ref{app:spectral_changes} for derivation). 
 Eq.~\eqref{eq:change_spectral} demonstrates that the system evolution changes in time at least as much as the spectrum of the evolution operator does.
 Thus $\lVert e^{t_1\L}-e^{t_2\L}\rVert>0$ for $t_1\neq t_2$, unless $\lambda_k=0$ or $-\infty$ for all $k$, that is the dynamics is trivial with $\L=0$ or infinitely fast $(\I-\Pss)\L=-\infty$~\footnote{Non-zero changes are simply consequence of the Markovianity of the dynamics as the system always changes with time when not initialised in a stationary state; see  Appendix~\ref{app:spectral_nonzero}.}.

\subsubsection{Initial and final regimes}
Using Eq.~\eqref{eq:change_spectral}, we can now characterise timescales of the dynamics in terms of the master operator spectrum.
First, for all $k$ we have ($t_1=t$ and $t_2=0$)
\begin{equation}\label{eq:change_spectral_0}
	|e^{t\lambda_k}-1|\leq   \lVert e^{t\L}-\I\rVert.
\end{equation}
Second, for  $k> m_\text{ss}$ ($t_1=t$ and $t_2\rightarrow \infty$)
\begin{equation}\label{eq:change_spectral_ss}
	|e^{t\lambda_k}|\leq   \lVert e^{t\L}-\Pss\rVert.
\end{equation}

From Eq.~\eqref{eq:change_spectral_0}, we obtain that in the initial regime of Eq.~\eqref{eq:initial} all eigenvalues of the evolution operator 
are approximated by $1$. As $1-e^{t\lambda_k^R}=1-|e^{t\lambda_k}|\leq	|e^{t\lambda_k}-1|$, we have $t_{0}(\C)(-\lambda^R_{D^2})\leq -\ln (1-\C)$. Furthermore, as $|\sin(t\lambda_k^I)| e^{t\lambda_k^R}\leq |e^{t\lambda_k}-1|$, we also have $t_0(\C) \max_{k}|\lambda_k^I|\leq \arcsin[\C/(1-\C)]$ (cf.~Appendix~\ref{app:single1}).

Similarly, from Eq.~\eqref{eq:change_spectral_ss}, in the final regime of Eq.~\eqref{eq:final} all eigenvalues of the evolution operator 
are approximated by $0$. As $e^{t\lambda_k^R}=|e^{t\lambda_k}|$, we have $t_\text{ss}(\C)(-\lambda^R_{m_\text{ss}+1})\geq -\ln (\C)$.

Finally, for the shortest and longest timescales of the dynamics we analogously obtain 
[cf.~Eqs.~\eqref{eq:tau_0} and~\eqref{eq:tau_ss}]
\begin{equation}\label{eq:spectral_tau}
	\tau_0\,(-\lambda^R_{D^2})\leq 1\quad \text{and}\quad	\tau_\text{ss}\,(-\lambda^R_{m_\text{ss}+1})\geq 1.
\end{equation}
It can also be shown that $-\lambda^R_{D^2}\leq \max_{k}|\lambda_{k}|\leq \lVert \L \rVert$.

	\subsection{Spectral decomposition of metastability} \label{sec:spectral_separation}

We now characterise the spectrum of the evolution operator during a time regime where approximate stationarity occurs and show it is found within the overlap of either initial or final regime for the dynamics of its eigenvalues. 
We then relate a non-trivial separation in the spectrum of the evolution operator to a non-trivial separation in the spectrum of the master operator and the phenomenon of metastability (see Fig.~\ref{fig:separation}).

\subsubsection{Separation in evolution operator spectrum}
From Eq.~\eqref{eq:change_spectral}, for a time regime in Eq.~\eqref{eq:meta_cond} and $C_\Delta(t'',t')\leq 1/4$ we have~\cite{Note3}
\begin{subequations}\label{eq:lambda_spectral}
	\begin{align}
			\label{eq:lambda_spectral+}
		e^{t'\lambda_k^R }&\geq 
		E_+^2[C_\Delta(t'',t')]
		\quad \text{or}\quad\\
		\label{eq:lambda_spectral-}
	e^{t'' \lambda_k^R}   &\leq E_-[C_\Delta(t'',t')]
	\end{align}
\end{subequations}
[note that $e^{t\lambda_k^R }-e^{2t\lambda_k^R }\leq |e^{t\lambda_k}-e^{2t\lambda_k}|$ and consider $t''\leq t\leq t'/2$].
Therefore, as the eigenvalues are ordered in decreasing real part, for $C_\Delta(t'',t')< 1/4$
there exists $ m_\text{ss}\leq m \leq D^2$ such that [cf.~Eq.~\eqref{eq:t_0_ss2} and Fig.~\ref{fig:separation}(b)]
\begin{subequations}\label{eq:meta_lambda}
	\begin{align}
		\label{eq:meta_lambda'}
		t' (-\lambda_k^R) &\leq -\ln\{E_+^2[C_\Delta(t'',t')]\},
		&\quad k\leq m,\\
				\label{eq:meta_lambda''}
		t'' (-\lambda_k^R) &\geq  - \ln\{E_-[C_\Delta(t'',t')]\},
		&\quad k>m.
	\end{align}
\end{subequations}
Furthermore, we also have
\begin{eqnarray}
	\label{eq:meta_lambda'''}
(t'-t'') |\lambda_k^I| &\leq& \arcsin\bigg\{\frac{\Cd(t'',t')}{E_{+}[\Cd(t'',t')]}\bigg\},
\quad k\leq m.\qquad
\end{eqnarray}
(note that $|\sin[(t_1-t_2) \lambda_k^I]| e^{t_2\lambda_k^R }\leq |e^{t_1\lambda_k}-e^{t_2\lambda_k}|$ and consider $t_1=t'$ and $t_2=t''$; cf.~Appendix~\ref{app:single2}).
Therefore, when Eq.~\eqref{eq:meta_Delta2} holds, the  time regime for which the approximate stationarity occurs belongs to the initial regimes for the eigenvalues with $k\leq m$ and the final regimes for the eigenvalues with $k>m$ (cf.~Appendix~\ref{app:single1}).

\subsubsection{Separation in master operator spectrum}
When the approximate stationarity corresponds to $m_\text{ss}<m<D^2$ in Eqs.~\eqref{eq:meta_lambda} and~\eqref{eq:meta_lambda'''}, there exists a non-trivial separation in the master operator spectrum with the eigenvalues for $k\leq m$ being negligible in comparison to the real part of eigenvalues with $k>m$ [cf.~Fig.~\ref{fig:separation}(b)]. Indeed,  
\begin{subequations}\label{eq:separation}
	\begin{align}
		\label{eq:separation_R}
		\frac{\lambda_{m}^R}{\lambda_{m+1}^R}  &\leq  \frac{2t''}{t'} \frac{\Cd(t'',t')}{-\ln[\Cd(t'',t')]}(1+...),\quad\\
		\label{eq:separation_I}
		\frac{|\lambda_{k}^I|}{-\lambda_{m+1}^R}  &\leq \frac{t''}{t'-t''}\frac{\Cd(t'',t')}{-\ln[\Cd(t'',t')]}(1+...),\quad k\leq m,
	\end{align}
\end{subequations}
where $2t''/t'',t'/(t'-t'')\leq 1$  [cf.~Eq.~\eqref{eq:meta_cond}]. Here, we expanded up to multiplicative corrections linear in $\Cd(t'',t')$  [cf.~Eq.~\eqref{eq:meta_Delta2}].

 It is important to note that a separation in Eqs.~\eqref{eq:meta_lambda} and~\eqref{eq:meta_lambda'''} with $m_\text{ss}<m<D^2$ cannot occur in the initial and final regimes [which imply $m=D^2$ and $m=m_\text{ss}$, respectively; cf.~Eqs.~\eqref{eq:change_spectral_0} and~\eqref{eq:change_spectral_ss}], so that in this case the approximate stationarity necessarily corresponds to the metastability [cf.~Fig.~\ref{fig:separation}(a)]. Actually, it follows that the bounds in Eqs.~\eqref{eq:dist_0} and~\eqref{eq:dist_ss} can be observed in an experiment by measuring observables related to left (generalised) eigenmodes $k\leq m$ and $k> m$, respectively [cf.~Eq.~\eqref{eq:change_spectral} and see Appendix~\ref{app:spectral_changes}].

Finally, from Eqs.~\eqref{eq:change_spectral} and~\eqref{eq:lambda_spectral}, the timescales of the initial relaxation before the metastable regime [Eq.~\eqref{eq:tau''}] and of the long-time dynamics afterwards [Eq.~\eqref{eq:tau'}]  can be bounded as   [cf.~Eq.~\eqref{eq:spectral_tau} and see Fig.~\ref{fig:separation}(b)]
 \begin{eqnarray}\label{eq:spectral_tau2}
 	\tau''(-\lambda^R_{m+1})\geq1\quad&\text{and}&\quad
 	\tau'(-\lambda^R_{m})\leq1.	
 \end{eqnarray}
Here, the former bound holds when $m<D^{2}$, while the latter when $m>m_\text{ss}$.

\section{Relation to spectral theory of metastability} \label{sec:spectral_theory}

In the previous section, we discussed how metastability relates to a separation in the spectra of evolution and master operators. Such a separation is a prominent feature of the spectral theory of metastability introduced in Ref.~\cite{Macieszczak2016a}. We now clarify how the operational approach established in this work connects to the spectral theory and typically implies its validity.

\subsection{Spectral theory of metastability}

The spectral theory of metastability outlined in Ref.~\cite{Macieszczak2016a} discusses open quantum systems with a large separation in the real part of the spectrum of the master operator, 
\begin{equation}\label{eq:ratio_R}
	\frac{{\lambda}_{m}^{R}}{{\lambda}_{m+1}^{R}}\ll 1
\end{equation}
for $m_\text{ss}<m<D^2$ [cf.~Eq.~\eqref{eq:separation_R}]. 
For correspondence to the definition of metastability introduced in this work, we will also assume that the imaginary parts of the eigenvalues with $k\leq m$ are negligible  [cf.~Eq.~\eqref{eq:separation_I}]
\begin{equation}\label{eq:ratio_I}
	\frac{|{\lambda}_k^{I}|}{{\lambda}_{m+1}^{R}}\ll 1, \quad k\leq m.
\end{equation} 
We refer to (generalised) eigenmodes with $k\leq m$ and $k> m$ as \emph{slow} and \emph{fast} (generalised) eigenmodes, respectively. 

The central assumption of the spectral theory is the existence of a time regime $t''\leq t\leq t'$ during which a system state can be approximated by the projection on the slow eigenmodes [cf.~Eq.~\eqref{eq:rho_t_spectral}],
\begin{equation}\label{eq:rho_t_P}
	\rho_t=\sum_{k=1}^m \Tr(L_k \rho_0) \,{R}_{k}+...\equiv\P(\rho_0)+...,
\end{equation}
in analogy to a stationary  state determined by the projection on the zero eigenmodes in Eq.~\eqref{eq:rho_ss_spectral}. 
That is,  errors in this approximation,
\begin{eqnarray}\label{eq:C_P}
	\C_\P(t'',t')&\equiv&\sup_{\rho_0} \sup_{t''\leq t\leq t'} \lVert\rho_t-\P(\rho_0) \rVert\\\nonumber
	&=&  \sup_{t''\leq t\leq t'}\lVert e^{t\L} -\P\rVert,
\end{eqnarray}
are assumed negligible,
\begin{eqnarray}\label{eq:C_P2}
	\C_\P(t'',t')&\ll& 1.
\end{eqnarray}

The goal of the spectral theory  is to understand the structure of the projection $\P$, which, in turn, determines the structure of system states during the considered time regime as well as their dynamics afterwards.  
In particular, the spectral theories of bimodality and classical metastability in open quantum systems are formulated in Refs.~\cite{Rose2016,Macieszczak2020}, respectively. For proximity to dissipative phase transitions, see also~\cite{Minganti2018}.

\subsection{Operational approach vs.  spectral theory}

\begin{figure*}[t!]
	\begin{center}
		\includegraphics[width=\textwidth]{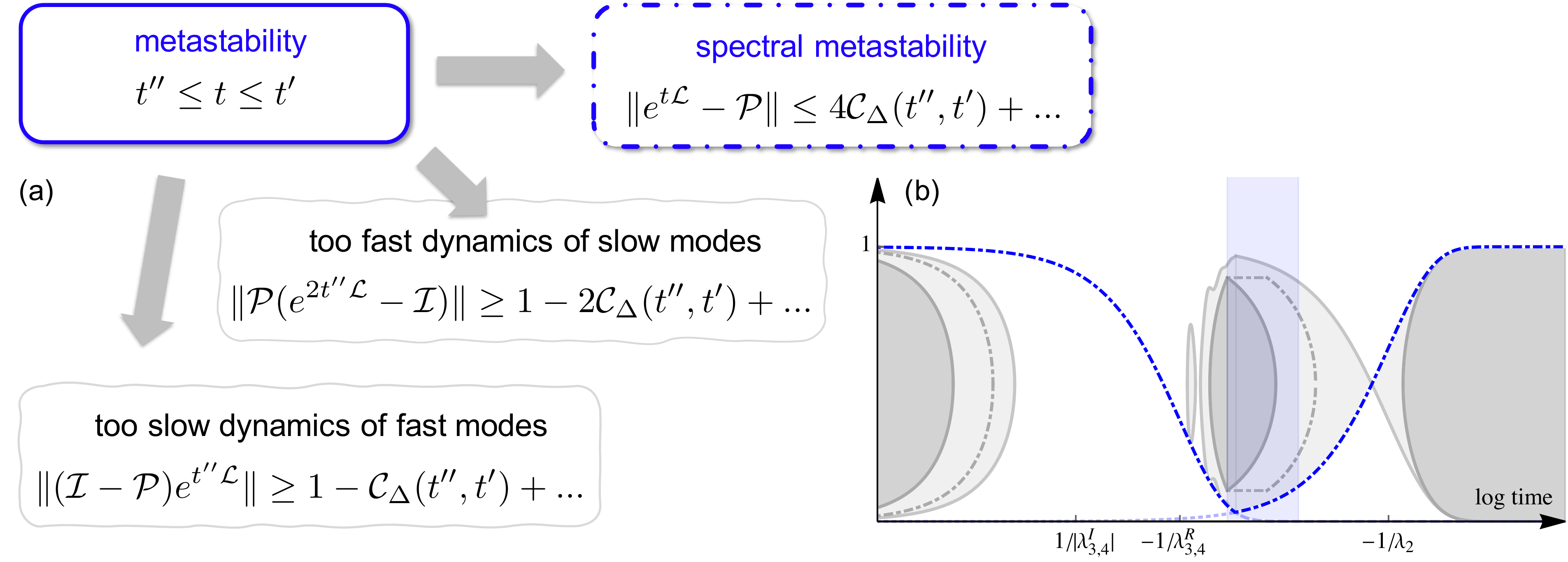} 
		\caption{
			\textbf{Spectral metastability}. \textbf{(a)} When metastability takes place in the dynamics of an open quantum system, that is, there exists a  time regime $t''\leq t\leq t'$ such that Eqs.~\eqref{eq:meta_cond},~\eqref{eq:meta_Delta2}, and~\eqref{eq:meta_cond2} hold, there exists a separation in the spectrum of the corresponding evolution operator [cf.~Eqs.~\eqref{eq:meta_lambda} and~\eqref{eq:meta_lambda'''}]. When this separation is non-trivial, one can consider approximating slow (generalised) eigenmodes by their initial contribution and neglecting fast (generalised) eigenmodes due to their asymptotic decay. Such an approximation is assumed to be valid in the spectral theory of metastability [cf.~Eqs.~\eqref{eq:rho_t_P}-\eqref{eq:C_P2}]. We show that it is indeed valid unless the dynamics of slow  (generalised) eigenmodes or fast (generalised) eigenmodes contributes significantly [see~Eqs.~\eqref{eq:meta_cond3},~\eqref{eq:meta_cond4}, and~\eqref{eq:C_P_better}, and cf.~Eq.~\eqref{eq:C_P4}]. \textbf{(b)} For the model in Fig.~\ref{fig:example} in the limit $\kappa/\gamma\ll 1$, the dynamics is found in proximity to a phase transition  with two disjoint stationary states, $\mathds{1}/2\pm S_z$. Therefore, during the metastable regime (area shaded blue), the dynamics is approximated by the projection on the slow eigenmodes with $k\leq m=2$ [blue dot-dashed; cf.~Eq.~\eqref{eq:rho_t_P}]. This approximation depends on the negligible contributions from the dynamics of the slow eigenmodes (light blue dotted) and from the fast eigenmodes (light blue dashed). In fact, here,   $\lVert e^{t\L}-\P\rVert=\max[\lVert \P(e^{t\L}-\I)\rVert,\lVert(\I -\P)e^{t\L}\rVert]=\max(1-e^{\lambda_2t},e^{\lambda_{3,4}^R t})$,  so that the errors are bounded by $E_-(\lVert e^{t\L}-e^{2t\L} \rVert)$ [$E_\pm(\lVert e^{t\L}-e^{2t\L} \rVert)$ shown as light grey solid; see Fig.~\ref{fig:example}(b)]; cf.~Eqs.~\eqref{eq:meta_lambda},~\eqref{eq:P_better} and~\eqref{eq:IP_better}. To compare with the changes in the dynamics during the metastable regime, we show $E_\pm\{\Cd[t,(t'/t'')t]\}$ (dark grey solid), which are weaker bounds [cf.~Fig.~\ref{fig:example}(c)], but nevertheless valid for time intervals between $t$ and $[t'/(2t'')]t$ rather than only $t$ (e.g., in the metastable regime between $t''$ and $t'/2$; shown as grey dot-dashed). 
		}\vspace*{-7mm}
		\label{fig:spectral}
	\end{center}
\end{figure*}

The spectral theory describes a phenomenon distinct from the initial and final regimes of dynamics. Indeed, using the methods introduced in this work, we show in Appendix~\ref{app:spectral_theory} that 
\begin{eqnarray}\label{eq:dist_0_P}
	\lVert e^{t\L} -\I \rVert&\geq&1-\C_\P(t'',t'),
	\\\label{eq:dist_ss_P}
	\lVert e^{t\L} -\Pss \rVert&\geq&1-\C_\P(t'',t')
\end{eqnarray}	
[cf.~Eqs.~\eqref{eq:initial} and~\eqref{eq:final}], as well as 
\begin{subequations}\label{eq:spectral_P}
	\begin{align}\label{eq:spectral_P1}
		t'' (-\lambda_{m+1}^R) &\geq  - \ln[C_\P(t'',t')],\\
		\label{eq:spectral_P2}
		t' (-\lambda_{m}^R) &\leq -\ln[1-C_\P(t'',t')],
	\end{align}
\end{subequations}
so that $\tau_0\leq t''$ and $t'\leq \tau_\text{ss}$ from Eq.~\eqref{eq:spectral_tau} [cf.~Eq.~\eqref{eq:meta_cond2}]. 
While this phenomenon bears close similarity to the metastability defined in this work, we now argue that the assumption in Eq.~\eqref{eq:C_P2} is \emph{highly non-trivial}. 

For a generalised right eigenmode $R_k$ with $k> m_\text{ss}$ it is always possible to consider a time regime such that a coefficient $e^{t\lambda_k}\Tr(L_k \rho_0)$ in the decomposition of the system state in Eq.~\eqref{eq:rho_t_spectral} is approximated by $\Tr(L_k \rho_0)$ or $0$, with errors negligible in comparison to $\sup_{\rho_0 }|\Tr(L_k \rho_0)|\leq\lVert L_k\rVert_{\max}$ (with the equality for Hermitian $L_k$; see Appendix~\ref{app:norm_def}). Indeed, this can be achieved by considering $t$ within the initial or final regimes of the single-mode dynamics $e^{t\lambda_k}$. When the separation in the spectrum of the master operator given by Eqs.~\eqref{eq:ratio_R} and~\eqref{eq:ratio_I} is present, there exists a time regime corresponding to the overlap of the initial regimes for $k\leq m$ and the final regimes for $k>m$. In that case, individual coefficients in Eq.~\eqref{eq:rho_t_spectral} can be approximated by $1$ and $0$, respectively. The number of (generalised) eigenmodes simultaneously approximated in Eq.~\eqref{eq:rho_t_P}  is, however, given by $D^2-m_\text{ss}$, where the Hilbert space dimension $D$ scales exponentially with size for many-body systems (e.g., for spin chains exponentially in their length). Furthermore, the norms of right (generalised) eigenmodes also need to be considered in this approximation, as $\lVert L_k\rVert_{\max}\lVert R_k\rVert$ is not determined by the condition $\Tr(L_k R_k)=1$ (especially when $R_k\neq L_k^\dagger$, as generally is the case for $\L\L^\dagger\neq\L^\dagger\L$, such a product of norms is not bounded from above; cf.~Refs.~\cite{Song2019,Mori2020}).\\

Nevertheless, we now demonstrate how closely the operational approach and the spectral theory of metastability are related (see Fig.~\ref{fig:spectral}). 
In particular, we give conditions on how when Eqs.~\eqref{eq:meta_cond},~\eqref{eq:meta_Delta2}, and~\eqref{eq:meta_cond2} are fulfilled, the central assumption in Eq.~\eqref{eq:C_P2} also follows.
For simplicity, we refer to the case when Eqs.~\eqref{eq:meta_cond},~\eqref{eq:meta_Delta2}, and~\eqref{eq:meta_cond2} are fulfilled, as the \emph{operational metastability}, while to the case when  Eq.~\eqref{eq:C_P2} holds as the \emph{spectral metastability}.

\subsubsection{When spectral metastability implies operational metastability}

We first show that  when $ \lceil t''/(t'-t'')\rceil \C_\P(t'',t')\ll 1$, the spectral theory of metastability in fact describes the phenomenon of metastability defined in this work. Indeed,
by the triangle inequality we have
\begin{eqnarray}\label{eq:C_P3}
	\Cd(t'',t')\leq 2\C_\P(t'',t'),
\end{eqnarray}
so that Eq.~\eqref{eq:meta_Delta2} follows from Eq.~\eqref{eq:C_P2}.
When $ \lceil t''/(t'-t'')\rceil \C_\P(t'',t')\ll 1$, the considered time regime can be further extended to fulfil the condition in Eq.~\eqref{eq:meta_cond} [cf.~Eq.~\eqref{eq:meta_lin}], and thus the approximate stationarity is implied. Finally, as the separation in the master operator is non-trivial,  the approximate stationarity corresponds to the operational metastability.

In particular, it follows that operational metastability is present in the dynamics of  open quantum systems perturbed away from a static phase transition at a finite size (so called class A of systems displaying the spectral metastability in Ref.~\cite{Macieszczak2016a}; see Figs.~\ref{fig:example},~\ref{fig:separation}(b), and~\ref{fig:spectral}(b) for a minimal model). That is, for an open quantum system with the master operator $\L^{(0)}$ [cf.~Eq.~\eqref{eq:master}] featuring degenerate stationary states,  $m_\text{ss}^{(0)}>1$ [cf.~Eqs.~\eqref{eq:rho_t_spectral}], small enough perturbations in its Hamiltonian and jump operators give rise to a perturbed master operator $\L$, and lead to the operational metastability with $m=m_\text{ss}^{(0)}$ occurring at times longer than the final relaxation time of $\L^{(0)}$ (cf.~Supplemental Material of Ref.~\cite{Macieszczak2016a}).

\subsubsection{When operational metastability implies spectral metastability}

We now discuss when the operational approach to metastability leads to the assumption in Eq.~\eqref{eq:C_P2} being fulfilled. We also characterise the system dynamics during and after the metastable regime in this case.\\

We begin by showing that  the spectral metastability in a time regime $t''\leq t\leq t'$ is in fact equivalent to two approximations. First, when Eq.~\eqref{eq:C_P2} holds, the dynamics of the slow (generalised) eigenmodes can be neglected  [cf.~Eq.~\eqref{eq:spectral_P2}],
\begin{eqnarray}\label{C_P_P}
	\lVert \P(e^{t\L}-\I)\rVert \leq [1+\C_\P(t'',t')]\,\C_\P(t'',t'),
\end{eqnarray}
while the fast (generalised) eigenmodes no longer contribute as a consequence of their decay [cf.~Eq.~\eqref{eq:spectral_P1}],
\begin{eqnarray}\label{C_P_IP}
	\lVert (\I-\P) e^{t\L}\rVert \leq [2+\C_\P(t'',t')]\,\C_\P(t'',t').
\end{eqnarray}
Thus, the considered regime simultaneously belongs the initial regime of the dynamics restricted to the slow (generalised) eigenmodes, $k\leq m$,  and the final regime of dynamics restricted to the fast (generalised) eigenmodes, $k> m$; see Appendix~\ref{app:spectral_theory} for derivations. Second, as $	\lVert e^{t\L}-\P\rVert\leq\lVert \P(e^{t\L}-\I)\rVert+ \lVert (\I-\P) e^{t\L}\rVert$, Eq.~\eqref{eq:C_P2} is implied by
\begin{equation}\label{eq:C_P4} 
\lVert \P(e^{t\L}-\I)\rVert\ll 1\quad \text{and}\quad 	\lVert (\I-\P) e^{t\L}\rVert\ll 1.
\end{equation} \\

We are now ready to present the \emph{central result} of this section (see Fig.~\ref{fig:spectral}). Namely, for the operational metastability leading to the separation in the real part of the spectrum of the evolution operator in Eq.~\eqref{eq:lambda_spectral} with $m_\text{ss}<m< D^2$, and thus to the corresponding separation 
in the real part spectrum of the master operator in Eq.~\eqref{eq:separation}, Eq.~\eqref{eq:C_P2} holds 
provided that 
\begin{align}
	\label{eq:meta_cond3}
			\lVert \P (e^{2t''\L}-\I)\rVert &\leq E_+ [2C_\Delta(t'',t')]=1- 2C_\Delta(t'',t')+...,\quad\\
		\label{eq:meta_cond4}
	\lVert (\I-\P) e^{t''\L}\rVert &\leq E_+[C_\Delta(t'',t')]=1-C_\Delta(t'',t')+....\nonumber\\
	\end{align}
Here, we consider $C_\Delta(t'',t')\leq 0.0997...$ and the metastable regime such that $t'\geq 4t''$, which can be achieved by doubling the length of the metastable regime [cf.~Eqs.~\eqref{eq:meta_lin} and~\eqref{eq:meta_cond}]; we then expand up to linear order in $C_\Delta(t'',t')$ using Eq.~\eqref{eq:meta_Delta2}. Note that in Eqs.~\eqref{eq:meta_cond3} and~\eqref{eq:meta_cond4} \emph{finite} contributions from the dynamics of the (generalised) eigenmodes with $k\leq m$ and  $k>m$ are allowed in contrast to Eq.~\eqref{eq:C_P4}~\footnote{In particular, Eq.~\eqref{eq:meta_cond3} holds when $\lVert \P(e^{t''\L}-\I)\rVert\leq E_+ [2C_\Delta(t'',t')]$, as 
	$\lVert \P(e^{2t\L}-\I)\rVert\leq 2\lVert \P(e^{t\L}-\I)\rVert$; cf.~Eq.~\eqref{eq:0_lin}.}.

Indeed, for this case  in Appendix~\ref{app:spectral_proof}  we show that 
\begin{eqnarray}\label{eq:Pnorm2}
	\lVert P\rVert&\leq& 1+E_- [C_\Delta(t'',t')]+E_-[2C_\Delta(t'',t')]\\\nonumber&=&1+3C_\Delta(t'',t')+....
\end{eqnarray}
This further leads to [cf.~Eqs.~\eqref{eq:all} and~\eqref{eq:ss_IP}, and see Fig.~\ref{fig:spectral}(b)]
\begin{eqnarray}	\label{eq:P_better}
	\lVert \P (e^{t\L}\!-\!\I) \rVert&\leq&\! E_-[\lVert \P\rVert\Cd\!(t''\!,t')]\leq \Cd\!(t''\!,t')\!+\!...,\\
	\label{eq:IP_better}
	\lVert (\I\!-\!\P) e^{t\L} \rVert&\leq& \!E_-[\lVert \I\!-\!\P\rVert\Cd\!(t''\!,t')]\leq 2\Cd\!(t''\!,t')\!+\!...\,\,\,\,\,\,\,\,\,\,\,
\end{eqnarray}
for $t''\leq t\leq t'/2$, while for $t'/2< t\leq t'$, $E_-[\lVert \P\rVert\Cd(t'',t')]$ is replaced by $E_-[\lVert \P\rVert\Cd(t'',t')]+\lVert \P\rVert\Cd(t'',t')=2\Cd(t'',t')+...$.

Therefore, Eq.~\eqref{eq:C_P4}  is implied by Eq.~\eqref{eq:meta_Delta2}, and thus the assumption in  Eq.~\eqref{eq:C_P2} of the spectral theory of metastability is indeed fulfilled.
In fact, we obtain  (cf.~Fig.~\ref{fig:spectral}).
\begin{eqnarray}\label{eq:C_P_better}
	\C_{\P}(t'',t')&\leq& E_-[\lVert \P\rVert\Cd(t'',t')]\!+\! E_-[\lVert \I\!-\!\P\rVert\Cd(t'',t')]\qquad\\\nonumber
	&&+\Cd(t'',t')
	\leq4\Cd(t'',t')+.... \qquad
\end{eqnarray}

We also find that the corrections in Eq.~\eqref{eq:rho_t_P} \emph{increase approximately linearly}  before and during the metastable regime for the (generalised) eigenmodes with $k\leq m$ [cf.~Fig.~\ref{fig:spectral}(b)] since 
 \begin{equation}\label{eq:0_exp_P}
 	2t\lVert \P\L\rVert-	e^{t\lVert \P\L\rVert}+1  \leq \lVert \P (e^{t\L}-\I) \rVert\leq e^{t\lVert \P\L\rVert}-1
 \end{equation}
  [cf.~Eqs.~\eqref{eq:0_exp} and~\eqref{eq:0_exp2}] as well as 
\begin{eqnarray}\label{eq:P_lin3}
	&&(t'-t'')\lVert  \P\L \rVert\leq E_2[\lVert \P\rVert\Cd(t'',t')]=2\Cd(t'',t')+... \qquad\,\,
\end{eqnarray}
for $\Cd\leq0.0837...$ [cf.~Eq.~\eqref{eq:all_lin3} and see Appendix~\ref{app:spectral_proof}].  Due to the contractivity of the dynamics,  the contribution from the (generalised) eigenmodes with $k>m$ decreases at all times, but during the metastable regime and afterwards it \emph{decays at least exponentially}  as 
\begin{equation}\label{eq:ss_exp_P}
	\lVert (\I-\P) e^{nt\L} \rVert\leq 	\lVert (\I-\P) e^{t\L} \rVert^n
\end{equation}
[cf.~Eq.~\eqref{eq:ss_exp} and Fig.~\ref{fig:spectral}(b)].
These results could be used to aid the search of a metastable regime in the dynamics of an open quantum system, with Eq.~\eqref{eq:ss_exp_P} limiting $t''$ from below, while Eqs.~\eqref{eq:0_exp_P} and~\eqref{eq:P_lin3} restricting $t'$ from above (cf.~Fig.~\ref{fig:example}).
\\

While we demonstrate here that the operational metastability with the conditions in Eqs.~\eqref{eq:meta_cond3} and~\eqref{eq:meta_cond4} corresponds to the spectral metastability,  we would like to emphasize that we do not argue that these conditions can be broken by the operational metastability. In fact, they appear as a consequence of considering dynamics with respect to a general projection $\P$ that commutes with the master operator. Only when the conditions are fulfilled, $\P$ is determined as the projection in Eq.~\eqref{eq:rho_t_P}  (see Appendix~\ref{app:spectral_proof}). 

On a related note, if the operational metastability corresponded to  $m=D^2$ or $m=m_\text{ss}$ in Eq.~\eqref{eq:lambda_spectral} (cf.~Fig.~\ref{fig:separation}), we would have $\P=\I$ or $\P=\Pss$, respectively. In such cases, however, the  approximation in Eq.~\eqref{eq:rho_t_P} would be in the contradiction with Eq.~\eqref{eq:dist_0} or Eq.~\eqref{eq:dist_ss}, respectively [cf.~Eqs.~\eqref{eq:C_P} and~\eqref{eq:C_P2}].

\section{Conclusions and outlook}\label{sec:conclusion}

In this work, we  introduced an operational approach to measuring  changes in the evolution of a finitely dimensional Markovian open quantum system by considering how averages of system observables change with time. This led to the changes quantified by the distance in the induced trace norm between the corresponding evolution operators. This further allowed us to consider the initial and final regimes of the dynamics when the system is approximately stationary as a consequence of being approximated either by its initial or asymptotic states. We also characterised the shortest and longest timescales in the dynamics. Finally, we linked the changes in the system to the experimentally observable changes in the evolution operator spectrum, which in turn enabled us to limit  the initial and final regimes, as well as, the dynamics timescales, in terms of the master operator eigenvalues. These results presented in Secs.~\ref{sec:initial_final} and~\ref{sec:spectral_mode}, with corresponding proofs in Appendixes~\ref{app:dynamics} and~\ref{app:spectral}, are valid for any finitely dimensional Markovian open quantum dynamics.

The central goal of this work, however, was to understand how negligible changes in the dynamics can arise beyond its initial and final regimes. This was achieved by establishing an analogy of open quantum dynamics to the single-mode dynamics, which motivated the introduction of non-perturbative methods for the investigation of  changes in the dynamics with respect to the logarithmic rather than the linear scale of time. Using those methods, we showed that there may exist a distinct time regime during which an open quantum system is approximately stationary with its states different both from initial and asymptotic states. 
We recognised this phenomenon as metastability, which implies the existence of quasi-conserved observables. 
We also characterised the dynamics leading towards the metastable regime and taking place afterwards. Finally, we connected metastability to a separation in the spectra of evolution and master operators. We also discussed general conditions for the validity of the spectral theory of metastability~\cite{Macieszczak2016a}, typically associated with such a separation. While we considered dynamics governed by a master operator in Eq.~\eqref{eq:L},  only its positivity, instead of the complete positivity, was exploited in derivations; this renders all results valid for positive dynamics (see, e.g., Ref.~\cite{Idel2013}).

Our approach to  quantifying changes in the system states with respect to the  trace norm arises naturally from considering changes in the averages of observables. It is important to note, however, that it lends itself to yet another operational interpretation. Namely, it determines the minimal average error of distinguishing the system states at different times during a considered time regime as at least $1/2-\Cd(t'',t')/4$
via the Holevo-Helstrom theorem (see, e.g., Ref.~\cite{Nielsen2010}). In particular, during the metastable regime, the error approaches that of a random guess, so that system states at different times are virtually indistinguishable. This can also be confirmed by considering the fidelity between states  instead of the trace norm, as the minimal fidelity for a given regime is bounded from below by $1-\Cd(t'',t')/2$ and from above by $1-[\Cd(t'',t')/2]^2$
	 by the Fuchs-van de Graaf inequalities (see, e.g., Ref.~\cite{Nielsen2010}). Furthermore, this operational approach translates directly to Markovian dynamics of probability distributions describing classical systems with finite number of configurations (with the trace and max norms replaced by $l_1$ and $l_{\max}$ vector norms, respectively),  and thus also to the results of Refs.~\cite{Gaveau1987,Gaveau1998,Bovier2002,Gaveau2006,Kurchan2016} on metastability in classical stochastic dynamics. It would be interesting to see how this approach can be further adapted for infinitely dimensional quantum and classical systems.

We considered here Markovian open quantum systems,  whose asymptotic states were independent from time.  
Notions of approximate stability allowing for the general case of unitary dynamics of asymptotic states and generalised metastability, in relation to Refs.~\cite{Bellomo2017,Buvca2021}, will be investigated in the future. Finally, a more general property of the dynamics crucially exploited in this work was its Markovianity and it remains to be seen whether a similar approach can be also used to understand how metastability may arise in non-Markovian dynamics,  e.g., how prethermalisation may occur in the dynamics of a part of a larger closed quantum system~\cite{Gring2012}.

\bigskip

\acknowledgments
K.M.\ thanks R.~L.~Jack for comments and gratefully acknowledges support from a Henslow Research Fellowship. 


\input{MetastabilityOperational_arXiv2.bbl}

\begin{appendix}

\section*{Appendix}

In this Appendix, we provide complementary results and derivations to the main text. In Appendix~\ref{app:single}, we discuss the dynamics and approximate stationarity of a complex mode. In Appendix~\ref{app:norm}, we review the definitions of trace and max norms for operators, as well as, of the corresponding induced norms for superoperators. We derive the results on general open quantum dynamics  in Appendix~\ref{app:dynamics} and give proofs relevant to the phenomenon of metastability in Appendix~\ref{app:meta}. We then extend the results of the main text to the dynamics in the Heisenberg picture in Appendix~\ref{app:Heisenberg}. Finally, the derivations relevant for the spectral decomposition of the open quantum dynamics can be found  in Appendix~\ref{app:spectral}, while the proofs in relation to the spectral theory of metastability are given in Appendix~\ref{app:spectral_relation}.

\section{Single-mode dynamics}\label{app:single}

In analogy to the dynamics of a real mode considered in Sec.~\ref{sec:single}, we discuss here the dynamics of a complex mode, which describes a spiral motion in a plane.

\subsection{Dynamics}\label{app:single1}

We consider the dynamics given by $e^{t\lambda}$, where  a complex rate $\lambda=\lambda^R+i\lambda^I$ features a negative real part,  $\lambda^R<0$.
In analogy to accessing a state of the quantum system by measuring observables, we consider the real part of the function $e^{t\lambda+i\phi}$, i.e., $e^{t\lambda_R}\cos(t\lambda^I +\phi)$. A real $\phi$ corresponds to the choice of a \emph{real reference frame} as it determines the axis on which $e^{t\lambda}$ is projected (the real axis rotated clockwise by $\phi$); cf.~Fig.~\ref{fig:spiral}(a).

Two timescales in the dynamics are $-1/\lambda^R$ and  $1/|\lambda_I|$, which describe the amplitude decay and  the period of oscillations, respectively.
For a given $\phi$,  in analogy to Eqs.~\eqref{eq:c} and~\eqref{eq:t_0_ss}, we introduce initial and final regimes with respect to the initial and asymptotic values $\cos(\phi)$ and $0$, respectively. As the functions $|e^{t\lambda^R}\!\cos(t\lambda^I +\phi)-\cos(\phi)|$ and  $e^{t\lambda^R}|\cos(t\lambda^I\!+\!\phi)|$ are not monotonic in time $t$, however, the initial and final regime need to be further specified as $t\leq t_\text{0}(c;\phi)$ and $t\geq t_\text{ss}(c;\phi)$, respectively, where $t_\text{0}(c;\phi)$ is  the shortest time such that $|e^{t_\text{0}(c;\phi)\lambda^R}\!\!\cos[t_\text{0}(c;\phi)\lambda^I \!\!+\!\phi]-\cos(\phi)|=c$, while $t_\text{ss}(c;\phi)$ is  the longest time such that $e^{t_\text{ss}(c;\phi)\lambda^R}\!|\!\cos[t_\text{ss}(c;\phi)\lambda^I \!\!+\!\phi]|=c$. Note that in the initial regime both the decay and oscillations are negligible. 

The \emph{overlap} of the initial regimes or the final regimes for all $\phi$ yields [cf.~Eq.~\eqref{eq:t_0_ss}]
\begin{equation}\label{eq:t_0_ss_C}
	|e^{t\lambda}\!-\!1| \leq c\quad\text{or}\quad|e^{t\lambda}|\leq c,
\end{equation}
where the initial regime is specified as times $t\leq t_0(c)$, where $t_0(c)$ is the \emph{shortest} time such that $|e^{t_0(c)\lambda}\!-\!1|=c$ [when  $|e^{t_\text{ss}(c)\lambda}|=c$, 
$|e^{t\lambda}|<c$ holds for $t\geq t_\text{ss}(c)$ as $|e^{t\lambda}|=e^{t\lambda^R}$ decays with time]; see Fig.~\ref{fig:spiral}. 
Indeed, Eq.~\eqref{eq:t_0_ss_C} follows from
\begin{equation}\label{eq:phi_max}
	\sup_\phi\, [e^{t\lambda^R}\!\!\cos(t_1\lambda^I \!\!+\!\phi)\!-\!e^{t_2\lambda^R}\!\!\cos(t_2\lambda^I\!\! +\!\phi)]=|e^{t_1\lambda}-e^{t_2\lambda}|.
\end{equation}
Here, we normalise differences in the function values by the maximum over all $\phi$ equal $1$, which coincides with the induced norm for axis projections of complex numbers.

From Eq.~\eqref{eq:t_0_ss_C}, it follows that 	[note that $1-|e^{t\lambda}|\leq|e^{t\lambda}\!-\!1|$ and cf.~Eq.~\eqref{eq:t_0_ss2}]
\begin{eqnarray}\label{eq:t_0_ss_C2}
	&t(-\lambda^R)\leq -\ln(1-c)=c+...\,\,\text{or}\,\, t	(-\lambda^R)\geq-\ln(c).\qquad
\end{eqnarray} 
For the initial regime,  it also follows that [note that the difference to the initial value in the reference frame with $\phi=-t\lambda^I-\pi/2$ is $e^{t\lambda^R}|\sin(t\lambda^I)|\geq (1-c)|\sin(t\lambda^I)|$]
\begin{eqnarray}\label{eq:t_0_C3}
	& t|\lambda^I|\leq \arcsin\left(\frac{c}{1-c}\right)=c+....\quad
\end{eqnarray} 
Here, the inequality is valid for $c\leq1/2$ and we expanded up to linear in $c$ using Eq.~\eqref{eq:c}.
Therefore, the initial regime of approximate stationarity is in general shorter than for a real mode, while  the final regime remains the same; cf.~Figs.~\ref{fig:exp}(b) and~\ref{fig:spiral}(b). As a result, the regimes defined in Eq.~\eqref{eq:t_0_ss_C} can be distinct also for $c>1/2$ [for $c<e^{T\lambda^R}$, where $T$ is the shortest time such that $2e^{T\lambda^R}\!\!\cos(T\lambda^I)=1$].

We note that by considering real reference frames, we access the dynamics of the \emph{complex mode} $e^{t\lambda}$, which describes the spiral motion with the radius decaying  at the rate $\lambda^R$ and the period equal $1/|\lambda_I|$. Despite the presence of two  timescales in the dynamics, the initial and final regimes in Eq.~\eqref{eq:t_0_ss_C} are the only distinct regimes of approximate stationarity, as  the counterpart of the dynamics related to the imaginary part of the rate does not feature any asymptotic limit (see also Appendix~\ref{app:single2}).

\begin{figure}[t!]
	\begin{center}
		\includegraphics[width=\columnwidth]{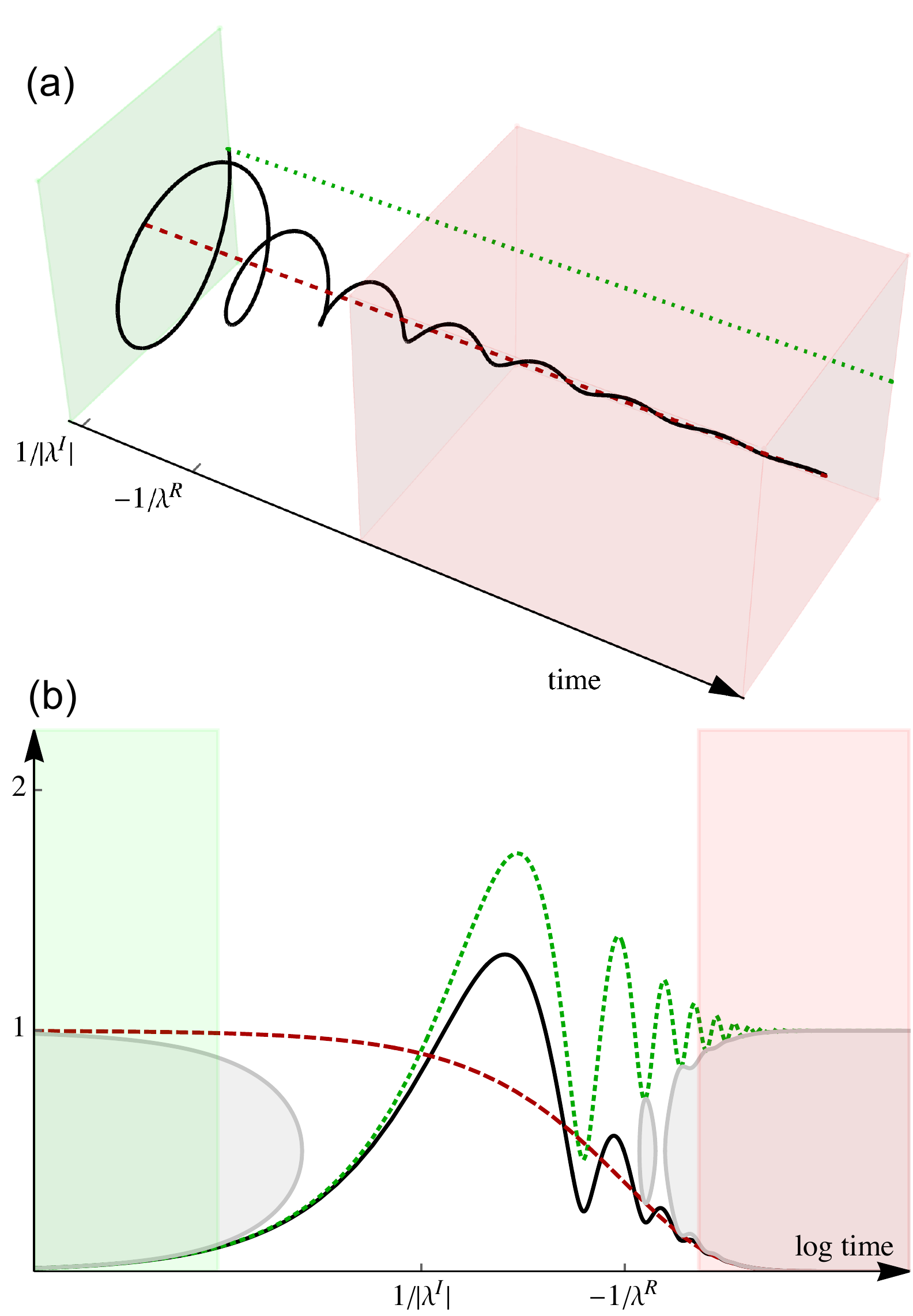} 
		\caption{
			\textbf{Complex mode and its approximate stationarity}. \textbf{(a)} A complex single-mode dynamics described by $e^{t\lambda}$, where $\lambda=\lambda^R+i\lambda^I$ with $\lambda_R<0$, corresponds to a spiral motion in the complex plane with the period $1/|\lambda^I|$	and its radius decaying at the rate $\lambda^R$. As in the real case, the initial and asymptotic values are $1$ (green dotted) and $0$ (red dashed), respectively, but the initial regime is shorter, while the final regime remains  unchanged [areas shaded green and red, respectively, for $c=0.1$ in Eq.~\eqref{eq:t_0_ss_C}; $\lambda^I\!/(-\lambda^R) =10$]. \textbf{(b)} While the distance to the asymptotic limit (red dashed) decays exponentially at the rate $\lambda^R$ , the distance to the initial value (green dotted) is non-monotonic, and so are changes on the logarithmic scale, e.g., $|e^{t\lambda}-e^{2t\lambda}|$ (black solid). Nevertheless, the dynamics changes negligibly only in the initial and final regimes. For $c_\Delta=|e^{t\lambda}-e^{2t\lambda}|\leq 1/4$, $E_+(c_\Delta)$ and $E_-(c_\Delta)$ (grey solid) provide lower and upper bounds on the distance to the initial and asymptotic value at time $t$ (excluded values shaded grey) [this follows from $|e^{t\lambda}|(1-|e^{t\lambda}|), |e^{t\lambda}-1|(1-|e^{t\lambda}-1|)\leq c_\Delta$; cf.~Eqs.~\eqref{eq:change2} and~\eqref{eq:change2_+-}].
		}\vspace*{-7mm}
		\label{fig:spiral}
	\end{center}
\end{figure}

\subsection{Changes in time} \label{app:single2}

We now consider how the complex mode changes with time, that is,  the distance between its values at different times, $|e^{t_1\lambda}-e^{t_2\lambda}|$ and $t_1\neq t_2$. This approach  has been already used with respect to initial and asymptotic function values to define the initial and final regimes in Eq.~\eqref{eq:t_0_ss_C}. As a consequence, within those regimes the changes negligible, as they are bounded by $2c$, $|e^{t_1\lambda}-e^{t_2\lambda}|	\leq |e^{t_1\lambda}-1|+|e^{t_2\lambda}-1| $ and $|e^{t_1\lambda}-e^{t_2\lambda}|	\leq |e^{t_1\lambda}|+ |e^{t_2\lambda}|$. 

Similarly, as in the case of a real mode, the dynamics is contractive
$|e^{ (t_1+\delta t)\lambda}-e^{(t_2+\delta t)\lambda}|=|e^{\delta t\lambda}| |e^{t_1 \lambda }-e^{t_2\lambda}|\leq |e^{t_1 \lambda }-e^{t_2\lambda}|$
for $\delta t\geq 0$, which motivates considering changes with respect to the logarithmic rather than the linear scale.  
We show below that in this case the complex mode changes negligibly only within the initial and final regimes [see~Fig.~\ref{fig:spiral}(b)].

Analogously to Eq.~\eqref{eq:change2}, we can consider changes between $t$ and $2t$ bounded by $c_{\Delta}$ in each reference frame, which requires [cf.~Eq.~\eqref{eq:phi_max}]
\begin{equation}\label{eq:change2_C}
	|e^{t\lambda}-e^{2t\lambda}|\leq   c_{\Delta}.
\end{equation}
As 	$|e^{t\lambda}|-|e^{2t\lambda}|\leq |e^{t\lambda}-e^{2t\lambda}|$, we obtain 
\begin{equation}\label{eq:change2_R} 
	e^{t\lambda^R}-e^{2t\lambda^R}\leq c_\Delta,
\end{equation}
which for $c_\Delta\leq 1/4$ in turn implies [cf.~Eq.~\eqref{eq:change2_+-}]
\begin{equation}\label{eq:t_0_ss_R}
	e^{t\lambda^R}\geq E_+(c_\Delta)\quad\text{or}	\quad e^{t\lambda^R}\leq E_-(c_\Delta).
\end{equation}

As $e^{t\lambda^R}$ decays with time $t$, in the latter case all times between $t$ and $2t$ belong to the final regime in Eq.~\eqref{eq:t_0_ss_C} with $c=E_-(c_\Delta)$.
In order to show that the former case corresponds to the initial regime, 
we further assume that also changes for time differences $0\leq \delta t\leq t$ are bounded by $c_{\Delta}$, that is,  
\begin{equation}
	\label{eq:change2_C2}
	|e^{t\lambda}-e^{(t+\delta t)\lambda}|\leq   c_{\Delta}
\end{equation}
[note that $|e^{t\lambda}-e^{2t\lambda}|$ is the maximal change for times between $t$ and $2t$ only for $t\leq T_0$, where $|e^{T_0\lambda}-1|=\sup_{t\geq 0}|e^{t\lambda}-1|$; otherwise the maximal change is $|e^{t\lambda}-e^{(t+T_0)\lambda}|$].
It then follows that 
\begin{equation}\label{eq:t_0_R2}
	t|\lambda^I|	\leq \arcsin\left[\frac{c_\Delta}{E_+(c_\Delta)}\right]=c_\Delta+...,\quad
\end{equation}
where we expanded up to linear order in $c_{\Delta}$  using Eq.~\eqref{eq:c_delta}  [note that the change in the reference frame with $\phi=-(t+\delta t)\lambda^I-\pi/2$  is $e^{t\lambda^R}|\sin(\delta t\lambda^I)|\geq E_+(c_\Delta)|\sin(\delta t\lambda^I)|$].
Therefore, for $0\leq \delta t\leq t$ we have 
$|e^{\delta t \lambda}-1|\leq e^{t|\lambda|}-1\leq 
e^{\arcsin[c_\Delta/E_+(c_\Delta)] }/E_+(c_\Delta)-1=
2c_\Delta+...$, while 
$|e^{(t+\delta t) \lambda}-1|\leq 
|e^{ t \lambda}-1| +c_\Delta\leq3 c_\Delta+...$ [cf.~Eq.~\eqref{eq:change2_C2}].
Thus,  times between $t$ and $2t$ belong to the initial regime in Eq.~\eqref{eq:t_0_ss_C}  with $c=3c_\Delta+...$, which is distinct from the final regime with $c=E_-(c_\Delta)$ already when $c_\Delta<0.185...$.

\section{Trace and max norms}\label{app:norm}

Here, we review definitions of the trace and max norms for operators acting on a Hilbert space of a finite dimension.  We also discuss the relation between the corresponding induced norms for Hermiticity-preserving superoperators, with a special focus on the master operator and superoperators related to observables. 

\subsection{Definitions}\label{app:norm_def}

The \emph{trace norm} is defined for an operator $X$ as the sum of its singular eigenvalues, i.e., the eigenvalues of $\sqrt{X^\dagger X}$,
\begin{equation}\label{eq:norm_trace}
	\lVert X \rVert \equiv\Tr(\sqrt{X^\dagger X}).
\end{equation}
For Hermitian operators the trace norm is equal to the sum of absolute values of its eigenvalues.
In particular, for a positive operators, the trace norm equals their trace, so that for a density matrix $\rho$ we have $\lVert \rho\rVert =1$. \\

The \emph{max norm} is defined as the maximal singular eigenvalue
\begin{equation}\label{eq:norm_max}
	\lVert X \rVert_{\max} \equiv \sup_\rho \Tr(\sqrt{X^\dagger X}\rho)= \sup_\rho \sqrt{\Tr(X^\dagger X\rho)}.
\end{equation}
Since $ |\Tr(X\rho)|^2 \leq \Tr(X^\dagger X\rho) \Tr(\rho)=\Tr(X^\dagger X\rho)$ from the Cauchy-Schwarz inequality with respect to $\rho^{1/2}X^\dagger$ and $\rho^{1/2}$, we have
\begin{equation}\label{eq:norm_max2}
	\lVert X \rVert_{\max} \geq \sup_\rho |\Tr(X\rho)|,
\end{equation}
where the supremum is achieved for a pure state as $|\Tr\{X [p\rho+(1-p)\rho']\}|\leq p|\Tr(X\rho)|+(1-p)|\Tr(X\rho')|\leq \max[|\Tr(X\rho)|,|\Tr(X\rho')|]$ for $0\leq p\leq 1$. In particular, 	when $X$ is a Hermitian operator, the inequality in Eq.~\eqref{eq:norm_max2} is saturated for $\rho$ supported in the eigenspace of $X$ with the eigenvalue whose absolute value equals $\lVert X \rVert_{\max} $.\\

Both norms can be  related by the \emph{von Neumann trace inequality}~\cite{vonNeumann1937,Mirsky1975,Grigorieff1991}
\begin{equation}\label{eq:vN}
	|\Tr(XY)|=|\Tr(YX)|\leq\sum_{n=1}^{D}\sigma_n^{(X)}\sigma_n^{(Y)} ,
\end{equation}
where $\sigma_n^{(X)}$ denote singular eigenvalues of $X$ ordered decreasing in value, and $D$ is the dimension of the space on which $X$ and $Y$ act. As  $\sum_n\sigma_n^{(X)}=\lVert X\rVert$ and $\sigma_n^{(X)}\leq \lVert X\rVert_{\max}$, we have
\begin{eqnarray}\label{eq:vN2}
	|\Tr(XY)|&\leq&  \min\left(\lVert X\rVert_{\max}\lVert Y\rVert,\lVert X\rVert\lVert Y\rVert_{\max} \right).
\end{eqnarray}

\subsection{Induced norms and their correspondence}\label{app:norm_ind}

\subsubsection{Definitions}
For a superoperator $\mathcal{X}$ acting linearly on operators,  the norm induced by the trace norm is 
\begin{equation}\label{eq:norm_trace_ind}
	\lVert \mathcal{X} \rVert \equiv\sup_{X\neq 0}\frac{\lVert \mathcal{X}(X) \rVert }{\lVert X \rVert }.
\end{equation}
In this work, we consider Hermiticity-preserving superoperators, $[\mathcal{X}(X)]^\dagger=\mathcal{X}(X^\dagger)$, which we restrict to the space of Hermitian operators. In this case, the norm of a superoperator can be shown to be achieved for a pure state, $\rho=\rho^2$ (or, more generally, a rank-one Hermitian operator); cf.~Ref.~\cite{Watrous2005}. Indeed, let $X$ be a Hermitian operator with eigenvalues $x_n$ (possibly degenerate) and projections on the corresponding one-dimensional eigenspaces denoted as $\rho_n$, $X=\sum_{n} x_n \rho_n$. For a superoperator $\mathcal{X}$, we have $\lVert\mathcal{X}(X)\rVert/\lVert X\rVert\leq \sum_n |x_n| \lVert\mathcal{X}(\rho_n)\rVert /(\sum_n |x_n| )\leq \max_n \lVert\mathcal{X}(\rho_n)\rVert$.

Analogously to Eq.~\eqref{eq:norm_trace_ind}, the norm induced by the max norm is defined as 
\begin{equation}\label{eq:norm_max_ind}
	\lVert \mathcal{X} \rVert \equiv\sup_{X\neq 0}\frac{\lVert \mathcal{X}(X) \rVert_{\max} }{\lVert X \rVert_{\max} }.
\end{equation}

\subsubsection{Correspondence}
Below, we argue that for a Hermiticity-preserving superoperator $\mathcal{X}$ restricted to the space of Hermitian operators we have
\begin{equation}\label{eq:norm_trace_max}
	\lVert \mathcal{X}\rVert =	\lVert \mathcal{X}^\dagger\rVert_{\max}.
\end{equation}
Here, $\mathcal{X}^\dagger$ is the Hermitian conjugate of $\mathcal{X}$.\\

Indeed, from Eq.~\eqref{eq:vN2} we have
\begin{eqnarray}
	&&	\lVert \mathcal{X}\rVert = \sup_\rho 	\lVert \mathcal{X}(\rho)\rVert\geq \sup_\rho \sup_{X\neq 0}	\frac{|\Tr[X\mathcal{X}(\rho)]|}{\lVert X\rVert_{\max}} \\\nonumber
	&&=\sup_{X\neq 0}\sup_\rho	\frac{|\Tr[\mathcal{X}^\dagger(X)\rho]|}{\lVert X\rVert_{\max}}=\sup_{X\neq 0}\frac{\lVert \mathcal{X}^\dagger(X)\rVert_{\max}}{\lVert X\rVert_{\max}}=\lVert \mathcal{X}^\dagger\rVert_{\max},
\end{eqnarray}
where $\rho$ is a density matrix and $X$ is a Hermitian operator. Furthermore, this inequality is saturated as 
\begin{eqnarray}
	&&	\sup_{X\neq 0}	\frac{|\Tr[X\mathcal{X}(\rho)]|}{\lVert X\rVert_{\max}} =\lVert \mathcal{X}(\rho)\rVert,
\end{eqnarray}
which follows by considering $X=\Pi_+-\Pi_-$, where $\Pi_+$ and $\P_-$ are projections on the direct sum of positive and negative eigenspaces of $ \mathcal{X}(\rho)$, respectively, so that $\Tr[X\mathcal{X}(\rho)]=\lVert \mathcal{X}(\rho)\rVert$ and $\lVert X\rVert_{\max}=1$.

\subsection{Induced norms of master operator}\label{app:norm_master}

\subsubsection{Master operator and continuity of dynamics}
Below, we show that (also proven in Ref.~\cite{Watrous2005})
\begin{equation}\label{eq:norm_master}
	\lVert e^{t\L}\rVert =	\lVert e^{t\L^\dagger}\rVert_{\max} =1,
\end{equation}
which is a consequence of the positivity and  trace preservation of the dynamics. 
We further show that the dynamics is (uniformly) continuous,  
\begin{equation}\label{eq:master_cont}
	\lim_{dt\rightarrow0} \lVert e^{(t+dt)\L}-e^{t\L}\rVert  =0.
\end{equation}	
As a result, its distance to any superoperator $\T$, e.g., $\I$ or $\Pss$,  is continuous  as well,
\begin{equation}\label{eq:master_distance}
	\lim_{dt\rightarrow0} \lVert e^{(t+dt)\L}-\T\rVert  =\lVert e^{t\L}-\T\rVert .
\end{equation}\\

As the master operator $\L$ is Hermiticity-preserving, so is $e^{t\L}$ for any time $t$, and thus the first equality in Eq.~\eqref{eq:norm_master} follows from Eq.~\eqref{eq:norm_trace_max}. The second inequality follows from the fact that $e^{t\L}$ is trace preserving and (completely) positive, so that it maps density matrices to density matrices, and thus $\lVert e^{t\L}\rVert =\sup_\rho \lVert e^{t\L}(\rho)\rVert =1$. 

Furthermore, $\lVert e^{(t+dt)\L}-e^{t\L}\rVert \leq e^{dt \lVert \L\rVert}-1\rightarrow 0$ when $dt\rightarrow 0$, which gives Eq.~\eqref{eq:master_cont}.  Eq.~\eqref{eq:master_distance} then follows by the triangle inequality,   $\lVert e^{t\L}-\T\rVert-\lVert e^{(t+dt)\L}-e^{t\L}\rVert  \leq 	\lVert e^{(t+dt)\L}-\T\rVert  \leq \lVert e^{t\L}-\T\rVert+\lVert e^{(t+dt)\L}-e^{t\L}\rVert  $.

\subsubsection{Projections}
As Eq.~\eqref{eq:norm_master} holds for any time $t$, we also obtain that for  the projection $\Pss$ in Eq.~\eqref{eq:rho_ss_spectral}
\begin{equation}\label{eq:norm_Pss}
	\lVert \Pss\rVert =	\lVert \Pss^\dagger\rVert_{\max} =1.
\end{equation}

In fact, for a Hermiticity-preserving projection $[\P(X)]^\dagger=\P(X^\dagger)$ and $\P^2=\P$, we always have 
\begin{equation}\label{eq:norm_P}
	\lVert \P\rVert =	\lVert \P^\dagger\rVert_{\max} \geq1,
\end{equation}
with the norm equal $\lVert \P^\dagger(\mathds{1})\rVert_{\max}$ for a positive projection [so that
the bound is saturated for a positive and trace-preserving projection; cf.~Eq.~\eqref{eq:norm_Pss}].  There exists, however, no upper bound for the norm of a general projection. 

An analogous result also holds for a superoperator $\mathcal{X}$ which is Hermiticity-preserving and conserves an operator $X$, $\mathcal{X}^\dagger(X)=X$ (e.g., $X=\mathds{1}$ for a trace-preserving $\mathcal{X}$), 
\begin{equation}\label{eq:norm_X}
	\lVert \mathcal{X}\rVert =	\lVert \mathcal{X}^\dagger\rVert_{\max} \geq1.
\end{equation}\\

As $\Pss$ is Hermiticity and trace preserving, as well as  positive, Eq.~\eqref{eq:norm_Pss} follows analogously to Eq.~\eqref{eq:norm_master}.  

For a projection $\P$ we have $\lVert \P(\rho)\rVert\geq \Tr[X\P(\rho)]/\lVert X\rVert_{\max}$ [cf.~Eq.~\eqref{eq:vN2}], while $\Tr[X\P(\rho)]=\Tr[\P^\dagger(X) \P(\rho)]$, so that $\lVert \P(\rho)\rVert\geq \Tr[X\P(\rho)]/\lVert \P^\dagger(X)\rVert_{\max}$. As $\P$ is Hermiticity-preserving, there exists a Hermitian $X_\rho$ such that  $\lVert \P(\rho)\rVert=\Tr[X_\rho\P(\rho)]/\lVert X_\rho\rVert_{\max}$, e.g., a difference between the orthogonal projections on the sums of $\P(\rho)$ eigenspaces corresponding to its positive and negative eigenvalues. It then follows $\lVert \P^\dagger(X)\rVert_{\max}\geq \lVert X\rVert_{\max}$, which gives the inequality in Eq.~\eqref{eq:norm_P}. In fact,  $\lVert \P\rVert=\lVert \P^\dagger(X_\rho)\rVert_{\max}/ \lVert X_\rho\rVert_{\max}$, where $\rho$ is such that $\lVert \P\rVert=\lVert \P(\rho)\rVert$. Therefore, when $\P$ is positive, $\lVert \P(\rho)\rVert=\Tr[\P(\rho)]=\Tr[\mathds{1}\P(\rho)]$, 
so that $\lVert \P\rVert=\lVert \P^\dagger(\mathds{1})\rVert_{\max}$.

We now show that the norm of a general projection is not bounded from above. Indeed, consider $\P(\rho)\equiv R\,\Tr(L\rho)/\Tr(LR)$, where $L$ and $R$ are Hermitian operators such that $\Tr(L R)\neq 0$. We have that $\P$ is Hermiticity-preserving and $\lVert\P\rVert=\lVert L\rVert_{\max}\lVert R\rVert/|\Tr(LR)$| as  $\lVert\P(\rho)\rVert=|\Tr(L\rho)| \lVert R\rVert/| \Tr(LR) |$  with the maximum value achieved by  considering $\rho$ in the eigenspace of $L$ corresponding to $\lVert L\rVert_{\max}$. For example, when $L=\Pi_R+ c R^\perp$, where $\Pi_R$ is a difference between projections on the positive and negative eigenspaces of $R$, so that $\Tr(\Pi_R R)=\lVert R\rVert$ and $\lVert \Pi_R \rVert_{\max}=1$, while $R^\perp$ is orthogonal to $R$ $\Tr(R^\perp R)=0$, we have that $\lVert\P\rVert=\lVert L\rVert_{\max}\geq |c|\lVert R^\perp \rVert-1$ diverges when $|c|\rightarrow \infty$.

Finally, for an operator $X$ conserved by a superoperator $\mathcal{X}$, we have $\lVert \mathcal{X}^\dagger(X)\rVert_{\max}=\lVert X\rVert_{\max}$, which for a Hermitian $X$ leads to the inequality in  Eq.~\eqref{eq:norm_X} [cf.~Eq.~\eqref{eq:norm_max_ind}], while otherwise $(X+X^\dagger)/2$ and $(X-X^\dagger)/(2i)$ can be considered instead of $X$, as these Hermitian operators are also conserved by  Hermiticity-preserving $\mathcal{X}$.

\subsection{Induced norms of superoperators related to observables}\label{app:norm_obs}

We first discuss norms for superoperators corresponding to von Neumann and POVM measurements. We then consider superoperators encoding observable correlations.

\subsubsection{Von Neumann measurement}
Let $X$ be a Hermitian  operator, $X=\sum_{n} x_n P_n$, where $P_n$ is an orthogonal projection on the eigenspace of $X$ with a eigenvalue $x_n$ (we assume $x_n\neq x_m$ for $n\neq m$). We define a superoperator $\mathcal{X}(\rho)\equiv \sum_n x_n \,P_n \rho P_n$. This corresponds to the measurement of $X$ on a state $\rho$ with $\mathcal{X}(\rho)$ equal to the average of measurement outcomes with the conditional state of the system. Below, we show that  
\begin{equation}\label{eq:norm_vN}
	\lVert \mathcal{X}\rVert =	\lVert \mathcal{X}\rVert_{\max}=\lVert X\rVert_{\max}.
\end{equation}

We note that $ \mathcal{X}$ is Hermiticity preserving as $P_n=P_n^\dagger$, which also gives $\mathcal{X}^\dagger=\mathcal{X}$, and thus via Eq.~\eqref{eq:norm_trace_max} we obtain the first equality in Eq.~\eqref{eq:norm_vN}. Furthermore, since $P_n \rho P_n\geq 0$ for $\rho\geq 0$, we have $\lVert \mathcal{X}(\rho) \rVert\leq \sum_n |x_n| \lVert P_n \rho P_n  \rVert =\sum_n |x_n| \Tr (P_n \rho)\leq \max_{n} |x_n| \sum_n \Tr (P_n \rho )=\lVert X\rVert_{\max} \Tr(\rho)$. The inequality is saturated for $\rho$ chosen supported within the eigenspace of $X$ that corresponds to $\max_{n} |x_n|$, which gives the second equality in Eq.~\eqref{eq:norm_vN}.

\subsubsection{POVM}

Let us consider a superoperator $\mathcal{X}(\rho)\equiv \sum_n x_n \,P_n \rho P_n^\dagger$, where $\sum_{n} P_n^\dagger P_n=\mathds{1}$ and $x_n$ are real. This describes the average of outcomes with the conditional state of the system in an imperfect measurement. In this case [cf.~Eq.~\eqref{eq:norm_vN}]
\begin{equation}\label{eq:norm_POVM}
	\lVert \mathcal{X}\rVert =	\lVert \mathcal{X}^\dagger\rVert_{\max}\leq \lVert X \rVert_{\max},
\end{equation}
where $X\equiv\sum_n|x_n| P_n^\dagger P_n$, so that  $\lVert X \rVert_{\max}\leq \max_{n} |x_n|$.
For example, for a superoperator describing an observable measurement followed by the system evolution, $\mathcal{X}_t\equiv e^{t\L}\mathcal{X}_0$, where $\mathcal{X}_0$ corresponds to a (von Neumann) measurement of an observable $X_0$, we have $\lVert \mathcal{X}_t\rVert\leq \lVert X_0 \rVert_{\max}$.
Finally, the inequality in Eq.~\eqref{eq:norm_POVM} is saturated whenever  there exists  $\rho$ supported in the eigenspace of $X$ with the absolute value of the  eigenvalue equal $\lVert X \rVert_{\max}$ such that $P_n \rho P_n$  for all $n$ are either pairwise orthogonal, or pairwise parallel with the same sign of $x_n$.

Indeed, $\mathcal{X}$ is Hermiticity preserving, so that the first equality in Eq.~\eqref{eq:norm_POVM} follows from Eq.~\eqref{eq:norm_trace_max}. Furthermore, $\lVert \mathcal{X}(\rho) \rVert\leq \sum_n |x_n| \lVert P_n \rho P_n^\dagger  \rVert =\sum_n |x_n| \Tr (P_n^\dagger P_n \rho   )=  \Tr (\sum_n |x_n| P_n^\dagger P_n \rho   )$.

\subsubsection{Correlator}

For a Hermitian  operator $X$, we define a superoperator $\mathcal{X}(\rho)\equiv(X\rho+\rho X)/2$. This superoperator allows for encoding symmetrised correlations of $X$ [cf.~Eq.~\eqref{eq:meta_corr}]. We have 
\begin{equation}\label{eq:norm_corr}
	\lVert \mathcal{X}\rVert =	\lVert \mathcal{X}\rVert_{\max}= \lVert X \rVert_{\max}.
\end{equation}

As $\mathcal{X}$ is Hermiticity preserving and $\mathcal{X}^\dagger= \mathcal{X}$,we obtain the first equality in Eq.~\eqref{eq:norm_corr}. 
Considering $\rho$ corresponding to an eigenstate of $X$ with an eigenvalue $x_n$, we have $\mathcal{X}(\rho)=x_n\rho$, which gives $\lVert \mathcal{X}\rVert \geq \lVert X \rVert_{\max}$.
Finally, we have $\lVert\mathcal{X}(\rho)\rVert\leq (\lVert X\rho\rVert+\lVert X\rho\rVert)/2=\lVert X\rho\rVert$, while for a pure state $\rho=\rho^2$, by definition of the trace norm in Eq.~\eqref{eq:norm_trace}, we have $\lVert X\rho\rVert=\sqrt{\Tr(\rho X^2)}\leq \sqrt{\lVert X^2\rVert_{\max}}=\lVert X\rVert_{\max}$.

\section{Dynamics of open quantum systems} \label{app:dynamics}

\subsection{Distance between initial and stationary states} \label{app:dynamics0}

Here, we derive the bounds in Eq.~\eqref{eq:IPss}. The lower bound is derived in two ways. First, we use the structure of the stationary states of completely-positive dynamics, which also delivers a lower bound dependent on the rank of stationary states. Second, we use the methods introduced in Sec.~\ref{sec:single}.
	
	
	\subsubsection{Upper bound}

By the triangle inequality we obtain (cf.~Appendix~\ref{app:norm_master})
\begin{equation}\label{eq:IPss2}
	\lVert \I-\Pss \rVert\leq 	\lVert \I\rVert+\rVert\Pss \rVert=2.
\end{equation}

When there exists a decay subspace in the dynamics, by definition, there exists a state $\rho_0$  which does not share support with any of stationary states. In particular, for the stationary state $\rhoss$ corresponding to $\rho_0$ we have $\lVert \rho_0-\rhoss \rVert=2$ and the inequality in Eq.~\eqref{eq:IPss2}  \emph{saturates}. 

\subsubsection{Lower bound from structure of stationary states}
	
	First, we review the structure of stationary states. We also recall the structure of the corresponding projection in the case without decay. We then derive the lower bound in Eq.~\eqref{eq:IPss} and also give an improved  lower bound dependent on the rank of stationary states.\\
	
	\emph{Stationary states}.
In general, a stationary state $\rhoss$ is of the form (see, e.g., Ref.~\cite{Wolf2012})
\begin{equation}\label{eq:rho_ss2}
	\rhoss	=\sum_{k} p_k \omega_k\otimes \rhoss^{(k)},
\end{equation}
where $p_k$ is a probability distribution, $\omega_k$ is an arbitrary state supported on $\mathcal{K}_k$, $\rhoss^{(k)}$ is a fixed state supported $\mathcal{H}_k$. Here, subspaces $\mathcal{K}_k\otimes\mathcal{H}_k$ are orthogonal. Let $d_k$ and $D_k$ denote the dimensions of $\mathcal{K}_k$ and $\mathcal{H}_k$, respectively. It follows  $m_\text{ss}=\sum_k d_k^2$ [cf.~Eq.~\eqref{eq:rho_ss_spectral}].\\

When there is no decay,  the system Hilbert space decomposes as $\mathcal{H}=\oplus_k \mathcal{K}_k\otimes\mathcal{H}_k$, so that $D=\sum_k d_k D_k$ where $d_k$ and $D_k$ are dimensions of $\mathcal{K}_k$ and $\mathcal{H}_k$, respectively. In this case, the stationary state depends on the initial condition as  [cf.~Eq.~\eqref{eq:rho_ss2}]
\begin{equation}\label{eq:Pss2}
	\Pss(	\rho_0)	=\sum_{k} \Tr_k(\mathds{1}_{\mathcal{K}_k}\otimes\mathds{1}_{\mathcal{H}_k}\rho_0 \mathds{1}_{\mathcal{K}_k}\otimes\mathds{1}_{\mathcal{H}_k})\otimes \rhoss^{(k)},
\end{equation}
where $\Tr_k$ denotes the partial trace over $\mathcal{H}_k$ on $\mathcal{K}_k\otimes\mathcal{H}_k$.  \\

\emph{Derivation of the lower bound in Eq.~\eqref{eq:IPss}}.
In the case without decay, let $\rho_0^{(k)}$ correspond to an eigenstate in $\rhoss^{(k)}$ with an eigenvalue $p_\text{ss}^{(k)}\leq 1/D_k$. From Eq.~\eqref{eq:Pss2}, the stationary state corresponding to the initial state  $\rho_0=\omega_k\otimes\rho_0^{(k)}$ is $\rhoss=\omega_k\otimes \rhoss^{(k)}$ (see, e..g. Ref.~\cite{Wolf2012}), so that $\lVert \rho_0-\rhoss \rVert=\lVert \rho_0^{(k)}-
\rhoss^{(k)}\rVert=2(1-p_\text{ss}^{(k)})$. As the minimal eigenvalue $\rhoss^{(k)}$ is less than $ 1/D_k$, we obtain
\begin{equation}\label{eq:IPss3}
	\lVert \I-\Pss \rVert\geq 	\max_k 2\left(1- \frac{1}{D_k}\right).
\end{equation}
Therefore, 
\begin{equation}\label{eq:IPss4}
	\lVert \I-\Pss \rVert\geq 	1
\end{equation}
unless $D_k=1$ for all $k$. Even in this case, however,  by considering an initial state $\rho_0$  corresponding to an equal superposition of pure states in $\mathcal{K}_k\otimes\mathcal{H}_k$ and $\mathcal{K}_{k'}\otimes\mathcal{H}_{k'}$, where $k\neq k'$, we have $\lVert \rho_0-\rhoss \rVert=1$ [note that $k'$ exists, as for $D_k=1$, $\mathcal{H}= \mathcal{K}_k\otimes\mathcal{H}_k$ implies that the dynamics is trivial, that is, $\L= 0$ as $\I=\Pss$; cf.~Eq.~\eqref{eq:Pss2}].  Furthermore, the inequality in Eq.~\eqref{eq:IPss4} actually \emph{saturates} in this case when $D=2$.

\subsubsection{Lower bound from structure of dynamics}
	
	Let us consider time $t$ such that both $t$ and $2t$ belong to the initial regime in Eq.~\eqref{eq:initial} for $\C\leq1/4$. From the contractivity of the dynamics, $\C_\Delta(t,2t)\leq \C$ [cf.~Eq.~\eqref{eq:meta_Delta1}]. From Eq.~\eqref{eq:ss_IP}, we then have $\lVert e^{t\L}-\Pss\rVert\leq E_-[\C_\Delta(t,2t)]\leq E_-(\C)$ or $\lVert e^{t\L}-\Pss\rVert\geq E_+[\C_\Delta(t,2t)]\geq E_+(\C)$. Considering the limit $\C\rightarrow 0$ leads to $t\rightarrow 0$, so that 
	from the continuity of the induced norm (see Appendix~\ref{app:norm_master}), we arrive at $\lVert \I-\Pss\rVert=0$, which corresponds to the trivial dynamics with $\L=0$, or
	\begin{equation}
		\lVert \I-\Pss\rVert\geq 1.
	\end{equation}

\subsection{Logarithmic scale of time}\label{app:dynamics2}

We now show that the condition in Eq.~\eqref{eq:meta_cond} excludes approximate stationarity inherited from the initial regime That is, when the initial regime is shifted to obtain a time regime $t''\leq t\leq t'$, $\lceil t''/(t'-t'')\rceil\Cd(t'',t')$ is finite whenever both $\lVert e^{t''\L}-\I\rVert$ and $1-\lVert e^{t''\L}-\I\rVert$ are (positive and) finite.  A similar result holds for a metastable regime (see Appendix~\ref{app:def}). \\

Let us consider the initial regime in Eq.~\eqref{eq:initial}, that is, times $t\leq t_0(\C)$. Shifting  the regime forward 
we obtain $t'-t''=t_0(\C)$ and
	\begin{eqnarray}
		\Cd(t'',t')&\geq& \lVert e^{t''\L}-e^{t'\L}\rVert\\\nonumber
		&\geq&  	\lVert e^{t_0(\C)\L}-\I\rVert -\lVert e^{t''\L}-\I\rVert \Vert e^{t_0(\C)\L}-\I\rVert  \qquad\\\nonumber 
		&\geq&   \C (1-\lVert e^{t''\L}-\I\rVert).  \qquad
	\end{eqnarray}
	Furthermore, 
	from Eq.~\eqref{eq:0_lin}, for $t''=nt_0(\C)$, $n$ is bounded from below as $n\geq \lVert e^{t''\L}-\I\rVert/\C$, so that $[t''/(t'-t'')]\Cd(t'',t')\geq  (1\!-\!\lVert e^{t''\L}\!-\I\rVert)\lVert e^{t''\L}\!-\I\rVert$.
	For general $t''$, we then obtain
	\begin{eqnarray}
		\left	\lceil \frac{t''}{t'-t''}\right\rceil \Cd(t'',t')&\geq&  (1\!-\!\lVert e^{t''\L}\!-\I\rVert)\lVert e^{t''\L}\!-\I\rVert\qquad\\\nonumber
		&&-\C (1-\C), 
	\end{eqnarray}	
	as 	$\lVert e^{ \lceil t''/(t'-t'')\rceil t_0(\C)\L}\!-e^{t''\L}\rVert\leq \C$.
	Thus for finite $\lVert e^{t''\L}-\I\rVert$ and $1-\lVert e^{t''\L}-\I\rVert$, we obtain that  
	$[t''/(t'-t'')]\Cd(t'',t')$ is finite as well.

\subsection{Changes in time}\label{app:dynamics1}

Here we derive Eqs.~\eqref{eq:change2_all}, ~\eqref{eq:change2_ss}, and~\eqref{eq:all_lin}.

\subsubsection{Derivation of Eq.~\eqref{eq:change2_all}}
We now prove 
\begin{eqnarray}\label{eq:change_all}
	\lVert e^{\delta t \L}-\I\rVert  \left(1-	\lVert e^{t\L}-\I\rVert   \right)&\leq& 	\lVert e^{t\L}-e^{(t+\delta t)\L}\rVert,\qquad
\end{eqnarray}
where  $t,\delta t\geq 0$. By considering $\delta t=t''$, we thus obtain  Eq.~\eqref{eq:change2_all}.\\

We have
\begin{eqnarray}\label{eq:change_all_app}
	&& \lVert e^{\delta t\L}\!- \I \rVert -\lVert e^{t\L}\! -e^{(t+\delta t)\L} \rVert\\\nonumber
	&&\leq \lVert e^{(t+\delta t)\L}\! -e^{t\L} \!-e^{\delta t\L}\!+ \I \rVert \\\nonumber
	&&=\lVert (e^{\delta t\L}\!-\I) (e^{t\L}\!-\I)\rVert\leq 	\lVert e^{\delta t\L}\!-\I\lVert\, \rVert e^{t\L}\!-\I\rVert
\end{eqnarray}
and Eq.~\eqref{eq:change_all} follows by rearranging the terms.

\subsubsection{Derivation of Eq.~\eqref{eq:change2_ss}}

We now prove 
\begin{eqnarray}
	\label{eq:change_ss}
	\lVert e^{t\L}-\Pss\rVert \left[1-\lVert e^{\delta t\L}-\Pss\rVert \right]&\leq& 	\lVert e^{t\L}-e^{(t+\delta t)\L}\rVert,\qquad
\end{eqnarray}
where  $t,\delta t\geq 0$. By considering $\delta t=t$, we thus obtain  Eq.~\eqref{eq:change2_ss}.\\

We have $ \rho_{t}-\rho_{t+\delta t}= (\I-\Pss) (\rho_{t}-\rho_{t+\delta t})$ [cf.~Eqs.~\eqref{eq:rho_t_spectral} and~\eqref{eq:rho_ss_spectral}], 
or equivalently, $(\I-\Pss) [e^{t\L}-e^{(t+\delta t)\L}]=e^{t\L}-e^{(t+\delta t)\L}$. Therefore,  
\begin{equation}\label{eq:ss_IP00}
	\lVert (\I-\Pss) [e^{t\L}-e^{(t+\delta t)\L}]\rVert= \lVert e^{t\L}-e^{(t+\delta t)\L}\rVert.
\end{equation}
Furthermore, for $t_1\geq t_2$,we have
\begin{eqnarray}\label{eq:ss_IP01}
	&&\lVert (\I-\Pss) e^{t\L}\rVert\big[1-\lVert (\I-\Pss) e^{\delta t\L}\rVert\big]	\\\nonumber
	&&\leq 	\lVert (\I-\Pss) e^{t\L}\rVert-\lVert (\I-\Pss) e^{(t+\delta t)\L}\rVert\\\nonumber &&\leq  \lVert (\I-\Pss) [e^{t\L}-e^{(t+\delta t)\L}]\rVert
\end{eqnarray}
and Eq.~\eqref{eq:change_ss} follows by rearranging the terms.

\subsubsection{Derivation of Eq.~\eqref{eq:all_lin}}

From the triangle inequality
\begin{eqnarray}
	\lVert e^{t\L}\!-\! e^{(t+\delta t)\L}\rVert
	&\geq& \delta t \lVert \L e^{t\L}\rVert \!-\! \lVert (e^{\delta t\L}\!-\!\I \!-\! \delta t\L)e^{t\L} \rVert,\,\, \qquad
\end{eqnarray}
where
\begin{eqnarray}
	\delta t \lVert \L e^{t\L}\rVert
	&\geq&(1\!-\!\lVert e^{t\L}\!-\!\I\rVert) \delta t \lVert \L \rVert .
\end{eqnarray}
Furthermore, from the contractivity of the dynamics
\begin{eqnarray}
	\lVert (e^{\delta t\L}\!-\!\I \!-\! \delta t)e^{t\L} \rVert &\leq&\lVert e^{\delta t\L}\!-\!\I \!-\! \delta t\L\rVert
\end{eqnarray}
and  by considering  the series for $e^{\delta t\L}$,
\begin{eqnarray}
	\lVert e^{\delta t\L}\!-\!\I \!-\! \delta t\L \rVert
	&\leq & e^{\delta t\lVert \L\rVert }-1-\delta t\lVert \L\rVert.
\end{eqnarray}

\section{Metastability, initial relaxation, and long-time dynamics}\label{app:meta}

\subsection{Definition of metastability}\label{app:def}

Here, we discuss the validity of the definition of metastability introduced in Sec.~\ref{sec:meta_def}. 
First, we discuss when Eqs.~\eqref{eq:dist_0} and~\eqref{eq:dist_ss} imply Eq.~\eqref{eq:meta_cond2}. Second, we derive bounds on the the values of $\C_\Delta(t'',t')$ [Eq.~\eqref{eq:meta_Delta}] for which Eqs.~\eqref{eq:dist_0} and~\eqref{eq:dist_ss} are equivalent to Eq.~\eqref{eq:meta_cond2}. Finally, we show that the condition in Eq.~\eqref{eq:meta_cond} excludes approximate stationarity inherited from the metastable regime.

\subsubsection{Metastable regime}

First, we have that $E_+[\Cd(t'',t')]>1-1/e$ for $\Cd(t'',t')<(1-1/e)/e=0.233...$ so that Eq.~\eqref{eq:dist_0} for $t=t''$ implies that 
\begin{equation}
	t''> \tau_0. 
\end{equation}
Note however, that we need $\Cd<(-1+\sqrt{2})/2=0.207...$, for the bound $\lVert e^{t\L}-\I\rVert\geq E_+[\Cd(t'',t')]-\Cd(t'',t')$ to hold for $t'/2< t\leq t'$, as assumed in the definition of metastability. 

Second, we have that $ E_+[\Cd(t'',t')]-\Cd(t'',t')>1/e$ for $\Cd(t'',t')<(\sqrt{e}-1)/e=0.239...$.
But we need  $\Cd(t'',t')\leq -2+\sqrt{5}=0.236...$ to have the bound $\lVert e^{t\L}-\Pss\rVert\geq E_+[\Cd(t'',t')]-\Cd(t'',t')$ for $t=t'$ (and, more generally, for $t'/2< t\leq t'$), which leads to
\begin{equation}
	t'> \tau_\text{ss}. 
\end{equation}

\subsubsection{Equivalent definition of metastability}

First, we show that for $\Cd(t'',t')\leq[3\ln(3/2)-1]/2=0.108...$, the condition
\begin{equation}\label{eq:meta_cond2_0}
	t''>\tau_0
\end{equation}
implies Eq.~\eqref{eq:dist_0}. 
Indeed, for $\Cd(t'',t')<(-1+\sqrt{2})/2=0.207...$, from Eq.~\eqref{eq:all}  for all $t''\leq t\leq t'$ we have that  $\lVert e^{t\L}-\I\rVert\geq E_+[\Cd(t'',t')]-\Cd(t'',t')$ or $\lVert e^{t\L}-\I\rVert\leq E_-[\Cd(t'',t')]+\Cd(t'',t')$.		
Assuming that the latter bound holds, from Eq.~\eqref{eq:all_lin3}  we obtain for $\Cd(t'',t')\leq[3\ln(3/2)-1]/2$ that the considered regime belongs to the initial regime with	$\C=\max\{e^{E_2[\Cd(t'',t')]}\!-\!1, E_-[\Cd(t'',t')]+\Cd(t'',t')\}\leq 1/2<1-1/e$, so that we have $t''\leq t'< \tau_0$ and, thus, we arrive at the contradiction with Eq.~\eqref{eq:meta_cond2_0}.

Second, for $\Cd(t'',t')\leq[3\ln(3/2)-1]/2=0.108...$, the condition
\begin{equation}\label{eq:meta_cond2_1}
	t''\leq \tau_0
\end{equation}
implies Eq.~\eqref{eq:all-} and the initial regime  in Eq.~\eqref{eq:initial} with $\C\leq1/2$. Indeed, from Eq.~\eqref{eq:meta_cond2_1} we have 
$\lVert e^{t''\L}-\I\rVert\leq1-1/e$ [cf.~Eq.~\eqref{eq:tau_0}]. Therefore, for $\Cd(t'',t')<(1\!-\!1/e)/e=0.233...$ when $E_+[\Cd(t'',t')]>1-1/e$, Eq.~\eqref{eq:all-} necessarily holds for $t''\leq t\leq t'/2$. Furthermore, when $\Cd(t'',t')<(-1+\sqrt{2})/2=0.207...$, we have that $\lVert e^{t\L}-\I\rVert\leq E_-[\Cd(t'',t')]+\Cd(t'',t')$ for $t'/2< t\leq t'$.	
Therefore, for $\Cd(t'',t')\leq[3\ln(3/2)-1]/2$, the regime $t''\leq t\leq t'$  belongs to the initial regime $\C=\max\{e^{E_2[\Cd(t'',t')]}\!-\!1, E_-[\Cd(t'',t')]+\Cd(t'',t')\}$ [cf.~Eq.~\eqref{eq:all_lin3}].

Third, 
we show that for $\Cd(t'',t')\leq(1-1/e)/e=0.233...$
the condition
\begin{equation}\label{eq:meta_cond2_2}
	t'	< \tau_\text{ss}
\end{equation}
implies Eq.~\eqref{eq:dist_ss}. From Eq.~\eqref{eq:meta_cond2_2}, for $t''\leq t\leq t'$  it follows that $\lVert e^{t\L}-\Pss\rVert\geq\lVert e^{t'\L}-\Pss\rVert> 1/e$ [cf.~Eq.~\eqref{eq:tau_ss}]. Therefore, for $\Cd(t'',t')\leq(1-1/e)/e$, when $E_-[\Cd(t'',t')]\leq1/e$, from Eq.~\eqref{eq:ss_IP} we obtain that Eq.~\eqref{eq:dist_ss} holds for $t''\leq t\leq t'/2$, while for $t'/2< t\leq t$ we have $\lVert e^{t\L}-\Pss\rVert\geq E_+[\Cd(t'',t')]-\Cd(t'',t')$ as $E_+[\Cd(t'',t')]-\Cd(t'',t')>E_-[\Cd(t'',t')]$ already for $\Cd(t'',t')<-2+\sqrt{5}=0.236...$.

Finally,  we show that for $\Cd(t'',t')< -2+\sqrt{5}=0.236...$, the condition
\begin{equation}\label{eq:meta_cond2_3}
	t'	\geq \tau_\text{ss}
\end{equation}
implies Eq.~\eqref{eq:ss_IP-} and thus the final regime  in Eq.~\eqref{eq:final} with $\C=E_-[\Cd(t'',t')]$. Indeed, for $\Cd(t'',t')<(\sqrt{e}-1)/e=0.239...$ 
we have $ E_+[\Cd(t'',t')]> E_+[\Cd(t'',t')]-\Cd(t'',t')>1/e$, while from Eq.~\eqref{eq:meta_cond2_3} we have $\lVert e^{t'\L}-\Pss\rVert\leq 1/e$. Thus, from Eq.~\eqref{eq:ss_IP}, we obtain $\lVert e^{t\L}-\Pss\rVert\leq E_-[\Cd(t'',t')]$
that for $\tau_\text{ss}\leq t\leq t'$. Furthermore, as $E_+[\Cd(t'',t')]-\Cd(t'',t')>E_-[\Cd(t'',t')]$ for $\Cd(t'',t')< -2+\sqrt{5}$, in that case we obtain that  $\lVert e^{t\L}-\Pss\rVert\leq E_-[\Cd(t'',t')]$ for $t''\leq t\leq t'$.

	\subsubsection{Logarithmic scale of time}

	We now show that when a metastable regime $t''\leq t\leq 2t''$ [cf.~Eq.~\eqref{eq:meta_cond2}] is first extended to $t'>2t''$ such that $\C(t'',t')=2\C(t'',2t'')$ and then shifted by $\delta t$, $\lceil (t''+\delta t)/(t'-t'')\rceil\Cd(t''+\delta t,t'+\delta t)$ is finite whenever  $\lVert e^{(t''+\delta t)\L}-e^{t''\L}\rVert$ and  $1-2\lVert e^{(t''+\delta t)\L}-e^{t''\L}\rVert$ are (positive and) finite (cf.~Appendix~\ref{app:dynamics2}). An analogous result holds for a general metastable regime, as from Eq.~\eqref{eq:meta_cond} it features the metastable regime considered here. \\

	We have
	\begin{eqnarray}\label{eq:meta_shift}
		&&\lVert e^{(t_1+\delta t)\L}-e^{(t_2+\delta t)\L}\rVert \geq \\\nonumber
		&&	\lVert e^{(t''+t_1)\L}-e^{(t''+t_2)\L}\rVert -\lVert e^{\delta t\L}-e^{ t''\L} \rVert \Vert e^{t_1\L}-e^{t_2\L}\rVert .
	\end{eqnarray}
	Let $t'> 2t''$ be the shortest time such that $\lVert e^{t''\L}-e^{t' \L}\rVert=2	\Cd(t'',2 t'')$. From  Eq.~\eqref{eq:meta_lin}, we have $t'\geq 3t''$.
	We then obtain $\lVert e^{t\L}-e^{t'\L}\rVert\geq 2	\Cd(t'',2t'')-\lVert e^{t''\L}-e^{t \L}\rVert\geq \Cd(t'',2t'')$ for $t''\leq t\leq 2t''$, while for $2t''\leq t\leq t'$, $\lVert e^{(t-t'')\L}-e^{(t'-t'') \L}\rVert<\Cd(t'',t')=2	\Cd(t'',2t'')$ holds.
	Therefore, considering $t_1=t''$ and $t_2=t'-t''$ in Eq.~\eqref{eq:meta_shift} we arrive at
	\begin{eqnarray}
		\Cd(t''\!+\!\delta t,t'\!+\!\delta t)&>&  \Cd(t'',2t'') (1\!-\!2\lVert e^{\delta t\L}\!-\!e^{t''\L}\rVert)\qquad\,\,\\\nonumber
		&=&\Cd(t'',t') \bigg(\frac{1}{2}\!-\!\lVert e^{\delta t\L}\!-\!e^{t''\L}\rVert\bigg).
	\end{eqnarray}
	Furthermore, from Eq.~\eqref{eq:'_lin}, for $\delta t= n (t'-t'')- t''$, we have $n\C_\Delta(t'',t')+(n-1)\C_\Delta(t'',2t'')\geq  \lVert e^{n(t'-t'')\L}-e^{t''\L}\rVert\geq  \lVert e^{\delta t\L}-e^{t''\L}\rVert-\C_\Delta(t'',2t'')$, so that $[(t''+\delta t)/(t'-t'')]\Cd(t'',t')\geq (1\!-\!2\lVert e^{\delta t\L}\!-e^{t''\L}\rVert)\lVert e^{\delta t\L}\!-e^{t''\L}\rVert/3$. For general $\delta t$,
	we then obtain
	\begin{eqnarray}\nonumber
		&&\left\lceil \frac{t''+\delta t}{t'-t''}\right\rceil\! \Cd(t'',t')\!\geq\!  \left(\!1\!-\!2\lVert e^{\delta t\L}\!-\!e^{t''\L}\rVert\!\right)\!\frac{\lVert e^{\delta t\L}\!-\!e^{t''\L}\rVert}{3}\\
		&&\qquad\!-\frac{1}{3}\Cd(t'',t') [1+\lVert e^{\delta t\L}\!-\!e^{t''\L}\rVert -\Cd(t'',t')],
	\end{eqnarray}	
	as 	$\lVert e^{ \lceil (t''+\delta t)/(t'-t'')\rceil (t'-t'')\L}\!-e^{(t''+\delta t)\L}\rVert\leq \Cd(t'',t')$.

\subsection{Initial relaxation}\label{app:time''}

Here, we derive Eq.~\eqref{eq:''_exp}. \\

We have
\begin{eqnarray}\label{eq:''_exp1}
	\lVert e^{2 t\L}-e^{t''\L}\rVert & \leq &	\lVert (e^{ t\L}-e^{t''\L})^2\rVert  +\lVert e^{ (t+t'')\L}-e^{2t''\L}\rVert\qquad\\\nonumber&&+\lVert e^{ (t+t'')\L}-e^{t''\L}\rVert
	\\\nonumber&\leq&	\lVert e^{ t\L}-e^{t''\L}\rVert^2+2\Cd(t'',t'),
\end{eqnarray}
where in the last line we assumed $|t''-t| \leq t'- t''$ (e.g., $t \leq t''$).
Analogously, for $n>2$, we have
\begin{eqnarray}\label{eq:''_exp2}
	&&\lVert e^{n t\L}-e^{t''\L}\rVert  \leq 	\lVert (e^{(n-1) t\L}-e^{t''\L})(e^{ t\L}-e^{t''\L})\rVert  \qquad\\\nonumber&&\qquad\qquad\quad+\lVert e^{ [(n-1)t+t'']\L}-e^{2t''\L}\rVert+\lVert e^{ (t+t'')\L}-e^{t''\L}\rVert
	\\\nonumber&&\qquad\qquad\leq	\lVert e^{(n-1) t\L}-e^{t''\L}\rVert\lVert e^{ t\L}-e^{t''\L}\rVert +2\Cd(t'',t'),
\end{eqnarray}
where in the last line we assumed $|t''-(n-1)t| \leq t'- t''$. By induction, we then obtain 
\begin{eqnarray}\label{eq:''_exp3}
	\lVert e^{n t\L}-e^{t''\L}\rVert  &\leq& \lVert e^{ t\L}-e^{t''\L}\rVert^n
	\\\nonumber&&+2\Cd(t'',t') \frac{1-\lVert e^{ t\L}-e^{t''\L}\rVert^{n-1}}{1-\lVert e^{ t\L}-e^{t''\L}\rVert}.
\end{eqnarray}
Thus,  Eq.~\eqref{eq:''_exp} follows for $\lVert e^{ t\L}-e^{t''\L}\rVert<1$.

\subsection{Long-time dynamics}\label{app:time'}

Here, we derive Eq.~\eqref{eq:'_lin} and~\eqref{eq:tau'2}.

\subsubsection{Derivation of Eq.~\eqref{eq:'_lin}}
By the triangle inequality, we have that 
\begin{eqnarray}	
	&&\lVert e^{ (t_1+t_2)\L}-e^{ t\L}\rVert\\\nonumber
	&&\leq 	\lVert e^{(t_1+t_2)\L}-e^{( t+t_2)\L} \rVert+	\lVert e^{( t+t_2)\L} -e^{ t\L} \rVert \\\nonumber
	&&\leq 	\lVert e^{(t_1+t_2)\L}-e^{( t+t_2)\L} \rVert+	\lVert e^{( t+t_2)\L} -e^{ 2t\L} \rVert\\\nonumber
	&&\quad+	\lVert e^{ t\L}- e^{2t\L} \rVert.
\end{eqnarray}
From the contractivity of the dynamics,
we arrive at 
\begin{eqnarray}	
	&&\lVert e^{ (t_1+t_2)\L}-e^{ t\L}\rVert\\\nonumber
	&&\leq 	\lVert e^{t_1\L}-e^{t\L} \rVert+	\lVert e^{t_2\L} -e^{ t\L} \rVert+	\lVert e^{t\L} -e^{ 2t\L} \rVert,
\end{eqnarray}
where $\lVert e^{t\L} -e^{ 2t\L} \rVert$ is bounded by $\Cd(t'',t')$ for $t''\leq t\leq t'/2$ [cf.~Eq.~\eqref{eq:meta_Delta}]. Eq.~\eqref{eq:'_lin} follows by induction.

\subsubsection{Derivation of Eq.~\eqref{eq:tau'2}}

We have [cf.~Eq.~\eqref{eq:'_lin}]
\begin{equation}\label{eq:'_lin2}
	\lVert e^{n t''\L}-e^{t''\L}\rVert  \leq ( n-1)\,\Cd(t'',t').
\end{equation}
Therefore, 
\begin{equation}
	\frac{\tau'}{t''}\geq\left\lfloor \frac{1-1/e - E_-[C_\Delta(t'',t')]}{C_\Delta(t'',t')}\right\rfloor+1.
\end{equation}

\section{Metastability in Heisenberg picture and quasi-conserved observables}\label{app:Heisenberg}

Here, we consider open quantum dynamics in the Heisenberg picture. First, we discuss how changes during a given time regime are related to the difference between observables in the max norm. Second, we show how the presence of metastability implies the existence of quasi-conserved observables. We also consider long-time dynamics in the Heisenberg picture.

\subsection{Dynamics of observables}

The dynamics of an observable $O_t$ at time $t$ in the Heisenberg picture is governed by the Hermitian conjugate of the master operator in Eq.~\eqref{eq:L},
\begin{equation}\label{eq:master_Heisenberg}
	\frac{d}{dt} O_t=\L^\dagger(O_t),
\end{equation}
so that [cf.~Eq.~\eqref{eq:rho_t}]
\begin{equation}\label{eq:O_t}
	O_t = e^{t\L^\dagger}(O_{0}).
\end{equation}
We then have $\Tr(O_t\rho_0)=\Tr(O_0\rho_t)$.

The asymptotic observable [cf.~Eq.~\eqref{eq:rho_ss}]
\begin{equation}\label{eq:O_ss}
	\Oss=\lim_{t\rightarrow\infty}O_t\equiv\Pss^\dagger (O_{0})
\end{equation}
is conserved by the dynamics, $\L^\dagger(\Oss)=0$ (due to the assumption of time-independent asymptotic states). In the generic case of a unique stationary state $\rhoss$, $\Pss^\dagger (O_{0})=\Tr(O_{0}\rhoss)\mathds{1}$.

\subsection{Changes in time}

For a given time regime, the maximal change in the observable averages corresponds to the distance between observables in the Heisenberg picture measured in the max norm. Indeed, 
from Eq.~\eqref{eq:meta_state_obs_Delta}, assuming that the system can be initialised in any state  (or, equivalently, $D^2$ linearly independent initial states can be prepared), 
\begin{eqnarray}\label{eq:meta_obs_Delta}
	\Cd(O_0;t'',t')&\equiv& \,\,\sup_{\rho_0}\,\, \Cd(\rho_0;O_0;t'',t')\\\nonumber
	&= &\sup_{t''\leq t_1,t_2\leq t' } \frac{\lVert O_{t_1}-O_{t_2}\rVert_{\max}}{\lVert O_0\rVert_{\max}},
\end{eqnarray}
where $O_t\equiv e^{t\L^\dagger}(O_0)$ as $|\Tr(O_0\rho_{t_1})-\Tr(O_0\rho_{t_2})|=||\Tr(O_{t_1}\rho_{0})-\Tr(O_{t_2}\rho_{0})|\leq \lVert O_{t_1}-O_{t_2}\rVert_{\max} \lVert \rho_0\rVert=\lVert O_{t_1}-O_{t_2}\rVert_{\max}$ with the equality achieved for $\rho_0$ chosen as the pure state corresponding to the eigenvector of $O_{t_1}-O_{t_2}$ with the extreme eigenvalues whose absolute value equals $ \lVert O_{t_1}-O_{t_2}\rVert_{\max}$.
Moreover, as the dynamics is \emph{contractive}  with respect to the max norm,  $\lVert O_{t_1+t}-O_{t_2+t}\rVert_{\max}\leq 
\lVert O_{t_1}-O_{t_2}\rVert_{\max}$, it follows that [cf.~Eqs.~\eqref{eq:meta_state_Delta1} and~\eqref{eq:meta_Delta1}]
\begin{eqnarray}
	\label{eq:meta_obs_Delta1}
	\Cd(O_0;t'',t')	&= &\sup_{t''\leq t\leq t' } \frac{\lVert O_{t''}-O_{t}\rVert_{\max}}{\lVert O_0\rVert_{\max}}.
\end{eqnarray}

The quantity in Eq.~\eqref{eq:meta_obs_Delta1} grows when the time regime is enlarged. Nevertheless, extending the length of the considered regime  leads to at most a linear increase [cf.~Eqs.~\eqref{eq:meta_state_lin} and~\eqref{eq:meta_lin}]
\begin{eqnarray}\label{eq:meta_obs_lin}
	&&\Cd[O_0;t'',t''+n(t'-t'')]\leq n \,\Cd(O_0;t'',t'),
\end{eqnarray}
while for any time regime we have [cf.~Eq.~\eqref{eq:meta_state_Delta3}]
\begin{eqnarray}\label{eq:meta_obs_Delta3}
	\Cd(O_0;t'',t')
	&\leq&\frac{ \sup_\rho\Tr(O\rho)-\inf_\rho \Tr(O\rho)}{\lVert O\rVert_{\max}}\leq 2,\qquad
\end{eqnarray}
where the bound equals the difference of the observable averages for states supported in the eigenspaces of $O$ corresponding to its maximal and minimal eigenvalues, respectively.

Finally, assuming that changes with respect to all observables can be considered we obtain [cf.~Eq.~\eqref{eq:meta_Delta}]
\begin{eqnarray}\label{eq:meta_obs_Delta4}
	\sup_{O_0}\Cd(O_0;t'',t')=\Cd(t'',t'),
\end{eqnarray}
as the norms induced by the trace and max norms are equal, so that
$\lVert e^{t_1\L^\dagger}-e^{t_2\L^\dagger}\rVert_{\max}=\lVert e^{t_1\L}-e^{t_2\L}\rVert$ (see~Appendix~\ref{app:norm_ind}).
In particular,  Eqs.~\eqref{eq:meta_lin} and~\eqref{eq:meta_Delta3} can be seen to follow from Eqs.~\eqref{eq:meta_obs_lin}  and~\eqref{eq:meta_obs_Delta3}, respectively.

\subsection{Quasi-conserved observables}\label{app:Heisenberg_meta}

When the metastability is present [Eqs.~\eqref{eq:meta_cond},~\eqref{eq:meta_Delta2}, and~\eqref{eq:meta_cond2}],  in the Heisenberg picture, all observables change negligibly during the metastable regime with respect to the norm of the initial observable [cf.~Eq.~\eqref{eq:meta_obs_Delta4}]. But there exist observables for which those changes are negligible also with respect to their norm within the metastable regime. In particular, \emph{quasi-conserved observables}  change negligibly  for all times up to the end of metastable regime, $t\leq t'$. From  Eq.~\eqref{eq:dist_0} they form a non-trivial subset of all observables. Furthermore, while they include any observable negligibly close to the subspace of conserved observables,  we show below that from Eq.~\eqref{eq:dist_ss} there exist quasi-conserved observables which are non-negligibly different.

For example, for a chosen time $t''\leq t_0\leq t'/2$, consider $O_0'$ such that $\lVert O_{t_0}'-\Oss'\rVert_{\max}/\lVert O_0'\rVert_{\max}=\lVert e^{t\L} -\Pss \rVert$. Then, for $O_0=O_{t_0}'-\Oss'$ and $ t\leq t'$ we have 
$\lVert O_{t}-O_0\rVert_{\max}/\lVert O_0\rVert_{\max}\leq 3 [\lVert O_0'\rVert_{\max}/\lVert O_0\rVert_{\max}]  \Cd(O_0';t'',t')\leq 3 \Cd(t'',t')/E_+[\Cd(t'',t')]$, where the last inequality follows from Eq.~\eqref{eq:dist_ss}.  In the model in Fig.~\ref{fig:example}, the magnetisation along $z$-axis is a quasi-conserved observable.

\subsection{Long-time dynamics}

Observable averages 
change negligibly in comparison to the initial observable norm when measured at times separated by intervals comparable to the length of the metastable regime [cf.~Eq.~\eqref{eq:meta_length_state}],
\begin{equation}\label{eq:meta_length_obs}
	\frac{	\lVert O_{t_1}-O_{t_2}\rVert_{\max}}{\lVert O_{0}\rVert_{\max}}\leq n\,\Cd(O_0;t'',t')
\end{equation}
for $t_1,t_2\geq t''$ such that $|t_1-t_2|\leq n(t'-t'')$ [cf.~Eq.~\eqref{eq:meta_obs_lin}].
In particular, it follows that the dynamics in the Heisenberg picture is simply approximated by the dynamics of the corresponding observable during the metastable regime, as 	$	\lVert O_t-O_{t+t''}\rVert_{\max}/\lVert O_0\rVert_{\max}\leq \Cd(O_0;t'',t')$ [here, $t''$ can be replaced by any time $\leq t'-t''$ or by $\leq t'$ when  $\C(O_0;t'',t')$ is replaced by  $2\C(O_0;t'',t')$].

\section{Spectral decomposition}\label{app:spectral}

Here, we consider changes in the dynamics with respect to changes in the dynamics of individual eigenmodes. We derive Eq.~\eqref{eq:change_spectral} and also discuss when changes in the dynamics are non-zero.

\subsection{Changes in the dynamics of eigenmodes}\label{app:spectral_changes}

Here, we derive Eq.~\eqref{eq:change_spectral}.

\subsubsection{Hermitian eigenmodes}

For a real $\lambda_k=\lambda_k^R$, we consider the observable given by the Hermitian left eigenmode $L_k$. The changes in the observable average are [cf.~Eq.~\eqref{eq:mode_real}]
\begin{eqnarray}\label{eq:change_spectral_R}
	&&\Tr(L_k\rho_{t_1})\!-\!\Tr(L_k\rho_{t_2})=	\Tr(L_k \rho_0)\,	(e^{t_1\lambda_k}\!\!-\!e^{t_2\lambda_k}),\,\qquad
\end{eqnarray}
while
\begin{eqnarray}\label{eq:change_spectral_R1}
	&&	\frac{|\Tr(L_k\rho_{t_1})\!-\!\Tr(L_k\rho_{t_2})|}{\lVert L_k \rVert_{\max}}
	\leq  \lVert \rho_{t_1}\!\!-\! \rho_{t_2}\rVert \leq \lVert e^{t_1\L}\!\!- \!e^{t_2\L}\rVert,\,\qquad
\end{eqnarray}
where the first inequality follows from Eq.~\eqref{eq:vN2}.
Considering the initial state $\rho_0$ supported in its eigenspace of $L_k$ with the absolute value of the eigenvalue equal $\lVert L_k \rVert_{\max}$, we arrive at Eq.~\eqref{eq:change_spectral}. 

Analogously, in the Heisenberg picture (see Appendix~\ref{app:Heisenberg}), for a Hermitian right eigenmode $R_k$, we have [cf.~Eqs.~\eqref{eq:change_spectral_R} and~\eqref{eq:change_spectral_R1}]
\begin{eqnarray}\nonumber
	&&\frac{|\Tr(O_0 R_k)|}{ \lVert O_{0}\rVert_{\max} \lVert R_k\rVert} 	|e^{t_1\lambda_k}-e^{t_2\lambda_k}|=\frac{|\Tr(O_{t_1} R_k)-\Tr(O_{t_2} R_k)|}{\lVert O_{0}\rVert_{\max} \lVert R_k\rVert}	\\\label{eq:change_spectral_R2}
	&&\leq  \frac{\lVert O_{t_1}- O_{t_2}\rVert_{\max}}{\lVert O_{0}\rVert_{\max} }\leq \lVert e^{t_1\L}\!\!- \!e^{t_2\L}\rVert.
\end{eqnarray}
Considering the initial observable $O_0$ being a difference of the projections on the sum of positive eigenvalue eigenspaces of $(e^{t_1\L}-e^{t_2\L})(R_k)$ and the sum of its negative eigenvalue eigenspaces,  we again arrive at Eq.~\eqref{eq:change_spectral}.

\subsubsection{Non-Hermitian eigenmodes}
For a complex $\lambda_k$,  by considering the observable $L_k(\phi)$ in Eq.~\eqref{eq:L_k_phi} 
we obtain [cf.~Eqs.~\eqref{eq:mode_complex}]
\begin{eqnarray}\label{eq:change_spectral_C}
	&&\Tr[L_k(\phi)\rho_{t_1}]-\Tr[L_k(\phi)\rho_{t_2}]=|\Tr(L_k \rho_0)| 
	\\\nonumber
	&&\,\times\{e^{t_1\lambda_k^R}\!\cos[t_1\lambda_I^R+\varphi_k(\rho_0)] -e^{t_2\lambda_k^R}\!\cos[t_2\lambda_I^R+\varphi_k(\rho_0)]\},
\end{eqnarray} 
where $\varphi_k(\rho_0)\equiv\phi-\phi_{k}(\rho_0)$ and  $\Tr(L_k \rho_0)=e^{i\phi_{k}(\rho_0)}|\Tr(L_k \rho_0)|$, while
\begin{eqnarray}\label{eq:change_spectral_C1}\
	&&\frac{|\Tr[L_k(\phi)\rho_{t_1}]-\Tr[L_k(\phi)\rho_{t_2}]|}{\lVert L_k (\phi)\rVert_{\max}}\leq  \lVert \rho_{t_1}- \rho_{t_2}\rVert \\\nonumber
	&&\leq   \lVert e^{t_1\L}-e^{t_2\L}\rVert.
\end{eqnarray} 
Noting that $\lVert L_k(\phi) \rVert_{\max}\leq \sup_{\rho_0'} |\Tr(L_k\rho_0')|$ and considering initial state $\rho_0$ corresponding to $\sup_{\rho'_0} |\Tr(L_k\rho'_0)|$ we then obtain
\begin{eqnarray}\nonumber
	&&\Big|e^{t_1\lambda_k^R}\!\cos[t_1\lambda_I^R+\varphi_k(\rho_0)] -e^{t_2\lambda_k^R}\!\cos[t_2\lambda_I^R+\varphi_k(\rho_0)]\Big|
	\\\label{eq:change_spectral_C2}
	&& \leq   \lVert e^{t_1\L}-e^{t_2\L}\rVert.
\end{eqnarray} 
Eq.~\eqref{eq:change_spectral} then follows from Eq.~\eqref{eq:phi_max} [or by choosing $\phi=-t\lambda_k^I-\phi_k(\rho_0)$].

In the Heisenberg picture (see Appendix~\ref{app:Heisenberg}), for a non-Hermitian  right eigenmode $R_k$, we consider 
\begin{equation}\label{eq:R_k_phi}
	R_k(\phi)\equiv e^{i\phi}R_k+e^{-i\phi}R_k^\dagger,
\end{equation}
so that $\Tr[	L_k(\phi)	R_k(\phi)]=1$ [cf.~Eq.~\eqref{eq:L_k_phi}]. We then have  [cf.~Eqs.~\eqref{eq:change_spectral_C} and~\eqref{eq:change_spectral_C1}]
\begin{eqnarray}\label{eq:change_spectral_C3}
	&&\Tr[O_{t_1} R_k(\phi)]\!-\!\Tr[O_{t_2} R_k(\phi)]=|\Tr(O_0 R_k)|	\\\nonumber &&\times\{e^{t_1\lambda_k^R}\!\cos[t_1\lambda_I^R\!+\!\varphi_k(O_0)] \!-\!e^{t_2\lambda_k^R}\!\cos[t_2\lambda_I^R\!+\!\varphi_k(O_0)]\},
\end{eqnarray}
where $\varphi_k(O_0)=\phi-\phi_k(O_0)$ and $\Tr(O_0 R_k )=e^{i \phi_k(O_0)}|\Tr(O_0 R_k )|$, while
\begin{eqnarray}\nonumber
	&&\frac{|\Tr[O_{t_1} R_k(\phi)]\!-\!\Tr[O_{t_2} R_k(\phi)]|}{\lVert O_{0}\rVert_{\max} \lVert R_k(\phi)\rVert}\leq  \frac{\lVert O_{t_1}\!-\! O_{t_2}\rVert_{\max}}{\lVert O_{0}\rVert_{\max} }\\
	\label{eq:change_spectral_C4}
	&&\leq   \lVert e^{t_1\L}-e^{t_2\L}\rVert.
\end{eqnarray}
Noting that $\lVert R_k(\phi) \rVert\leq \sup_{O_0'} |\Tr(O_0' R_k)|/\lVert O_0'\rVert$ and considering $O_0$  corresponding to $\sup_{O_0'} |\Tr(O_0' R_k)|/\lVert O_0'\rVert$ we obtain [cf.~Eq.~\eqref{eq:change_spectral_C2}]
\begin{eqnarray}\nonumber
	&&\Big|e^{t_1\lambda_k^R}\!\cos[t_1\lambda_I^R+\varphi_kO_0)] -e^{t_2\lambda_k^R}\!\cos[t_2\lambda_I^R+\varphi_k(O_0)]\Big|
	\\\label{eq:change_spectral_C5}
	&& \leq   \lVert e^{t_1\L}-e^{t_2\L}\rVert,
\end{eqnarray} 
so that Eq.~\eqref{eq:change_spectral} follows for $\phi=-t\lambda_k^I-\phi_k(O_0)$.

\subsubsection{Generalised eigenmodes}
Finally, for an eigenvalue corresponding to a Jordan block, the results in Eqs.~\eqref{eq:change_spectral_R},~\eqref{eq:change_spectral_C}, and~\eqref{eq:change_spectral_C1} hold for the last from the corresponding generalised left eigenmodes ($L_k$ such that $d_k=D_k>1$) and Eqs.~\eqref{eq:change_spectral_R2},~\eqref{eq:change_spectral_C3} and~\eqref{eq:change_spectral_C4} for the first from the corresponding generalised right eigenmodes ($R_k$ such that $d_k=1$ while $D_k>1$). 

\subsection{Non-zero changes}\label{app:spectral_nonzero}

We now argue that the system always changes with time unless initialised in a stationary state, $\rho_0=\rhoss$. Analogously, in the Heisenberg picture, any observables changes  unless it is conserved, $O_0= \Oss$ (cf.~Appendix~\ref{app:Heisenberg}). We also discuss how this results follows from the decomposition of a system state or an observable between right and left (generalised) eigenmodes of the master operator.

\subsubsection{Non-zero changes in system states}
Let $\rho_0$ be such that there exists $0\leq t_2<t_1<\infty$ such that $\rho_{t_1}=\rho_{t_2}=\rho$. Then $\rho$ is left unchanged by  $e^{(t_1-t_2)\L}$, that is, $e^{(t_1-t_2)\L}(\rho)=\rho$. As $\Pss=\lim_{n\rightarrow\infty} [e^{(t_1-t_2)\L}]$, we obtain that $\Pss(\rho)=\rho$ and thus $\rho$ is a stationary state of the dynamics. Furthermore, at finite time the evolution operator is invertible as it features no $0$ eigenvalues; cf.~Eq.~\eqref{eq:rho_t_spectral}. But it leaves a stationary state invariant, and thus $\rho_0=\rho$. 
Therefore, we obtain
\begin{equation}\label{eq:nonzero_state}
	\lVert \rho_{t_1}- \rho_{t_2}\rVert>0
	,\quad t_1\neq t_2
\end{equation}
unless $\rho_0=\rhoss$. 

Eq.~\eqref{eq:nonzero_state} also follows from Eqs.~\eqref{eq:change_spectral_R}, \eqref{eq:change_spectral_R1}, \eqref{eq:change_spectral_C}, and~\eqref{eq:change_spectral_C1}, unless $\Tr(L_k \rho)=0$ for a left eigenmode $L_k$ [for a non-Hermitian eigenmode, we consider $\phi=-t_1\lambda_k^I-\phi_k(\rho_0)$]. This also holds for the last generalise left eigenmodes in each Jordan block, i.e., $L_k$ with  $d_k=D_k>1$ unless $\Tr(L_k \rho)=0$, in which case the preceding eigenmode in the block can be considered instead.
Therefore, we again obtain Eq.~\eqref{eq:nonzero_state} unless $\Tr(L_k \rho)=0$  for all $k> m_\text{ss}$, which implies $\rho_0=\rhoss$ [cf.~Eqs.~\eqref{eq:rho_t_spectral} and~\eqref{eq:rho_ss_spectral}].

\subsubsection{Non-zero changes in observables}
Let $O_0$ be such that there exists $0\leq t_2<t_1<\infty$ such that $O_{t_1}=O_{t_2}=O$ in the Heisenberg picture. Then $O$ is left unchanged by  $e^{(t_1-t_2)\L^\dagger}$, that is, $e^{(t_1-t_2)\L^\dagger}(O)=O$. We thus obtain that $\Pss^\dagger(O)=O$ and thus $O$ is a conserved observable, and we also have $O_0=O$.  Therefore, we arrive at
\begin{equation}\label{eq:nonzero_obs}
	\lVert O_{t_1}- O_{t_2}\rVert>0,\quad t_1\neq t_2
\end{equation}
unless $O_0=O_\text{ss}$.

Eq.~\eqref{eq:nonzero_obs} also follows from Eqs.~\eqref{eq:change_spectral_R2}, \eqref{eq:change_spectral_C3}, and ~\eqref{eq:change_spectral_C4} [with $\phi=-t_1\lambda_k^I-\phi_k(O_0)$], unless $|\Tr(O R_k)|=0$ for all $k> m_\text{ss}$, in which case $O_0=\Oss$.

\section{Relation to spectral theory of metastability}\label{app:spectral_relation}

We now apply the methods introduced earlier in this work to  the spectral theory of metastability. In particular, we derive conditions on when its validity  is implied by Eqs.~\eqref{eq:meta_cond},~\eqref{eq:meta_Delta2} and~\eqref{eq:meta_cond2}.

\subsection{Spectral theory of metastability}\label{app:spectral_theory}

Here, we derive and further discuss the results in Eqs.~\eqref{eq:dist_0_P}-\eqref{eq:spectral_P}. We also prove Eqs.~\eqref{C_P_P} and~\eqref{C_P_IP}. 

\subsubsection{Distance to initial and final regimes}

We now derive Eqs.~\eqref{eq:dist_0_P} and~\eqref{eq:dist_ss_P}. To this aim, we first prove that [cf.~Eq.~\eqref{eq:IPss}]
\begin{eqnarray}\label{eq:I_P}
	&&1\leq	\lVert \I-\P \rVert  \leq 2+\C_\P(t'',t'),
	\\\label{eq:Pss_P}
	&&1\leq \lVert \P-\Pss \rVert  \leq 2+\C_\P(t'',t').
\end{eqnarray}
Then, by the triangle inequality,
\begin{eqnarray}
	\lVert e^{t\L} -\I \rVert&\geq&	\lVert \P-\I \rVert-	\lVert e^{t\L} -\P \rVert,
	\\
	\lVert e^{t\L} -\Pss \rVert&\geq&\lVert \P-\Pss \rVert	-	\lVert e^{t\L} -\P\rVert,
\end{eqnarray}
so that Eqs.~\eqref{eq:dist_0_P} and~\eqref{eq:dist_ss_P} follow from Eq.~\eqref{eq:C_P} and the lower bounds in Eqs.~\eqref{eq:I_P} and~\eqref{eq:Pss_P}.\\

\emph{Derivation of Eqs.~\eqref{eq:I_P} and~\eqref{eq:Pss_P}}.  The upper bounds follow from the triangle inequality as $\lVert \I-\P \rVert \leq 1 +\lVert \P \rVert$, $\lVert \P-\Pss \rVert \leq 	1 +\lVert \P \rVert$ and $	\lVert \P \rVert\leq 1+\C_\P(t'',t')$ (cf.~Appendix~\ref{app:norm_master}). The lower bounds follow from the definition of the induced norm  by considering $ (\I-\P)^\dagger$ and $(\P-\Pss)^\dagger$ acting on observables related to the master operator eigenmodes with $k> m$ and $m_\text{ss}<k\leq m$, respectively [that is, $L_k$ for a Hermitian (generalised) eigenmode or $L_k(\phi)$ in Eq.~\eqref{eq:L_k_phi} for a non-Hermitian (generalised) eigenmode].

\subsubsection{Separation in master operator spectrum}
We now derive Eq.~\eqref{eq:spectral_P}. 
Note that, analogously to Eq.~\eqref{eq:change_spectral},
we have [cf.~Eq.~\eqref{eq:lambda_spectral}]
\begin{subequations}\label{eq:change_spectral_P}
	\begin{align}
		|e^{t \lambda_k}-1|&\leq   \lVert e^{t\L}-\P\rVert,
		&\quad k\leq m,\\
		|e^{t\lambda_k}|&\leq   \lVert e^{t\L}-\P\rVert,
		&\quad k>m.
	\end{align}
\end{subequations}
Therefore, from Eq.~\eqref{eq:C_P} we obtain [cf.~Eq.~\eqref{eq:meta_lambda}]
\begin{subequations}\label{eq:meta_lambda_P}
	\begin{align}
		\label{eq:meta_lambda'_P}
		t' (-\lambda_k^R) &\leq -\ln[1-C_\P(t'',t')],
		&\quad k\leq m,\\
		\label{eq:meta_lambda''_P}
		t'' (-\lambda_k^R) &\geq  - \ln[C_\P(t'',t')],
		&\quad k>m,
	\end{align}
\end{subequations}
so that  Eq.~\eqref{eq:spectral_P} follows. We also obtain [cf.~Eq.~\eqref{eq:meta_lambda'''}]
\begin{eqnarray}
	\label{eq:meta_lambda'''_P}
	(t'-t'') |\lambda_k^I| &\leq& \arcsin[\C_\P(t'',t')]
	\quad k\leq m.\qquad
\end{eqnarray}

We note that from Eqs.~\eqref{eq:meta_lambda_P} and~\eqref{eq:meta_lambda'''_P} we specify the conditions in Eqs.~\eqref{eq:ratio_R} and~\eqref{eq:ratio_I} as  [cf.~Eq.~\eqref{eq:separation}]
\begin{subequations}\label{eq:separation_P}
	\begin{align}
		\label{eq:separation_R_P}
		\frac{\lambda_{m}^R}{\lambda_{m+1}^R}  &\leq  \frac{t''}{t'} \frac{\C_\P(t'',t')}{-\ln[\C_\P(t'',t')]}(1\!+\!...),\quad\\
		\label{eq:separation_I_P}
		\frac{|\lambda_{k}^I|}{-\lambda_{m+1}^R}  &\leq \frac{t''}{t'-t''}\frac{\C_\P(t'',t')}{-\ln[\C_\P(t'',t')]}(1\!+\!...),\quad k\leq m,\qquad\quad
	\end{align}
\end{subequations}
where we expanded in  $\C_\P(t'',t')$ using Eq.~\eqref{eq:C_P2}.
In particular, from Eq.~\eqref{eq:separation_R_P} we observe that when Eq.~\eqref{eq:C_P2} holds, the real parts of eigenvalues with $k\leq m$ are indeed negligible as $t''/t'<1$. The imaginary parts, however,  are implied by Eq.~\eqref{eq:separation_I_P} to be negligible only  when ${t''}/({t'-t''})\ll -\ln[\C_\P(t'',t')]/\C_\P(t'',t')$ [which is true, e.g., $[{t''}/({t'-t''})]\C_\P(t'',t')\ll 1$].

Finally, we note that since $t''/t'<1$,  we obtain from  Eq.~\eqref{eq:meta_lambda_P} that
\begin{equation}
	\frac{\ln[1-C_\P(t'',t')]}{\ln[C_\P(t'',t')]}\geq \frac{\lambda_{m}^R}{\lambda_{m+1}^R}
\end{equation}
and thus $C_\P(t'',t')$  can be bounded from below in terms of the separation of the eigenvalues in Eq.~\eqref{eq:ratio_R}.

\subsubsection{Dynamics of eigenmodes}
Here, we derive Eqs.~\eqref{C_P_P} and~\eqref{C_P_IP} using 
\begin{equation}\label{eq:Pnorm}
	\lVert \P\rVert\leq 1+\C_\P(t'',t'),
\end{equation}
which follows from Eq.~\eqref{eq:C_P} by the triangle inequality.\\

\emph{Derivation of Eq.~\eqref{C_P_P}}. We have
\begin{eqnarray}
	\lVert \P(e^{t\L}-\I)\rVert  = 	\lVert \P(e^{t\L}-\P)\rVert  
	&\leq& \lVert \P\rVert \lVert  e^{t\L}-\P\rVert.\qquad
\end{eqnarray}
Equation~\eqref{C_P_P} follows from Eqs.~\eqref{eq:C_P} and~\eqref{eq:Pnorm}.\\

\emph{Derivation of Eq.~\eqref{C_P_IP}}. We have
\begin{eqnarray}
	\lVert (\I-\P) e^{t\L}\rVert &=& 	\lVert (\I-\P) (e^{t\L}-\P)\rVert \\\nonumber
	&\leq& \lVert \I-\P\rVert \lVert  e^{t\L}-\P\rVert \\\nonumber
	&\leq& (1+\lVert \P\rVert)\, \lVert  e^{t\L}-\P\rVert.
\end{eqnarray}
Equation~\eqref{C_P_IP} follows from Eqs.~\eqref{eq:C_P} and~\eqref{eq:Pnorm}.

\subsection{Spectral theory from operational approach}\label{app:spectral_proof}

Here, we derive bounds on the dynamics restricted to (generalised) eigenmodes corresponding to a projection on Jordan blocks of the master operator, that is,  $\P=\P^2$ and $[\P,\L]=0$, in analogy to bounds in Eqs.~\eqref{eq:all} for the projection $\I$ on all (generalised) eigenmodes and Eq.~\eqref{eq:ss_IP} for the projection $\Pss$ on the stationary eigenmodes. In particular, the results here are applicable to  the projection on the slow (generalised) eigenmodes in the master operator spectrum defined in Eq.~\eqref{eq:rho_t_P}. 

\subsubsection{Changes in dynamics}

We begin by deriving the following bounds on changes in the dynamics. 
First, similarly to Eq.~\eqref{eq:change_all}, we have
\begin{eqnarray}\label{eq:change3_all}
	&& \lVert \P[e^{(t_1-t_2)\L}\!- \I] \rVert [1- \rVert\P[ e^{(t_2+t_3)\L}\!-\I]\rVert]\qquad\\\nonumber
	&& \leq	\lVert \P e^{t_3\L} \rVert\lVert e^{t_1\L}\! -e^{t_2\L}\rVert.
\end{eqnarray}
Second, similarly to Eq.~\eqref{eq:change_ss},  we have
\begin{eqnarray}\label{eq:change3_ss}
	&&\lVert (\I-\P) e^{(t_2+t_3)\L}\rVert\big[1-\lVert (\I-\P) e^{(t_1-t_2)\L}\rVert\big]\qquad	\\\nonumber
	&&\leq  \Vert (\I-\P) e^{t_3\L}\rVert \lVert e^{t_1\L}-e^{t_2\L}\rVert.
\end{eqnarray}\\

\emph{Derivation of Eqs.~\eqref{eq:change3_all}}. We have
\begin{eqnarray}
	&& \lVert \P[e^{(t_1-t_2)\L}\!- \I] \rVert -\lVert \P e^{t_3\L} \rVert\lVert e^{t_1\L}\! -e^{t_2\L} \rVert\\\nonumber
	&&\leq\lVert  \P[e^{(t_1-t_2)\L}\!- \I] \rVert -\lVert \P [e^{(t_1+t_3)\L}-e^{(t_2+t_3)\L}] \rVert
	\\\nonumber
	&&\leq\lVert  \P [e^{(t_1+t_3)\L}-e^{(t_2+t_3)\L}-e^{(t_1-t_2)\L}\!+ \I]\rVert
	\\\nonumber
	&&=\lVert  \P[e^{(t_1-t_2)\L}\!-\I][ e^{(t_2+t_3)\L}\!-\I] \rVert
	\\\nonumber
	&&\leq 	\lVert \P[e^{(t_1-t_2)\L}\!-\I]\lVert\, \rVert\P[ e^{(t_2+t_3)\L}\!-\I]\rVert
\end{eqnarray}
and Eq.~\eqref{eq:change3_all} follows by rearranging terms. \\

\emph{Derivation of Eq.~\eqref{eq:change3_ss}}. We have
\begin{eqnarray}
	&&\lVert (\I-\P) e^{(t_2+t_3)\L}\rVert\big[1-\lVert (\I-\P) e^{(t_1-t_2)\L}\rVert\big]\qquad	\\\nonumber
	&&\leq  \lVert (\I-\P)e^{(t_2+t_3)\L}]\rVert-\lVert (\I-\P)e^{(t_1+t_3)\L}\rVert\\\nonumber
	&&\leq  \lVert (\I-\P) [e^{(t_1+t_3)\L}-e^{(t_2+t_3)\L}]\rVert\\\nonumber
	&&\leq  \Vert (\I-\P) e^{t_3\L}\rVert \lVert e^{t_1\L}-e^{t_2\L}\rVert.
\end{eqnarray}

\subsubsection{Approximation of eigenmodes by dynamics in final regime}

From Eq.~\eqref{eq:change3_ss} we obtain ($t_1=3t$ and $t_2=t_3=t$)
\begin{eqnarray}\label{eq:change4_ss} 
	&&\lVert (\I-\P) e^{2t \L}\rVert\big[1-\lVert (\I-\P) e^{2t\L}\rVert\big]	\\\nonumber
	&&\leq  \Vert (\I-\P) e^{t\L}\rVert \lVert e^{t\L}-e^{3t\L}\rVert.
\end{eqnarray}

We now assume that  the metastable regime fulfils $t'\geq 3t''$ [cf.~Eq.~\eqref{eq:meta_cond}]. Similarly to Eq.~\eqref{eq:ss_IP}, we can show that
\begin{subequations}\label{eq:IP} 
	\begin{align}
		\nonumber
		\lVert (\I-\P) e^{t''\L}\rVert &\geq	\sqrt{E_+[C_\Delta(t'',t')]}
		\\\label{eq:IP2}
		&=1- \frac{C_\Delta(t'',t')}{2}+...\quad\text{or}
		\\	\nonumber
		\lVert (\I-\P) e^{2t''\L}\rVert&\leq  E_-\{C_\Delta(t'',t')\sqrt{E_+[C_\Delta(t'',t')]}\}\\
		\label{eq:IP1}
		&= \Cd(t'',t')+...
	\end{align}
\end{subequations}
for  $C_\Delta(t'',t')<1/4$.
Indeed, consider the case when $ \Vert (\I-\P) e^{t''\L}\rVert^2\leq E_+ [\Cd(t'',t')]\leq 1$, that is Eq.~\eqref{eq:IP2} does not hold. 
From Eq.~\eqref{eq:change4_ss}  we then obtain Eq.~\eqref{eq:IP1}  in analogy to Eq.~\eqref{eq:ss_IP-}; note that $\lVert (\I-\P) e^{2t''\L}\rVert\geq  E_+[C_\Delta(t'',t')E_+[\sqrt{C_\Delta(t'',t')}]\geq E_+[C_\Delta(t'',t')]$ would be in contradiction with the assumption as $\lVert (\I-\P) e^{2t\L}\rVert\leq \lVert (\I-\P) e^{t\L}\rVert^2$.

Finally, the condition in Eq.~\eqref{eq:meta_cond4} follows from Eq.~\eqref{eq:IP2} by noting that $\sqrt{E_+[C_\Delta(t'',t')]}> E_+[C_\Delta(t'',t')]$ for $C_\Delta(t'',t')>0$.

\subsubsection{Approximation of eigenmodes by dynamics in initial regime}

From Eq.~\eqref{eq:change3_all} we obtain ($t_1=3t$ and $t_2=t_3=t$)
\begin{eqnarray}\label{eq:change4_all} 
	&&\lVert \P (e^{2t \L}-\I)\rVert\big[1-\lVert \P (e^{2t\L}-\I)\rVert\big]	\\\nonumber
	&&\leq  \Vert \P e^{t\L}\rVert \lVert e^{t\L}-e^{3t\L}\rVert.
\end{eqnarray}

We again consider the metastable regime with $t'\geq 3t''$ and assume that $ \Vert (\I-\P) e^{t''\L}\rVert^2\leq E_+ [\Cd(t'',t')]$. 
By the triangle inequality, we then have	$	\lVert \P e^{t''\L}\rVert\leq  \lVert e^{t''\L}\rVert+ \lVert (\I-\P) e^{t''\L}\rVert		\leq 1+\sqrt{E_{+}[\Cd(t'',t')]}\leq 2$.
Thus, analogously to Eq.~\eqref{eq:all}, we obtain
\begin{subequations}\label{eq:P} 
	\begin{align}
		\nonumber
		\lVert \P( e^{2t''\L}-\I)\rVert &\geq	E_+(C_\Delta(t'',t')\{1+\sqrt{E_{+}[\Cd(t'',t')]}\})\\
		\label{eq:P2}
		&=1-2C_\Delta(t'',t')+... \quad\text{or}
		\\ 	\nonumber
		\lVert\P (e^{2t''\L}-\I)\rVert&\leq  E_-(C_\Delta(t'',t')\{1+\sqrt{E_{+}[\Cd(t'',t')]}\})\\
		\label{eq:P1}
		&=2C_\Delta(t'',t')+...
	\end{align}
\end{subequations}
for $\Cd(t'',t')\leq0.130...$.

Finally, the condition in Eq.~\eqref{eq:meta_cond3} follows from Eq.~\eqref{eq:P2} by noting that $E_+(C_\Delta(t'',t')\{1+\sqrt{E_{+}[\Cd(t'',t')]}\})> E_+[2C_\Delta(t'',t')]$ for  $C_\Delta(t'',t')>0$.

\subsubsection{Approximation by spectral theory of metastability}

We now discuss the case when both Eqs.~\eqref{eq:IP1} and~\eqref{eq:P1} hold. Beforehand, we note that as for a (generalised) left eigenmode $L_k$ corresponding to an eigenvalue $\lambda_k$ we have [cf.~Eq.~\eqref{eq:change_spectral_P}]
\begin{subequations}\label{eq:change_spectral_P2}
	\begin{align}
		|\Tr[\P^\dagger(L_k)]| |e^{t \lambda_k}-1|&\leq   \lVert \P(e^{t\L}-\I)\rVert,\\
		|\Tr[(\I-\P^\dagger)(L_k)]| |e^{t\lambda_k}|&\leq   \lVert e^{t\L}-\P\rVert.
	\end{align}
\end{subequations}
Equation~\eqref{eq:IP1} implies that if $\P^\dagger(L_k)=L_k$, it follows that $k\leq m$, while Eq.~\eqref{eq:P1}  ensures that if $\P^\dagger(L_k)=0$, $k>m$. Thus, $\P$ is determined as the projection on the (generalised) eigenmodes with $k\leq m$, where $m$ is given in Eq.~\eqref{eq:separation}. Therefore, this case corresponds to the spectral theory of metastability [cf.~Eqs.~\eqref{eq:ratio_R}-\eqref{eq:rho_t_P}].\\

From Eqs.~\eqref{eq:IP1} and~\eqref{eq:P1}, it follows that
\begin{eqnarray}\label{eq:P4}
	\lVert \P \rVert&\leq&\lVert \P e^{2t''\L}\rVert+\lVert \P (e^{2t''\L}-\I)\rVert\\\nonumber
	&\leq&   1+E_{-}\{\Cd(t'',t')\sqrt{E_+[C_\Delta(t'',t')]}\}
	\\\nonumber&&+E_-(C_\Delta(t'',t')\{1+\sqrt{E_{+}[\Cd(t'',t')]}\})\\\nonumber
	&=&1+3\Cd(t'',t')+....\leq 2,\quad
\end{eqnarray}
while $\lVert \I-\P \rVert\leq \lVert \I\rVert+ \lVert \P \rVert \leq 1+ \lVert \P \rVert \leq 3$. This gives  Eq.~\eqref{eq:Pnorm2}.
Furthermore, from Eq.~\eqref{eq:change3_all} ($t_1=2t$, $t_2=t$, $t_3=0$)
\begin{eqnarray}\label{eq:change5_all} 
	&&\lVert \P (e^{t \L}-\I)\rVert\big[1-\lVert \P (e^{t\L}-\I)\rVert\big]	\\\nonumber
	&&\leq  \Vert \P \rVert \lVert e^{t\L}-e^{2t\L}\rVert
\end{eqnarray}
and from Eq.~\eqref{eq:change3_ss} ($t_1=2t$, $t_2=t$, $t_3=0$)
\begin{eqnarray}\label{eq:change5_ss} 
	&&\lVert (\I-\P) e^{t \L}\rVert\big[1-\lVert (\I-\P) e^{t\L}\rVert\big]	\\\nonumber
	&&\leq  \Vert \I-\P\rVert \lVert e^{t\L}-e^{2t\L}\rVert.
\end{eqnarray}

We now consider the metastable regime with $t'\geq 4t''$. In analogy to Eq.~\eqref{eq:all} from Eq.~\eqref{eq:change5_all} we obtain for $t''\leq t\leq t'/2$ 
\begin{subequations}\label{eq:P+-} 
	\begin{align}
		\label{eq:P+}	
		&\lVert  \P(e^{t\L}-\I)\rVert \geq	E_+[ \Vert \P\rVert \Cd(t'',t')]
		\quad\text{or}
		\\
		\label{eq:P-}
		&\lVert   \P(e^{t\L}-\I)\rVert\leq  E_-[\Vert \P\rVert\Cd(t'',t')],
	\end{align}
\end{subequations}
while for $t'/2< t\leq t'$, $E_\pm[ \Vert \P\rVert \Cd(t'',t')]$ is replaced by $E_\pm[ \Vert \P\rVert \Cd(t'',t')]\mp\Vert \P\rVert \Cd(t'',t')$. We note however that Eq.~\eqref{eq:P+}	contradicts Eq.~\eqref{eq:P1} by considering $t=2t''\leq t'/2$, and thus Eq.~\eqref{eq:P-} holds [cf.~Eq.~\eqref{eq:P_better}]; from Eq.~\eqref{eq:P4}, the bounds are distinct for $\Cd(t'',t')<0.129...$.
Furthermore, from Eq.~\eqref{eq:change5_ss}, 	analogously to Eq.~\eqref{eq:ss_IP}, we obtain    for $t''\leq t\leq t'/2$ 
\begin{subequations}\label{eq:IP+-}
	\begin{align}
		\label{eq:IP+}
		&\lVert (\I-\P)e^{t\L}\rVert \geq	E_+[ \Vert \I-\P\rVert C_\Delta(t'',t')]
		\quad\text{or}
		\\	\label{eq:IP-}
		&\lVert   (\I-\P)e^{t\L}\rVert\leq  E_-[ \Vert \I-\P\rVert C_\Delta(t'',t')],
	\end{align}
\end{subequations}
while for $t'/2< t\leq t'$, $E_+[ \Vert \I-\P\rVert \Cd(t'',t')]$ is replaced by $E_+[ \Vert \I-\P\rVert \Cd(t'',t')]-\Vert\I- \P\rVert \Cd(t'',t')$; from Eq.~\eqref{eq:P4}, the bounds are distinct for $\Cd(t'',t')<0.0997...$. Similarly, as above, however, Eq.~\eqref {eq:IP+} contradicts Eq.~\eqref{eq:IP1} by considering $t=2t''\leq t'/2$, and thus Eq.~\eqref {eq:IP-} holds [cf.~Eq.~\eqref{eq:IP_better}].

Finally, from  Eqs.~\eqref{eq:P-} and~\eqref{eq:IP-}  we have
\begin{eqnarray}\label{eq:all_better}
	\lVert e^{t\L}\! -\!\P\rVert\!&\leq&\! E_-[\lVert \I\!-\!\P\rVert\Cd\!(t''\!,t')]\!+\!\!E_+[\lVert \P\rVert\Cd\!(t''\!,t')]\qquad\,\,\,
\end{eqnarray}
for $t''\leq t\leq t'/2$, while for $t'< t\leq t'$, the bound  increases by $\Cd(t'',t')$. Thus, we arrive at Eq.~\eqref{eq:C_P_better}.

\subsubsection{Linear corrections}
\label{app:lin}

We now prove that Eq.~\eqref{eq:P_lin3} follows from Eq.~\eqref{eq:P_better}. 
To this aim, we derive the following bound [cf.~Eq.~\eqref{eq:all_lin}]
\begin{equation}
	\label{eq:P_lin}
	\lVert \P[e^{t\L}\!-e^{(t+\delta t)\L}]\rVert\!\geq\! [2-\!\lVert \P(e^{t\L}\!-\!\I)\rVert]\delta t \lVert \P\L \rVert - e^{\delta t\lVert \P\L\rVert }\!+1.
\end{equation}
Therefore, for $t''\leq t\leq t'/2$ from Eq.~\eqref{eq:P_better}  we have [cf.~Eq.~\eqref{eq:all_lin2}]
\begin{equation}
	\label{eq:P_lin2}
	\lVert \P [e^{t\L}\!-\! e^{(t+\delta t)\L}]\rVert\geq \frac{3}{2}\,\delta t \lVert \P\L \rVert - e^{\delta t\lVert \P\L\rVert }+1,
\end{equation}
where the left-hand side is bounded from above by $ \lVert \P\rVert \lVert e^{t\L}\!-\! e^{(t+\delta t)\L}\rVert$. In particular, for $t=t''$ and $\delta t\leq t'-t''$ we obtain 
\begin{equation}
	\label{eq:P_lin4}
	\lVert \P\rVert\,\Cd(t'',t')\geq \frac{3}{2}\,\delta t \lVert \P\L \rVert - e^{\delta t\lVert \P\L\rVert }+1.
\end{equation}
Equation~\eqref{eq:P_lin3} follows for  $\lVert \P\rVert\,\Cd(t'',t')\leq [3\ln(3/2)-1]/2=0.108...$, which is guaranteed from Eq.~\eqref{eq:P4} for $\Cd(t'',t')\leq 0.0837...$  [cf.~Eq.~\eqref{eq:all_lin3}].\\

\emph{Derivation of Eq.~\eqref{eq:P_lin}}.
From the triangle inequality
\begin{eqnarray}
	\lVert \P[e^{t\L}\!\!-\! e^{(t+\delta t)\L}]\rVert
	&\geq& \delta t \lVert \P\L e^{t\L}\rVert \\\nonumber
	&&- \lVert \P(e^{\delta t\L}\!\!-\!\I \!-\! \delta t\L)e^{t\L} \rVert,\,\, \qquad
\end{eqnarray}
with
\begin{eqnarray}
	\delta t \lVert\P \L e^{t\L}\rVert&\geq& \delta t \lVert \P\L \rVert \!-\!\delta t\lVert\P\L (e^{t\L}\!-\!\I)\rVert \\\nonumber
	&\geq&[1\!-\!\lVert\P (e^{t\L}\!-\!\I)\rVert] \delta t \lVert \P\L \rVert .
\end{eqnarray}
Furthermore, from the contractivity of the dynamics
\begin{eqnarray}
	\lVert \P(e^{\delta t\L}\!-\!\I \!-\! \delta t)e^{t\L} \rVert &\leq&\lVert \P(e^{\delta t\L}\!-\!\I \!-\! \delta t\L)\rVert
\end{eqnarray}
and  by considering  the series for $\P e^{\delta t\L}$,
\begin{eqnarray}
	\lVert \P(e^{\delta t\L}\!-\!\I \!-\! \delta t\L )\rVert
	&\leq & e^{\delta t\lVert\P \L\rVert }-1-\delta t\lVert \P\L\rVert.\qquad
\end{eqnarray}

\end{appendix}

\end{document}

%% file: MetastabilityOperational_arXiv2.bbl
%